\PassOptionsToPackage{unicode}{hyperref}
\PassOptionsToPackage{hyphens}{url}
\PassOptionsToPackage{dvipsnames,svgnames,x11names}{xcolor}
\documentclass[
  12pt]{article}

\usepackage{amssymb}
\usepackage{mathrsfs}
\usepackage{iftex}
\usepackage{algpseudocode}
\usepackage{dsfont} 
\usepackage{graphicx,psfrag,epsf}
\usepackage{enumerate}
\usepackage{array}
\usepackage{url} 
\usepackage{longtable}
\usepackage{array} 
\usepackage{wrapfig}
\usepackage{amsmath}
\usepackage{tcolorbox}
\usepackage{adjustbox}
\usepackage{rotating} 
\usepackage{longtable}
\usepackage{subcaption}
\usepackage{bm}
\usepackage{hyperref}
\usepackage{arydshln}
\usepackage{algorithm}
\usepackage{enumitem}
\usepackage{caption}
\usepackage{xr}
\usepackage{wrapfig}
\usepackage[normalem]{ulem}

\newcommand{\vvec}{\mathrm{vec}}

\providecommand{\norm}[1]{\|#1\|}

\usepackage{booktabs}
\usepackage{multirow}
\newtheorem{assumption}{Assumption}

\let\hat\widehat
\let\tilde\widetilde

\newcommand{\EE}{\mathbb{E}}

\newcommand{\RR}{\mathbb{R}}

\def\T{{ \intercal }}

\ifx\BlackBox\undefined
\newcommand{\BlackBox}{\rule{1.5ex}{1.5ex}}  
\fi

\ifx\QED\undefined
\def\QED{~\rule[-1pt]{5pt}{5pt}\par\medskip}
\fi

\ifx\proof\undefined
\newenvironment{proof}{\par\noindent{\bf Proof\ }}{\hfill\BlackBox\\[2mm]}
\fi

\ifx\theorem\undefined
\newtheorem{theorem}{Theorem}
\fi
\ifx\example\undefined
\newtheorem{example}{Example}
\fi
\ifx\property\undefined
\newtheorem{property}{Property}
\fi
\ifx\lemma\undefined
\newtheorem{lemma}{Lemma}
\fi
\ifx\proposition\undefined
\newtheorem{proposition}{Proposition}
\fi
\ifx\remark\undefined
\newtheorem{remark}{Remark}
\fi
\ifx\corollary\undefined
\newtheorem{corollary}{Corollary}
\fi
\ifx\definition\undefined
\newtheorem{definition}{Definition}
\fi
\ifx\conjecture\undefined
\newtheorem{conjecture}{Conjecture}
\fi
\ifx\fact\undefined
\newtheorem{fact}{Fact}
\fi
\ifx\claim\undefined
\newtheorem{claim}{Claim}
\fi
\ifx\cond\undefined
\newtheorem{cond}{Condition}
\fi

\ifPDFTeX
  \usepackage[T1]{fontenc}
  \usepackage[utf8]{inputenc}
  \usepackage{textcomp} 
\else 
  \usepackage{unicode-math}
  \defaultfontfeatures{Scale=MatchLowercase}
  \defaultfontfeatures[\rmfamily]{Ligatures=TeX,Scale=1}
\fi
\usepackage{lmodern}
\ifPDFTeX\else  
\fi
\IfFileExists{upquote.sty}{\usepackage{upquote}}{}
\IfFileExists{microtype.sty}{
  \usepackage[]{microtype}
  \UseMicrotypeSet[protrusion]{basicmath} 
}{}
\makeatletter
\@ifundefined{KOMAClassName}{
  \IfFileExists{parskip.sty}{%
    \usepackage{parskip}
  }{
    \setlength{\parindent}{2pt}
    \setlength{\parskip}{6pt plus 2pt minus 1pt}}
}{
  \KOMAoptions{parskip=half}}
\makeatother
\usepackage{xcolor}
\makeatletter
\ifx\paragraph\undefined\else
  \let\oldparagraph\paragraph
  \renewcommand{\paragraph}{
    \@ifstar
      \xxxParagraphStar
      \xxxParagraphNoStar
  }
  \newcommand{\xxxParagraphStar}[1]{\oldparagraph*{#1}\mbox{}}
  \newcommand{\xxxParagraphNoStar}[1]{\oldparagraph{#1}\mbox{}}
\fi
\ifx\subparagraph\undefined\else
  \let\oldsubparagraph\subparagraph
  \renewcommand{\subparagraph}{
    \@ifstar
      \xxxSubParagraphStar
      \xxxSubParagraphNoStar
  }
  \newcommand{\xxxSubParagraphStar}[1]{\oldsubparagraph*{#1}\mbox{}}
  \newcommand{\xxxSubParagraphNoStar}[1]{\oldsubparagraph{#1}\mbox{}}
\fi
\makeatother

\usepackage{longtable,booktabs,array}
\usepackage{calc} 
\usepackage{etoolbox}
\makeatletter
\patchcmd\longtable{\par}{\if@noskipsec\mbox{}\fi\par}{}{}
\makeatother
\IfFileExists{footnotehyper.sty}{\usepackage{footnotehyper}}{\usepackage{footnote}}
\makesavenoteenv{longtable}
\usepackage{graphicx}
\makeatletter
\def\maxwidth{\ifdim\Gin@nat@width>\linewidth\linewidth\else\Gin@nat@width\fi}
\def\maxheight{\ifdim\Gin@nat@height>\textheight\textheight\else\Gin@nat@height\fi}
\makeatother
\setkeys{Gin}{width=\maxwidth,height=\maxheight,keepaspectratio}
\makeatletter
\def\fps@figure{htbp}
\makeatother

\makeatletter
\@ifpackageloaded{caption}{}{\usepackage{caption}}
\AtBeginDocument{%
\ifdefined\contentsname
  \renewcommand*\contentsname{Table of contents}
\else
  \newcommand\contentsname{Table of contents}
\fi
\ifdefined\listfigurename
  \renewcommand*\listfigurename{List of Figures}
\else
  \newcommand\listfigurename{List of Figures}
\fi
\ifdefined\listtablename
  \renewcommand*\listtablename{List of Tables}
\else
  \newcommand\listtablename{List of Tables}
\fi
\ifdefined\figurename
  \renewcommand*\figurename{Figure}
\else
  \newcommand\figurename{Figure}
\fi
\ifdefined\tablename
  \renewcommand*\tablename{Table}
\else
  \newcommand\tablename{Table}
\fi
}
\@ifpackageloaded{float}{}{\usepackage{float}}
\floatstyle{ruled}
\@ifundefined{c@chapter}{\newfloat{codelisting}{h}{lop}}{\newfloat{codelisting}{h}{lop}[chapter]}
\floatname{codelisting}{Listing}

\makeatother
\makeatletter
\@ifpackageloaded{caption}{}{\usepackage{caption}}
\@ifpackageloaded{subcaption}{}{\usepackage{subcaption}}
\makeatother

\ifLuaTeX
  \usepackage{selnolig}  
\fi
\usepackage[]{natbib}
\usepackage{bookmark}

\IfFileExists{xurl.sty}{\usepackage{xurl}}{} 
\hypersetup{
  pdftitle={Title},
  pdfauthor={Author 1; Author 2},
  pdfkeywords={3 to 6 keywords, that do not appear in the title},
  colorlinks=true,
  linkcolor={blue},
  filecolor={Maroon},
  citecolor={Black},
  urlcolor={Blue},
  pdfcreator={LaTeX via pandoc}}

\begin{document}

\def\spacingset#1{\renewcommand{\baselinestretch}%
{#1}\small\normalsize} \spacingset{1.2}

\title{A Latent Variable Framework for Scaling Laws \\in Large Language Models}

\author{Peiyao Cai\textsuperscript{1}\footnote{Co-first authors (alphabetical order).}, Chengyu Cui\textsuperscript{1}$^\ast$, Felipe Maia Polo\textsuperscript{1}$^\ast$, Seamus Somerstep\textsuperscript{1}$^\ast$\\Leshem Choshen\textsuperscript{2}, Mikhail Yurochkin\textsuperscript{3},\\Yuekai Sun\textsuperscript{1}, Kean Ming Tan\textsuperscript{1} and Gongjun Xu\textsuperscript{1}\vspace{0.5cm}\\\textsuperscript{1}{\it \small Department of Statistics, University of Michigan}\\\textsuperscript{2}{\it \small IBM Research and CSAIL, MIT}\\\textsuperscript{3}{\it \small Institute of Foundation Models, MBZUAI}}

    \date{}
  \maketitle











\begin{abstract}
We propose a statistical framework built on latent variable modeling for scaling laws of large language models (LLMs). 
Our work is motivated by the rapid emergence of numerous new LLM families with distinct architectures and training strategies, evaluated on an increasing number of benchmarks. This heterogeneity makes a single global scaling curve inadequate for capturing how performance varies across families and benchmarks.
To address this, we propose a latent variable modeling framework in which each LLM family is associated with a latent variable that captures the common underlying features in that family. An LLM’s performance on different benchmarks is then driven by its latent skills, which are jointly determined by the latent variable and the model’s own observable features.
We develop an estimation procedure for this latent variable model and establish its statistical properties. We also design efficient numerical algorithms that support estimation and various downstream tasks. Empirically, we evaluate the approach on 12 widely used benchmarks from the Open LLM Leaderboard (v1/v2).
\end{abstract}



\section{Introduction}\label{sec_intro}
Large language models (LLMs) now deliver state-of-the-art performance across tasks such as summarization, translation, question answering, and instruction following. This progress has been accompanied by an expanding ecosystem of benchmarks designed to probe different aspects of model capability. As model families proliferate, benchmark designers face increasing challenges in understanding how complex, multifaceted capabilities evolve with scale. At the same time, 
developers must decide how to allocate resources across model and data sizes to meet performance requirements across these criteria. Together, these trends have created a growing demand for principled LLM performance evaluation.
\subsection{Motivating Application and Dataset}
{ One concrete example is the Open LLM Leaderboard~\citep{beeching2023open}, a public platform for evaluating and comparing open LLMs. The leaderboard was first introduced to make scores reproducible and comparable across models by requiring submissions to be evaluated on the same questions in the same order, thereby reducing dependence on private prompts or hand-tuned evaluation setups. This original version, which we refer to as Leaderboard v1, became a widely used resource for identifying open-source LLM releases and benchmarking new pretraining or fine-tuning methods. As LLMs continued to improve, however, their performance began to saturate the original benchmark suite, calling for harder evaluations that better distinguish model capabilities. This motivated the current version, Leaderboard v2, which incorporates more challenging benchmarks covering instruction following, complex reasoning, mathematics, expert-level knowledge, long-context multistep reasoning, and robust multitask understanding. The resulting dataset records heterogeneous model performances across correlated but distinct capabilities, motivating more flexible scaling-law models that account for both family-level differences and benchmark-specific structure.}


Scaling laws are a popular framework for predicting LLM performance as a function of key resources, typically the number of parameters in the language model and the amount of training data. Early work by \citet{rosenfeld2019constructivepredictiongeneralizationerror} uncovers power-law relations between loss and architectural or data-related factors. \citet{kaplan2020scalinglawsneurallanguage} show that LLM validation loss follows a predictable power-law trend in both parameter count and training tokens, enabling principled compute allocation. This line of research led to the Chinchilla scaling laws of \citet{hoffmann2022training}, which reveal that optimal performance under fixed compute arises when model size and data scale proportionally. 

Formally, scaling laws posit a parametric relationship between model size $s$, training tokens $t$, and a performance metric $L(s,t)$ such as validation loss or perplexity. A common formulation \citep{hoffmann2022training,choshenhitchhiker} is
\begin{equation}\label{eq_scaling_base}
  L(s,t)\approx E+\frac{A}{s^{\alpha}}+\frac{B}{t^{\beta}},  
\end{equation}
with compute, number of floating point operations (FLOPs), often approximated by $6st$ for transformer models~\citep{kaplan2020scalinglawsneurallanguage} and $A=e^a,B=e^b$ to ensure positivity. Variants of this scaling law have been validated across several settings and remain a standard tool for forecasting performance and planning compute budgets \citep{kaplan2020scalinglawsneurallanguage,owen2024predictable,gadre2024languagemodelsscalereliably,choshenhitchhiker}.

These classical formulations, however, do not match the realities of contemporary LLM development. They focus almost exclusively on validation loss rather than on downstream performance across benchmarks that actually measure capabilities of interest. Meanwhile, hundreds of open models now exist, spanning diverse architectures, training pipelines, and design choices, and benchmarks probe distinct, yet correlated, facets of ability. In this setting, the simple power law above cannot capture the joint dependence structure induced by heterogeneous model families, multi-dimensional capabilities, and task-level performance. \citet{ruan2024observational} attempt to address this by embedding performance into a low-dimensional capability space and fitting family-specific scaling laws. But the approach assigns a separate scaling coefficient to each family–benchmark pair, producing a high-dimensional and difficult-to-interpret model; this formulation leads to poor fit and weak prediction performance. \citet{polo2024sloth} propose a different latent-skill framework with structural sharing across benchmarks, yet no consistency guarantees are provided for the resulting estimators, which limits their use in downstream statistical analysis. { Due to space limitations, we provide additional related work in Appendix A of the online supplementary material.}

\subsection{Our Contribution}
To overcome the limitation, we develop a latent-variable statistical model that captures the scaling behavior across heterogeneous model families and benchmarks. We assume that the performance across benchmarks is driven by a low-dimensional ability vector that scales predictably with LLM's observable features while allowing for heterogeneity across LLM families. This heterogeneity reflects systematic differences in architectures and training pipelines, which we capture through family-specific latent variables.
This construction yields a parsimonious representation of capability scaling and supports a coherent evaluation across heterogeneous benchmarks. Each dimension of the latent skill vector is intended to correspond to a specific, interpretable capability of practical interest, such as mathematical reasoning, instruction following, or logical reasoning. Each benchmark is embedded into the same latent space, with its loadings parameters indicating how performance on that benchmark depends on the corresponding latent abilities.

Within this framework, we introduce a likelihood-based estimator for the latent variable model with statistical guarantees. Under mild conditions, we establish its estimation consistency and asymptotic distribution. We further develop an efficient projected stochastic gradient ascent algorithm suited for large-scale datasets involving many LLM families and numerous benchmarks. We then demonstrate downstream tasks such as computing posterior distributions of the latent variables and constructing prediction intervals for the performance of unevaluated LLMs. Together, these results provide a statistically grounded framework for analyzing capability scaling in modern LLMs across different benchmarks. 

{ Using the Open LLM Leaderboard data described above, we validate our approach on benchmark performances from hundreds of LLMs. Our results show that scaling behavior is better characterized by skill-specific patterns that vary across benchmarks and model families, rather than by a single aggregate law. The estimated model supports a range of downstream analyses, including performance prediction, skill-level comparisons, and compute-optimal training strategies. Importantly, the framework provides a unified and statistically principled way to quantify relationships in heterogeneous leaderboard data, including patterns that align with existing research and empirical intuition.}



\vspace{0.2cm}

The rest of the paper is organized as follows.
Section~\ref{sec_method} presents the proposed latent-variable–based model for scaling law, describes its estimation framework, and establishes the statistical properties of the resulting estimators. 
Section~\ref{sec_computation} introduces algorithms for estimating the scaling model, the sampling procedure for estimating LLM latent abilities, and the construction of prediction intervals for LLM performance.
Section~\ref{sec_experiment} provides comprehensive empirical studies to evaluate LLM performances under the proposed framework.
Finally, we provide concluding remarks in Section~\ref{sec_conclude}. 
Additional experimental results, {simulation studies based on synthetic data}, and all proofs are provided in the online supplementary material.

\section{Methodology}\label{sec_method}

\subsection{Setup for the Latent Variable Model}\label{subsec_setup}
We begin by introducing our latent variable framework to model the scaling law for different LLMs and evaluation benchmarks. To avoid confusion with the large language models being evaluated, we refer to this statistical framework as the \textit{scaling model}.

Consider a setting of $N$ LLM families (e.g., LLaMa 3 \citep{dubey2024llama}, Qwen 2 \citep{yang2024qwen2}, etc), with each family $l$ containing $n_l$ LLM models. For each LLM $i$ in the $l$-th family, we observe its performance across $J$ benchmarks, represented by a $J$-dimensional response vector $Y_i^{(l)}$, whose components record the performance on the respective benchmarks. Existing scaling laws for LLMs predict performance based on observable features such as LLM's model size (i.e., number of parameters) and training data size (i.e., number of training tokens). However, with the rapid emergence of diverse LLM architectures, training strategies, data curation, and generation methodologies, observable features alone are no longer sufficient to accurately predict performance. Different LLMs can exhibit fundamentally different sensitivities to these features across different families. Crucially, these discrepancies arise from complex, family-specific latent features that are inherent in LLMs but not directly observed. Moreover, different benchmarks are designed to assess distinct facets of LLM ability that are difficult to capture using observable features alone.

To address the aforementioned problems, we propose to capture these hidden features using a $K$-dimensional latent variable $\alpha_l$ representing the family-wise common latent abilities of all LLMs within the family $l$; as in the field of stochastic frontier analysis \citep{kumbhakar2003stochastic}, $\alpha_l$ can be seen as the efficiency of a family in converting inputs into outputs. In this paper, $\alpha_l$ is treated as a random variable, and we use the same notation $\alpha_l$ for both the variable and its realization when the meaning is clear from context. Each dimension of $\alpha_l$ corresponds to a distinct facet of the LLM skill, such as mathematics, reasoning, and instruction. With this common latent ability that captures the family-wise effect, we assume that each $i$-th LLM in the $l$-th family has a $K$-dimensional latent ability $\theta_i^{(l)}$, given as
\begin{equation}\label{eq:model.specification}
    \theta_i^{(l)} ~=\;\alpha_l ~+~ \beta^\intercal x_i^{(l)}.
\end{equation}
We use $x_i^{(l)}\in\mathbb{R}^p$ to denote the observed features for this LLM $i$, such as size and training tokens of the LLMs,  and the parameter matrix $\beta \in \mathbb{R}^{p \times K}$ to represent the corresponding covariate effects. 
Within the same LLM family, $\beta$ characterizes the relationship between the observed features and the performance of the LLM, encoding the classical scaling law that quantifies the influence of the computational budget on performance~\citep{hoffmann2022training,choshenhitchhiker}. This parameter is therefore of substantial interest, as it provides key insights for downstream applications, such as designing optimal training strategies as in \citet{hoffmann2022training}.


In the literature, the response of LLM $i$ in the $l$-th family to the benchmark $j$, denoted by $Y_{ij}^{(l)}\in[0,1]$, is typically defined as the average of item-level correctness scores~\citep{owen2024predictable}, and thus we model it via a beta distribution. Given this $K$-dimensional latent ability $\theta_i^{(l)} = \alpha_l + \beta^\T x_i^{(l)}$, we assume that $Y_{ij}^{(l)}$ is distributed with density function
\begin{equation}{
    p\big(y\big|\alpha_l, \beta, \lambda_j, b_j, \gamma_j,\phi_j,x_i^{(l)}) \;=\; \mathrm{Beta}\Big(\phi_j\mu(\eta_{ij}^{(l)}), \phi_j\big\{1-\mu(\eta_{ij}^{(l)}) \big\}\Big),}\label{eq_model_setup}
\end{equation}
where $\mathrm{Beta}(\cdot,\cdot)$ is the density of the beta distribution and $\mu(\eta_{ij}^{(l)})$ is the conditional mean given as
\begin{equation*}
    \mu\big(\eta_{ij}^{(l)}\big) =\gamma_j + \frac{1-\gamma_j}{1+\exp\big(-\eta_{ij}^{(l)}\big)},\;\text{ with }\eta_{ij}^{(l)} = \lambda_j^\intercal \theta_i^{(l)} + b_j = \lambda_j^\T\big\{\alpha_l + \beta^\T x_i^{(l)}\big\} + b_j.
\end{equation*}
Here, $\lambda_j\in\mathbb{R}^K$ encodes the relationships between latent skills and the benchmark $j$, $\gamma_j$ is a guessing parameter reflecting the LLM’s baseline probability of correctly answering an item (e.g., by random guessing in multiple-choice settings), and $\phi_j$ controls the conditional variance, given by $\text{Var}\big(Y_{ij}^{(l)}\big | \eta_{ij}^{(l)},\phi_j\big) = {\mu(\eta_{ij}^{(l)})\big\{1-\mu(\eta_{ij}^{(l)})\big\}}/{(1+\phi_j)}$.
For simplicity, we also write the density as $p(y|\eta_{ij}^{(l)},\phi_j)$ when there is no ambiguity, i.e., 
\begin{equation*}
    \tag{\ref{eq_model_setup}$^{\prime}$}{p(y|\eta_{ij}^{(l)},\phi_j) ~=~ \mathrm{Beta}\Big(\phi_j\mu(\eta_{ij}^{(l)}), \phi_j\big\{1-\mu(\eta_{ij}^{(l)})\big\}\Big).}
\end{equation*}

Within each family, conditional on $\alpha_l$ and $\{\lambda_j,\phi_j,b_j\}_{j\in[J]}$, all benchmark responses from LLMs in that family are assumed to be mutually independent, which is a standard assumption in latent variable modeling \citep{skrondal2004generalized,reckase2009,bartholomew2011latent}. Across families, responses are also assumed to be independent given these parameters and the corresponding $\alpha_l$. { This parameterization is inspired by the beta regression framework proposed by \citet{ferrari2004beta} and can be seen as an extension that accommodates low-rank structure via latent variables.}

Throughout the paper, we treat $\gamma_j$ as fixed and known. This assumption is reasonable because these constants are typically specified in advance. For instance, for multiple-choice benchmarks such as MMLU \citep{hendrycks2020Measuring} with four answer options, the guessing parameter is conventionally taken to be $\gamma_j = 0.25$, corresponding to $100\%$ divided by the number of multiple-choice alternatives. For benchmarks with open-ended questions, $\gamma_j=0$ is assumed naturally.

\begin{remark}

Recently, there has been a surge of interest in applying psychometric approaches to evaluate large language models~\citep{demszky2023using,Sabour2024EmoBenchET}. Our work contributes to this line of research and is inspired by item response theory (IRT) models widely used in educational measurement, psychology, and behavioral science~\citep{B1981,reckase2009}. In IRT, the latent traits of the individuals, such as abilities, skills, or psychological attributes, are represented by latent variables, and their effects on responses are characterized by an item response function that also depends on item attributes.
Motivated by this perspective, we introduce a new latent variable framework designed to both the heterogeneous scaling behavior of LLMs and differences in benchmark characteristics, providing a unified scaling law framework across different families and benchmarks.
\end{remark}

\begin{remark}\label{rem2.2}
    { In our model, the LLM families should be viewed as a working grouping introduced to reflect shared LLM characteristics, rather than as a probability sample from a finite population of all possible families. Accordingly, the random variable $\alpha_l$ is introduced as a model-based tool to capture between-family heterogeneity, in the sense of a superpopulation~\citep{royall1970finite,ValliantDorfmanRoyall2000,little2004model}. Under this interpretation, the subsequent asymptotic theory is intended to measure the uncertainty under the fitted latent-variable model. A family typically consists of LLMs that share similar pre-training procedures, training data sources and architectural designs. This choice is motivated by the fact that LLMs sharing these designs may exhibit performance patterns that may be driven by similar underlying features, as captured by $\alpha_i$ in the scaling model.}
\end{remark}

\subsection{Estimation}\label{sec_marginal.estimation}

In this section, we discuss the estimation for the scaling model introduced in Section~\ref{subsec_setup}.
The common latent ability $\alpha_l$ is assumed to follow a $K$-variate zero-mean normal distribution with covariance matrix $\Sigma$, a standard assumption in latent variable modeling~\citep{reckase2009,bartholomew2011latent,chen2025item}. The density function is denoted as $\pi(x\mid \Sigma) = (2\pi)^{-K/2}\mathrm{det}(\Sigma)^{-1/2}\exp(-x^{\intercal}\Sigma^{-1}x/2)$,
where $\mathrm{det}(\cdot)$ denotes the determinant of a square matrix.
Let $$\xi := \{\vvec^\T(\Lambda), \vvec^\T(\beta), b^\T, \phi^\T, \mathrm{vec}^\T(\Sigma)\}^\T$$ be the parameters in the scaling model, where $\Lambda = (\lambda_1,\dots,\lambda_J)^\T$, $b = (b_1, \ldots, b_{{J}})^\T$, $\phi = (\phi_1, \ldots, \phi_{{J}})^\T$, and $\vvec(\cdot)$ denotes the vectorization operator, which stacks all columns of a matrix sequentially into a single vector.
Let $[J] = \{1,2,\ldots, J \}$.
The joint distribution of $Y_i^{(l)} = \{Y_{i1}^{(l)}, \ldots, Y_{i{J}}^{(l)}\}^\intercal$, with the realized value $y_i^{(l)} = \{y_{i1}^{(l)}, \ldots, y_{i{J}}^{(l)}\}^\T$, and the common ability $\alpha_l$ can be expressed as
\[
f_{l,i}\left\{y_i^{(l)},\alpha_l\big |\xi,x_i^{(l)}\right\} = \prod_{j\in[J]} p\left\{y_{ij}^{(l)}\big |\eta_{ij}^{(l)},\phi_j\right\}\pi (\alpha_l|\Sigma).
\]

We further denote the collection of responses as $Y = \{Y_{ij}^{(l)}\}_{l\in [N], i\in [n_l], j\in [J]}$, and covariates over all LLMs as ${X} = \{x_i^{(l)} \}_{l\in [N], i \in [n_l]}$, respectively.
Given the observed covariates ${X}$, the marginal log-likelihood function of $Y$ is given as
\begin{equation}
\label{eq: Likelihood}
    \mathcal{L} (Y|\xi,{X}) = \frac{1}{N}\sum_{l\in [N]}\log f_l\left\{Y^{(l)}\mid \xi,{X}^{(l)}\right\},
\end{equation}
where $Y^{(l)} = \{Y_{ij}^{(l)} \}_{i \in [n_l], j \in [J]}$, ${X}^{(l)} = \{x_i^{(l)} \}_{i \in [n_l]}$ and
\begin{align*}f_l\left\{y^{(l)}\mid \xi,{X}^{(l)}\right\}=&\,\int_{\alpha_l\in\mathbb{R}^K}\prod_{i\in [n_l]}f_{l,i}\left\{\alpha_l,y_i^{(l)}\mid\xi,x_i^{(l)}\right\}d\alpha_l\\=&\,\int_{\alpha_l\in\mathbb{R}^K}\prod_{i\in [n_l]}\prod_{j\in[J]} p\left\{y_{ij}^{(l)}\mid \eta_{ij}^{(l)},\phi_j\right\}\pi (\alpha_l|\Sigma)d\alpha_l.\end{align*}
For simplicity, we omit ${X}$ from the conditional distributions in the remainder of the paper. Directly maximizing the log-likelihood function in \eqref{eq: Likelihood} may lead to multiple solutions due to an identifiability issue. Specifically, for any invertible matrix $G$ and parameter $\xi$, define $\bar\xi = \{\vvec^\T(\Lambda G),\vvec^\T(\beta),b^\T,\phi^\T,\vvec^\T(G\Sigma G^\T)\}^\T$. It can be verified that $\mathcal{L}(Y|\xi) = \mathcal{L}(Y|\bar\xi)$ for any $Y$ and $X$. This type of identifiability issue has been a central concern in establishing model identifiability in similar settings~\citep{bai2012statistical,ouyang2024statistical}.

To address this issue, a common approach is to impose orthogonality constraints that $\Lambda^\T\Lambda$ is diagonal and $\Sigma = I_K$~\citep{fan2017sufficient,chen2019joint,zhang2020imputed}. However, this assumption typically does not lead to interpretable latent variables. To overcome this limitation, a widely used strategy is to designate anchor benchmarks that measure only a single latent dimension, a common practice in the literature to facilitate interpretability of the latent variables~\citep{andersonintroduction,bing2020adaptive,cui2025identifiability}. In particular, for the $k$-th component of the latent ability, we assume there exists a set of benchmarks $\mathcal{S}_k$ that measure only the $k$-th dimension, i.e., the corresponding loading vector $\lambda_{j}$ has a single nonzero entry at the $k$-th component for $j\in\mathcal{S}_k$. To remove sign indeterminacy, we require that for certain $j\in\mathcal{S}_k$, $\lambda_{jk}\ge 0$. The choice of these anchor benchmarks is typically guided by domain knowledge or prior studies employing exploratory tools such as rotation methods~\citep{browne2001overview,Rohe2020VintageFA,polo2024sloth}. 
To further address scale indeterminacy, we impose the constraint $\mathrm{diag}(\Sigma) = (1,\ldots,1)^\T$, where $\mathrm{diag}(\cdot)$ takes the diagonal elements of a square matrix into a vector. 



Now we define the feasible parameter space as follows:
 \begin{equation}\label{eq:feasible.set}
 \begin{aligned}
 \Xi(M) = \Big\{\xi \in \mathbb{R}^{\mathrm{dim}(\xi)}:&\,\|\xi\|_{\infty}\le M,~~\Lambda_{j,-k}=0 ~\mathrm{for~any~}j \in \mathcal{S}_k,~k \in [K],\\&\,\Sigma_{kk} = 1\text{ for }k\in[K],\text{ and }\Sigma = \Sigma^\T \Big\},\end{aligned}
 \end{equation}
 where $M$ is some constant that regulates the
magnitudes of the parameters, $\mathrm{dim}(x)$ denotes the dimension of vector $x$, $\|x\|_{\infty} = \max_{i\in[n]}|x_i|$ denotes the vector $\ell_{\infty}$ norm, and $\Lambda_{j,-k}$ denotes the $j$th row of $\Lambda$ with its $k$th entry removed.
 The marginal maximum likelihood estimator $\hat{\xi}$ is defined as the solution to
\begin{align}
    \hat \xi = &\,\mathop{\arg\max}_{\xi\in\Xi(M)}\mathcal{L} (Y|\xi)\label{eq_mmle}.
\end{align}
In the following, let $(\hat{\Lambda}, \hat{\beta}, \hat b,\hat\phi,\hat\Sigma)$ denote the components of $\hat\xi$ corresponding to $({\Lambda}, {\beta}, b, \phi, \Sigma)$.


\subsection{Theoretical Properties}\label{sec_theory}
In this section, we provide a theoretical analysis of the asymptotic behavior of the estimator $\hat{\xi}$ obtained from \eqref{eq_mmle}. 
We consider an asymptotic regime where $N$, the number of LLM families, diverges, while both the number of LLMs within each family and the number of benchmarks remain fixed.
Our theoretical analysis relies on a sequence of technical assumptions that are standard and commonly adopted for maximum likelihood estimators, which are deferred in Appendix B.1 of the online supplementary material.
Let $\norm{x} = \sqrt{\sum_{i = 1}^n x_i^2}$ denote the $\ell_2$ norm, we first establish the consistency of $\hat{\xi}$.

\begin{theorem}\label{thm:consistency}
  Under Assumption B.1 in Appendix B.1 of the online supplementary material, the estimator $\hat{\xi}$ obtained by~\eqref{eq_mmle} is consistent: $\norm{\hat{\xi} - \xi^*} = o_p(1)$, where $\xi^*$ is the vector of true parameters for the scaling model defined in Section~\ref{subsec_setup}.
\end{theorem}

The consistency result of $\hat{\xi}$ guarantees that  parameters for the scaling model can be estimated accurately, provided that a sufficiently large number of LLM families are available. Notably, unlike existing work on scaling laws~\citep{ruan2024observational,polo2024sloth}, our results are established using latent variables that can each be interpreted as a specific latent skill associated with the anchor benchmarks. We further provide theoretical guarantees for the estimated parameters, thereby establishing the scaling model as a reliable and interpretable tool for practitioners. In addition, the consistency results enable further analysis of the family-wise common abilities $\alpha_l$, which is treated as random variables in the estimation. In particular, recall that $\hat{\xi}$ is obtained through the maximum likelihood estimation problem~\eqref{eq_mmle}, where the random variable $\alpha_l$ is integrated out.
In practice, researchers may wish to study the distributional properties of $\alpha_l$ via posterior sampling, enabling comparisons of the inherent abilities across multiple LLM families.
In this scenario, the consistency of $\hat{\xi}$ plays a crucial role, as it ensures accurate approximation of the posterior distribution of $\alpha_l$.
We illustrate this application in Section~\ref{sec:post}.

Next we derive the asymptotic distribution of $\hat{\xi}$.
As defined in~\eqref{eq:feasible.set}, the feasible parameter space $\Xi$ fixes several components of $\xi$ at their true values.
We focus on the remaining free parameters for statistical inference.
For each benchmark $j \in [J]$, let $\bar{\lambda}_{j}$ denote the vector of free loading parameters to be estimated. Specifically, $\bar{\lambda}_j =\lambda_{jk}$ for $j \in \mathcal{S}_k,~k \in [K]$, and $\bar{\lambda}_{j}= \lambda_j$ otherwise.
Collecting all free loadings, define ${\lambda}_v = (\bar{\lambda}_{1}^\T, \dots, \bar{\lambda}_{J}^\T)^\T$.
The full vector of free parameters is then given by $\xi_f = \{\lambda_v^\T, \vvec^\T(\beta), b^\T, \phi^\T, \mathrm{vech}^\T(\Sigma)\}^\T$, where $\mathrm{vech}(\cdot)$ denotes the operator that stacks the strictly lower-triangular elements of a symmetric matrix into a vector.
The asymptotic distribution of these parameters is stated in the following theorem.

\begin{theorem}\label{thm:ci}
    Under Assumptions B.1--B.2 in Appendix B.1 of the online supplementary material, as $N\to\infty$,
    $\sqrt{N}(\hat\xi_f - \xi^*_f)\overset{d}{\to}\mathcal{N}(0,\Psi)$, where the asymptotic covariance matrix $\Psi$ is given in Appendix C of the online supplementary material.
\end{theorem}
In practice, the unknown covariance matrix $\Psi$ can be estimated by $\hat{\Psi}$, defined as the negative inverse of the likelihood Hessian.
Numerically, the estimator is obtained by automatic differentiation \citep{maclaurin2015autograd, paszke2019pytorch}.
Theorem~\ref{thm:ci} enables two key applications.
First, we show that under the proposed scaling model and estimation procedure, we are able to characterize the asymptotic distribution of the estimator $\hat{\xi}$. This result enables valid statistical inference on key parameters in the scaling model, thereby deepening our understanding of the underlying structure and guiding the optimal allocation of computational resources to achieve desired LLM latent abilities. For instance, one key component of $\hat{\xi}$ is the $\hat{\beta}$, which represents the estimated association between inputs and the latent abilities of LLMs. Moreover, the estimated loading parameters $\hat{\Lambda}$, which link the latent abilities to LLM performance, provide interpretable latent skills and valuable insights for benchmark design, offering practical guidance to developers.
Moreover, predicting LLM performance is of great interest in practice. The asymptotic normality of $\hat{\xi}$, combined with the posterior distribution of $\alpha_l$, enables the construction of prediction intervals for the response variable $Y^{(l)}$. We further demonstrate these applications later in Sections~\ref{sec:mcmc} and~\ref{sec:intervals}.

\subsection{Selection of Latent Dimension}\label{sec_infor}

The true latent dimension $K$ is typically unknown. The Akaike Information Criterion (AIC), originally proposed by \cite{akaike1998information}, has proven to be a highly versatile tool for model selection, with extensive applications in various areas.
Building on the demonstrated success of AIC across multiple domains, we develop an AIC-based procedure to determine the appropriate latent ability dimension $K$ for our scaling model.
Specifically, let $\xi_K$ be the parameter vector with $K$ latent dimensions, i.e., $\xi_K = \{\mathrm{vec}^\T(\Lambda_K), \mathrm{vec}^{\T}(\beta_K), b^\T, \phi^\T , \mathrm{vec}^\T(\Sigma)\}^\T$, where $\Lambda_K = (\lambda_1,\ldots , \lambda_{J}) \in \mathbb{R}^{K \times J}$ and $\beta_K \in \mathbb{R}^{p\times K}$. 
Moreover, let $\hat{\xi}_K$ be the corresponding estimator obtained from~\eqref{eq_mmle} under $K$ latent dimensions.
For now, the parameters are not subject to any identifiability constraints, since identifiability constraints do not affect the likelihood value and, hence, the model fit. Note that this selection step is carried out prior to estimation. Once the number of latent dimensions is determined, the estimation procedure and the associated identifiability constraints are applied as described in Section~\ref{sec_method}.

With $\mathcal{L}(\cdot)$ given in \eqref{eq: Likelihood}, the information criterion is defined as 
\begin{equation*}
    AIC(\xi_K) = -2\mathcal{L}(Y\mid \xi_K) + 2\mathrm{dim}(\xi_K).
\end{equation*}
Accordingly, the selected latent dimension $\hat{K}$ is obtained by the following rule
\begin{equation}\label{eq:k_hat}
    \hat K = \mathop{\arg\min}_{K \in \{ 1,2,\ldots,\bar K\}}AIC(\hat{\xi}_K),
\end{equation}
where $\bar {K}$ is a pre-determined upper-threshold of the latent dimension $K$.
The information criterion–based procedure for latent dimension selection has been extensively studied in the latent variable literature~\citep{bai2018consistency,chen2022determining} and has been shown in numerous studies to be successful in related models widely used in this area~\citep{sinha2021practitioner,cho2021gaussian,cui2024variational}.
\begin{algorithm}[H]
\caption{Projected Stochastic Gradient Ascent Algorithm}
\label{alg:psga}
\begin{algorithmic}[1]
\Require\\
\begin{itemize}
    \item Initial parameters $\xi^{(0)}=\{\Lambda^{(0)},\beta^{(0)},b^{(0)},\phi^{(0)},\Sigma^{(0)}\}$;
    \item Initial learning rate $\rho^{(0)}$, learning decay factor $\gamma$, and patience counter $T_\text{patience}$;
    \item Monte Carlo sample size $B$;
    \item Number of steps $T$ and averaging faction $c$;
    \item Data $\{Y_{ij}^{(l)}, x_i^{(l)}\}_{i \in [n_l],~j\in [J],~l \in [N]}$.
\end{itemize}

\For{$t = 1, 2, \dots, T$}
    \State Decompose $\Sigma^{(t-1)} = L^{(t-1)}(L^{(t-1)})^\intercal$, where $L^{(t-1)}$ is lower triangular.
    \State Draw $\{Z_k\}_{k=1}^B \overset{iid}{\sim} N(0,I_K)$ and compute:
    \[
    \tilde{\alpha}^{(k)} = L^{(t-1)}Z_k.
    \]
    \State Compute $\tilde{\xi}^{(t-1)} = \{\vvec^\T(\Lambda^{(t-1)}), \vvec^\T(\beta^{(t-1)}), (b^{(t-1)})^\T, (\phi^{(t-1)})^\T, \mathrm{vech}^\T(L^{(t-1)})\}^\T$.
    Compute gradient estimate
    \[
    g^{(t-1)} = \sum_{l=1}^{N} g^{(t-1)}_l,
    \]
    where $g^{(t-1)}_l$ is given as
    \begin{equation*}
      \nabla_{ \tilde{\xi}^{(t-1)}}\log\left(\sum_{k=1}^B\prod_{i\in [n_l],j \in [J]} f\left(Y_{ij}^{(l)}; \mu\{(\lambda_j^{(t-1)})^\intercal(\tilde{\alpha}^{(k)}+(\beta^{(t-1)})^\intercal x_i^{(l)})+b_j^{(t-1)}\},\phi^{(t-1)}\right)\right).
    \end{equation*}
    \State Compute
    \[
    \tilde{\xi}^{(t)} = \Pi_{\Xi(M)}\left(\textup{Adam}\big[\tilde{\xi}^{(t-1)},\rho^{(t-1)},\{g^{(t')}\}_{t'\leq t}\big]\right),
    \]
    where 
    $\Pi_{\Xi(M)}$ projects $\tilde{\xi}$ onto the feasible set $\Xi(M)$.
    \State If $t\geq T_\text{patience}$ and the likelihood function in~\eqref{eq: Likelihood} has not improved in the last $T_\text{patience}$ rounds, update $\rho^{(t)} = \gamma \cdot \rho^{(t-1)}$. Otherwise, set $\rho^{(t)} = \rho^{(t-1)}$.
    \State Obtain $\{\Lambda^{(t)}, \beta^{(t)},, b^{(t)}, \phi^{(t)} , L^{(t)}\}$ from corresponding components of $\tilde{\xi}^{(t-1)}$ and let $\Sigma^{(t)} = L^{(t)}(L^{(t)})^{\T}$.
    Let $\xi^{(t)} = \{\vvec^\T(\Lambda^{(t)}), \vvec^\T(\beta^{(t)}), (b^{(t)})^\T, (\phi^{(t)})^\T, \mathrm{vech}^\T(\Sigma^{(t)})\}^\T$. 
\EndFor

\Return Average of the last $\left\lceil c\cdot T \right\rceil$ realizations of $\xi^{(t)}$.
\end{algorithmic}
\end{algorithm}
\section{Computation}\label{sec_computation}
\subsection{Projected Stochastic Gradient Ascent Algorithm}\label{sec:psgd}
Obtaining a numerical solution $\hat{\xi}$ to the proposed maximum likelihood estimation  in~\eqref{eq_mmle} is computationally challenging, since problem \eqref{eq_mmle} involves a complicated integration component and does not admit a closed-form solution.
To address this issue,
existing literature typically uses the Expectation-Maximization (EM) algorithm \citep{B1981,mcculloch1997maximum}.
Some more recently proposed methods include Laplace approximation~\citep{wolfinger1993generalized}, Metropolis-Hastings Robbins-Monro (MHRM) \citep{C2010}, stochastic expectation maximization (SEM) algorithms~\citep{zhang2020improved}, and variational approximation methods~\citep{cho2021gaussian}. 
To accommodate our setting, where the likelihood involves a product over a large number of LLM families and benchmarks, we propose a projected stochastic gradient ascent algorithm to obtain $\hat{\xi}$. Our algorithm follows the general principle EM algorithm, replacing the maximization step with a more computationally efficient stochastic gradient update using the Adam optimizer \citep{kingma2014adam}, which allows it to scale to many LLMs and benchmarks.


The algorithm iteratively optimizes the parameter $\xi$ subject to the constraints defined by the feasible set $\Xi(M)$ given in \eqref{eq:feasible.set}. The optimization procedure is summarized in Algorithm~\ref{alg:psga}. 
To approximate the integral in~\eqref{eq_mmle}, we use Monte Carlo integration with a sample size of $B = 10^4$. The optimization is initialized with random Gaussian draws and is run for $T = 2 \cdot 10^4$ iterations and returns the average of the last $\lceil c\cdot T\rceil$ realizations of our estimates, for $c=20\%$, which, in our experiments, stabilizes the log-likelihood and produces small gradients. We experiment with initial learning rates in $\{0.1, 0.05, 0.01\}$ and learning rate decay factors in $\{0.999, 0.99\}$, running Algorithm~\ref{alg:psga} with five different random initializations for each configuration. The final parameter estimates are selected from the run that achieves the highest estimated log-likelihood.

\begin{remark}
To satisfy the condition that 
$\mathrm{diag}(\Sigma) = (1,\ldots, 1)^\T$ and $\Sigma = \Sigma^\T$,
we parameterize the covariance matrix using the Cholesky decomposition $\Sigma = L L^\intercal$, where $L$ is a lower triangular matrix and $\mathrm{diag}(L) = (1, \ldots, 1)^\T$, throughout the optimization routine.
For this parameter, the projection $\Pi_{\Xi(M)}$ simply maps $L^{(t-1)}$ to $L^{(t-1)} \mapsto L^{(t-1)}\mathrm{diag}\big\{L^{(t-1)}(L^{(t-1)}{})^\T\big\}^{-1/2}$, so that the rows have unitary Euclidean norm. Using the Cholesky factor in this way is a standard technique to ensure that $\Sigma^{(t)}$ remains a valid correlation matrix while maintaining numerical stability~\citep{lewandowski2009generating}. 
\end{remark}

\subsection{Sampling $\alpha_l$ from its Posterior Distribution}\label{sec:mcmc}
From the specification of the scaling model presented in Section~\ref{subsec_setup}, $\alpha_l$ is a random variable that captures the family-wise common abilities. 
This variable is of substantial practical interest for downstream analyses of LLMs such as investigating the strengths and weaknesses of a particular LLM family $l$ or making comparisons between multiple LLM families. To facilitate such analysis, we can construct posterior distributions for $\alpha_l$ based on the estimated parameters in the scaling models.
Specifically, we propose an algorithm to sample from posterior of $\alpha_l$ given the prior $\pi(\alpha|\Sigma)$, the observed data $\{Y_{ij}^{(l)}, x_i^{(l)}\}_{i \in [n_l],~j\in [J],~l \in [N]}$, and the estimator \eqref{eq_mmle}.
The posterior distribution of $\alpha_l$, given as 
\begin{equation}\label{eq_alpha_post}
    f\left\{\alpha_l|Y^{(l)},\xi\right\}\propto\prod_{i\in [n_l]}\prod_{j \in [J]} f\left[Y_{ij}^{(l)}:\mu\big\{(\lambda_j)^\intercal (\alpha_l+\beta^\T x_i^{(l)}) + b_j\big\},\phi\right]\pi (\alpha_l|\Sigma),
\end{equation}
is obtained by the Metropolis-Hastings algorithm~\citep{hastings1970monte}, using the estimate $\hat\xi$ as a proxy for the unknown $\xi^*$, with details included in Algorithm S1 in Appendix D of the online supplementary material. 

\subsection{Prediction Interval Construction}\label{sec:pred_int}
The parameter estimation approach in Section~\ref{sec_marginal.estimation} and the posterior sampling approach in Section~\ref{sec:mcmc} together enable the construction of prediction intervals for the performance of unevaluated language models on target benchmarks, in contrast to existing work that focuses solely on point predictions~\citep{owen2024predictable,polo2024sloth,ruan2024observational}.
This capability enables practitioners to accurately evaluate the precision of their predictions and subsequently derive relevant risk measures. For instance, after observing performance data from models within the Qwen 2 family with fewer than 8B parameters-where we consider Qwen 2 as the target family $l$ and let $i$ index specific language models, on certain benchmarks, we can construct a prediction interval for a specific benchmark of interest $j$ (e.g., MMLU).
See Section~\ref{sec_data_loading} for specific introductions of different LLM families and benchmarks.

We aim to construct an interval $\hat{I}_{i^{\prime}j}^{(l)}$ for the response of an unevaluated LLM $i^{\prime}$ in family $l$ to the benchmark $j\in [J]$.
The construction of the interval $\hat{I}_{i^{\prime}j}^{(l)}$ is based on~\eqref{eq_model_setup}, which depends on the unknown parameter $\xi$ and the latent ability of the family $\alpha_l$.
We draw $\xi_f$ from the asymptotic distribution established in Theorem~\eqref{thm:ci} and obtain posterior samples of $\alpha_l$ from~\eqref{eq_alpha_post}. 
Here, $\xi_f$ is draw from a Gaussian distribution with mean $\hat{\xi}_f$ and covariance $\hat{\Psi}/N$, denoted by $\hat{F}$. Taken together, we apply Algorithm~\ref{alg:pred_int} to compute the interval $\hat{I}_{i^{\prime}j}^{(l)}$.


\begin{algorithm}[h]
\caption{Prediction Interval Computation}
\label{alg:pred_int}
\begin{algorithmic}[1]
\Require\\
\begin{itemize}
    \item Data: responses $Y^{(l)}$, covariates $\{x_{i}^{(l)}\}_{i\in [n_l]}$ from training data;
    \item Estimator: approximate distribution $\hat{F}$ for $\hat{\xi}$;
    \item Covariate value: $x_{i^{\prime}}^{(l)}$ from prediction data;
    \item Benchmark specification $j \in [J]$;
    \item Sampling parameters: number of iterations $T$;
    \item Confidence level: $\tau\in (0,1)$.
\end{itemize}

\State {Initialize} $\mathcal{Y}=\{\}$

\For{$t=1 ,\dots,T$}
    \State {Sample} $\hat{\xi}^{(t)} \sim \hat{F}$.
    \State {Sample} $\alpha_l^{(t)}$ from Algorithm S1.
    \State Draw a sample from model~\eqref{eq_model_setup} given $x_{i^{\prime}}^{(l)}$ , $\alpha_l^{(t)}$ and $\hat{\xi}^{(t)}$, and add the sample to the collection $\mathcal{Y}$.
\EndFor   

\State \Return  $\hat{I}^{(l)}_{i^{\prime}j}= [\hat{Q}_{\frac{\tau}{2}}(\mathcal{Y}), \hat{Q}_{1 - \frac{\tau}{2}}(\mathcal{Y})]$, where $\hat{Q}_\tau(\mathcal{Y})$ denotes the $\tau$-level sample quantile of $\mathcal{Y}$.

\end{algorithmic}
\end{algorithm}

\section{Empirical Applications}\label{sec_experiment}

In this section, we present applications of the proposed scaling model framework\footnote{Our code and data can be found on \url{https://github.com/felipemaiapolo/statistical-scaling-law}.}. Our experiments cover 12 benchmarks and include various state-of-the-art LLM families such as LLaMa 3 \citep{dubey2024llama}, Qwen 2 \citep{yang2024qwen2}, and Yi 1.5 \citep{young2024yi}. In this experiment, we set the covariate associated with LLM $i$ in the $l$-th family as $x_i^{(l)}=[\log\{s_i^{(l)}\}, \log\{t_i^{(l)}\}, \log\{s_i^{(l)}\}\log\{t_i^{(l)}\}]^\intercal$ with $ s_i^{(l)} $ representing the size of the LLM $i$ (i.e., the number of parameters) and $t_i^{(l)}$ being its number of training tokens, following~\citep{polo2024sloth}. We investigate (i) how the LLMs' performance of each benchmark relies on the latent skills, as captured by the estimated $\Lambda$ (Section~\ref{sec_data_fit}); (ii) how model parameter size and training set size affect these skills, as captured by the estimated $\beta$ (Section~\ref{sec_data_fit}); (iii) the impact of fine-tuning (instruction tuning) in skills within a LLM family (Section~\ref{sec:post}); (iv) the accuracy of performance predictions for held-out LLMs (Section~\ref{sec:intervals}); and (v) the optimal scaling rules, that is, how a fixed computational budget should be allocated between parameter size and training tokens of LLMs (Section~\ref{sec_data_optimal}).

\subsection{Data and Benchmarks}\label{sec_data_loading}
Our dataset builds on those introduced by \citet{ruan2024observational} and \citet{polo2024sloth}, and we further augment it with additional models from both version 1 \citep{openllmldrboard} and version 2 \citep{open-llm-leaderboard-v2} of the HuggingFace Open LLM Leaderboard. This yields 168 LLMs from 75 distinct families in version 1 and 215 LLMs from 146 distinct families in version 2.  {In total, this gives us 285 unique LLMs from 170 distinct families.} Following the approach in \citet{polo2024sloth}, we treat fine-tuned LLMs as belonging to different families from their base LLMs; for example, we treat LLaMa 3 and LLaMa 3 Instruct as two different families\footnote{This is because it is hard to predict what effect fine-tuning can have on model behavior across families.}. 
We refer readers to Appendix E.1 of the online supplementary material for further details on the LLMs included in this work. 
In the main text, we combine the two leaderboards and conduct a singular analysis. 
{ We view this missingness as conditionally independent of the responses given the latent abilities, a common assumption adopted in the assessment literature~\citep{chen2023statistical}. In the presence of such missingness, our theoretical results can be extended to the maximum likelihood estimation based on the observed data likelihood. Moreover, the combined v1/v2 leaderboard setting may introduce block-structured missingness. The overlap between v1 and v2 may help mitigate the impact of such structured missingness. However, it does not fully rule out bias if missingness depends on unobserved benchmark-specific outcomes. Addressing such non-ignorable missingness would require more delicate methods to explicit modeling of the missingness mechanism, which we leave for future work.} In Open LLM Leaderboards v1 and v2, the benchmarks available are listed as follows, where the superscript (v1) denotes benchmarks in Leaderboard v1 and (v2) denotes Leaderboards v2.
\begin{itemize}
\item \textbf{MATH Lvl 5}$^{(\mathrm{v2})}$ \citep{hendrycks2021measuringmathematicalproblemsolving} Focuses exclusively on mathematical word problems and assesses numerical reasoning and problem solving.
\item \textbf{IFEval}$^{(\mathrm{v2})}$ \citep{zhou2023instructionfollowingevaluationlargelanguage}: Specializes in measuring instruction following by testing a model’s ability to adhere precisely to detailed, verifiable instructions.
\item \textbf{HellaSwag}$^{(\mathrm{v1})}$ \citep{zellers2019hellaswagmachinereallyfinish}: Challenges models with sentence completion in everyday scenarios, tapping into common sense reasoning.
\item \textbf{BBH}$^{(\mathrm{v2})}$ \citep{suzgun2022challengingbigbenchtaskschainofthought}: Presents logical deduction and linguistic puzzles, probing the interplay between logical reasoning and language proficiency.
\item \textbf{MMLU}$^{(\mathrm{v1})}$ \citep{hendrycks2020Measuring}: Includes mathematical challenges as well as advanced science and other academic questions.
\item \textbf{MMLU-PRO}$^{(\mathrm{v2})}$ \citep{wang2024mmluprorobustchallengingmultitask}: encompasses mathematical challenges alongside advanced science and academic questions, as a more challenging variant of MMLU.
\item \textbf{ARC}$^{(\mathrm{v1})}$ \citep{clark2018thinksolvedquestionanswering}: consists of grade school level science questions and evaluates fundamental scientific knowledge and reasoning.
\item \textbf{TruthfulQA}$^{(\mathrm{v1})}$ \citep{lin2021truthfulqa}: Gauges the truthfulness of responses, emphasizing the model’s ability to recall accurate information and maintain consistency.
\item \textbf{GSM8k}$^{(\mathrm{v1})}$ \citep{cobbe2021trainingverifierssolvemath}: Focuses exclusively on mathematical word problems, evaluating a model’s capacity for numerical reasoning and problem solving.
    \item \textbf{Winogrande}$^{(\mathrm{v1})}$ \citep{sakaguchi2019winograndeadversarialwinogradschema}: Centers on pronoun resolution tasks that require a deep understanding of everyday language context.
    \item \textbf{GPQA}$^{(\mathrm{v2})}$ \citep{rein2023gpqagraduatelevelgoogleproofqa}: Targets complex reasoning problems with nuanced question answering tasks at the graduate level.
    \item \textbf{MUSR}$^{(\mathrm{v2})}$ \citep{sprague2023musr}:  Assesses multi-step soft reasoning, where models must interpret and solve problems embedded in natural language narratives.
\end{itemize}

Aiming to increase the diversity of covered skills for any dimension $K$, we adopt the following ordering: MATH, IFEval, HellaSwag, BBH, MMLU, MMLU-Pro, ARC, TruthfulQa, GSM8k, Winogrande, GPQA, and MUSR.  This ordering reflects a progression from concrete mathematical reasoning to abstract language understanding and instruction following. In particular, the first six benchmarks cover a diversity of underlying cognitive skills as \textbf{MATH} captures symbolic and quantitative reasoning; \textbf{IFEval} isolates the ability to follow precise, verifiable instructions; \textbf{HellaSwag} tests common-sense inference in everyday contexts;  \textbf{BBH} focuses on multi-step logical reasoning and chain-of-thought processes; and \textbf{MMLU-Pro} and \textbf{MMLU} assess broad academic and factual knowledge across disciplines.
\subsection{Model Fitting and Interpretation}\label{sec_data_fit}
\label{subsec_modelfitting}
\begin{figure}[t]
    \centering
    \includegraphics[width=0.85\linewidth]{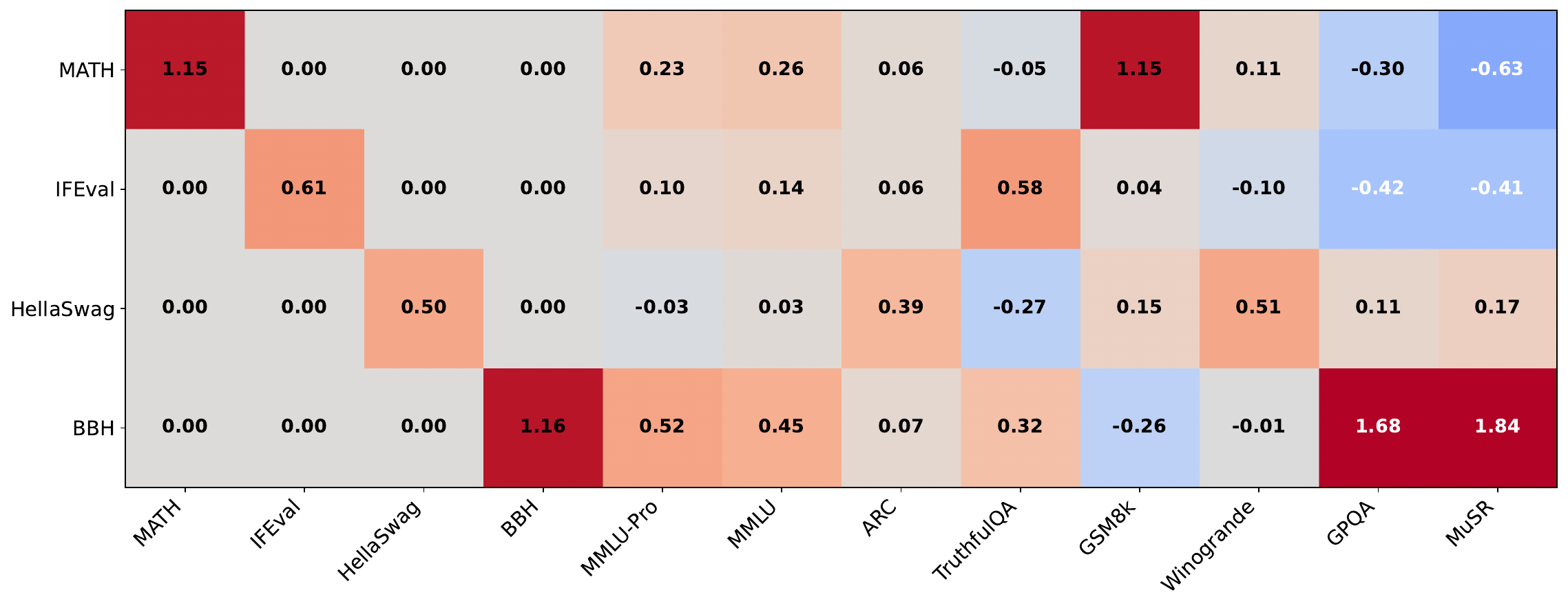}
    \caption{Estimated loadings for $K=4$. The $K$ skills in the vertical axis correspond to mathematical skills (MATH), instruction following (IFEval), common-sense reasoning (HellaSwag), and logical/linguistic reasoning (BBH).}
    \label{fig:loadings}
\end{figure}

To start, we select the dimension of our latent skill embeddings by fitting the scaling models with latent dimensions $K\in\{1,\cdots, 12\}$. Using the AIC criterion described in Section~\ref{sec_infor}, we observe that the AIC values (Figure S1 in Appendix E.2) of the online supplementary material decrease sharply from $K=1$ to $K=3$ and stabilize after $K=4$. Accordingly, we select $\hat K =4$ as the estimated number of latent dimensions. 
We have also presented the model fitting results for $K = 6$ in Appendix E.2 in the online supplementary material. To select the anchor benchmarks, we focus on diversity and interpretability in the underlying LLM skills. Accordingly, we choose MATH, IFEval, HellaSwag, and BBH, as each benchmark targets a distinct skill (e.g., instruction following), thereby enabling meaningful interpretation of the latent variable. { To assess the appropriateness of this choice and the sensitivity of our results, we apply rotation methods popularly used in the literature to detect sparsity structure in the loading matrix~\citep{browne2001overview,cui2026beyond}; see Appendix~E.5 of the online supplementary material. }

\begin{figure}[t]
    \centering
    \includegraphics[width=0.3\linewidth]{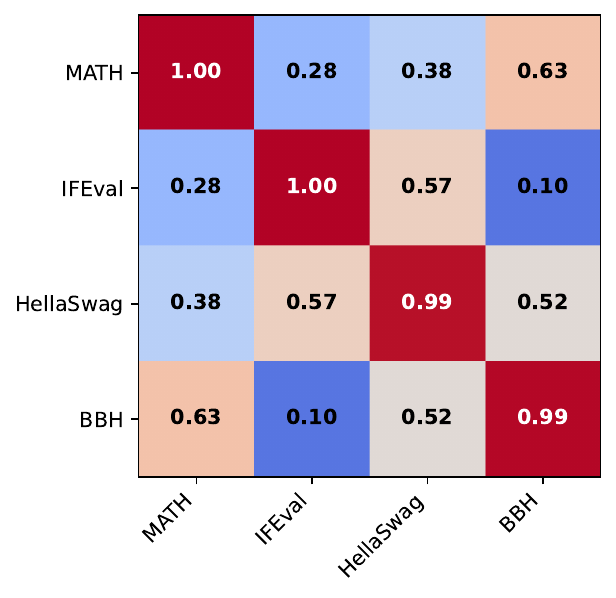}
    \caption{Correlation matrix of the latent random effects.}
    \label{fig:correlation}
\end{figure}

Figure \ref{fig:loadings} shows the estimated loadings, $\hat{\Lambda}$, of the model with $K=4$; this matrix reveals the relationship between benchmark performance and the latent abilities. For instance, because MATH predominantly assesses mathematical and reasoning abilities, the first ability is strongly associated with MMLU (which has STEM subsections) and GSM8K (another math benchmark), while being less related to ARC, TruthfulQA, Winogrande. IFEval measures instruction following, and its ability is strongly associated with TruthfulQA, which measures both the truthfulness and the informativeness of a model.  The ability connected to HellaSwag, a benchmark focused on common-sense natural language inference, aligns mostly with ARC and Winogrande, which are also focused on common-sense reasoning. Finally, the dimension defined by BBH exhibits strong positive relationships with other QA benchmarks such as MMLU/MMLU-Pro and GPQA (Figure \ref{fig:correlation}). 

Figure \ref{fig:correlation} presents the estimated correlation matrices $\hat\Sigma$ for the family-wise latent abilities, $\alpha_l$, associated with the anchor benchmarks. The matrix indicates that certain benchmarks share moderate to strong positive correlations that are induced by the latent abilities; for instance, BBH exhibits a correlation larger than 0.5 with MATH and HellaSwag, as does HellaSwag with IFEval. 
This analysis of correlation between LLM family-wise latent abilities is distinct from prior work. In \citet{ruan2024observational}, the latent skills are principal components, and thus orthogonal. \citet{polo2024sloth} study correlation among latent skills, which are a composition of family fixed effects and covariate effect.

\begin{table}[h]
\centering
\begin{tabular}{c c c c c}
\toprule
 & MATH & IFEval & HellaSwag & BBH \\
\midrule
\multirow{2}{*}{log(s)} & 0.432 & 0.190 & 0.840 & 0.656 \\
 & (0.051) & (0.055) & (0.064) & (0.107) \\
\midrule
\multirow{2}{*}{log(t)} & 0.774 & 0.324 & 0.384 & 0.445 \\
 & (0.086) & (0.088) & (0.076) & (0.154) \\
\midrule
\multirow{2}{*}{log(s)log(t)} & 0.026 & 0.154 & -0.039 & 0.010 \\
 & (0.024) & (0.034) & (0.015) & (0.050) \\
\bottomrule
\end{tabular}
\caption{Regression estimates for each one of the following skills (with standard errors in parentheses): mathematical skills (MATH), instruction following (IFEval), common-sense reasoning (HellaSwag), and logical/linguistic reasoning (BBH).}
\label{tab:beta}
\end{table}

Next, we examine how each covariate in $x = (\log (s),\log(t),\log(t)\log(s))^\T$ influences the LLM skill abilities. Table \ref{tab:beta} presents the corresponding $\beta$ estimates, with standard errors reported in parentheses, which are based on Theorem~\ref{thm:ci}. A key observation is that different skills scale differently. For instance, mathematical skills (MATH) appear to be strongly associated with data size, whereas common-sense reasoning (HellaSwag) and BBH abilities are more closely linked to model size. On the other hand, instruction following (IFEval) has a weaker association with both data and model size. This is unsurprising as instruction following is often endowed through a post-training process. Finally, the interaction term does not consistently play a significant role. Similar, but not one-to-one, analyses are conducted by \citet{polo2024sloth} and \citet{roberts2025compute}, where they show different skills scale differently. {The estimated loading patterns, correlations among latent abilities, and uncertainty quantification for the scaling slopes together provide a model-based empirical summary of how commonly used LLM benchmarks relate to latent skill dimensions. While some of these relationships are consistent with existing intuition about benchmark design and scaling behavior, they are not merely qualitative observations in our analysis. Rather, our framework offers a statistically principled way to quantify these relationships from heterogeneous leaderboard data.}

\subsection{Is a Certain Model Family Better 
than Another in Terms of a Skill of Interest?}\label{sec:post}
\begin{figure}[t]
    \centering
    \includegraphics[width=.85\linewidth]{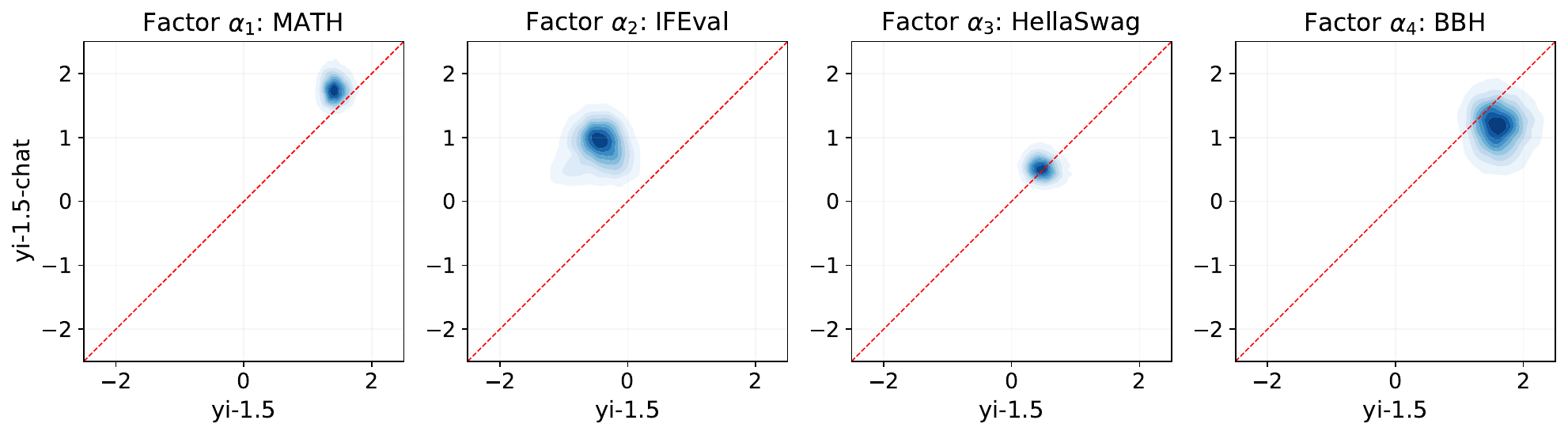}
    \caption{Joint density plot comparing the skills random intercepts ($\alpha_l$) of Yi-1.5 and Yi-1.5-chat models, highlighting the enhanced instruction following capability of the chat variants.}
    \label{fig:comparing_skills}
\end{figure}

Besides the relationship between benchmark performance and different latent skills, another interesting problem is to investigate the posterior joint distribution of the family-wise common abilities  ($\alpha_l$) for two different model families. This approach allows us to address questions such as, ``What is the probability that models from the LLaMa 3 family are inherently better than models from the Qwen 2 family at following instructions?'' To tackle such queries, we employ the methods described in Section \ref{sec:mcmc} to sample from the posterior distributions of $\alpha_l$ and $\alpha_{l'}$ for two distinct model families $l$ and $l'$. Once we obtain these samples, we plot their joint density. Figure \ref{fig:comparing_skills} illustrates this process using the families Yi-1.5 and Yi-1.5-chat. The Yi-1.5-chat models, fine-tuned from the base Yi-1.5 models, are specifically optimized for following instructions and engaging in conversations with humans. As shown, the chat models clearly exhibit superior performance in instruction following, while differences in other skill dimensions remain minimal. If one needs to compare specific models (and not families), we could alternatively plot their distribution of skills $\theta_i$'s. We can compare these results with the findings in \citet{polo2024sloth}. They find that instruct models are better at instruction following and worse at reasoning. We see here that instruction following might actually provide a minor help to mathematical reasoning. The finding that instruction tuning can unlock mathematical reasoning is also supported by \citet{zhang2024unveilingimpactcodingdata} for the case of instruction datasets with code examples. \citet{song2025dynamicsinstructionfinetuningchinese} also studies how benchmark scores and skills scale with \emph{instruction dataset} size. They find that for Grad-Math and biology, scores often remain constant, while scores do scale for humanities subjects such as history and creative writing. This is also reflected in Figure \ref{fig:comparing_skills}; instruction-tuning seems to have a small effect on STEM skills such as MATH and BBH.

\subsection{Prediction Intervals}\label{sec:intervals}
\begin{figure}
    \centering
    \includegraphics[width=1\linewidth]{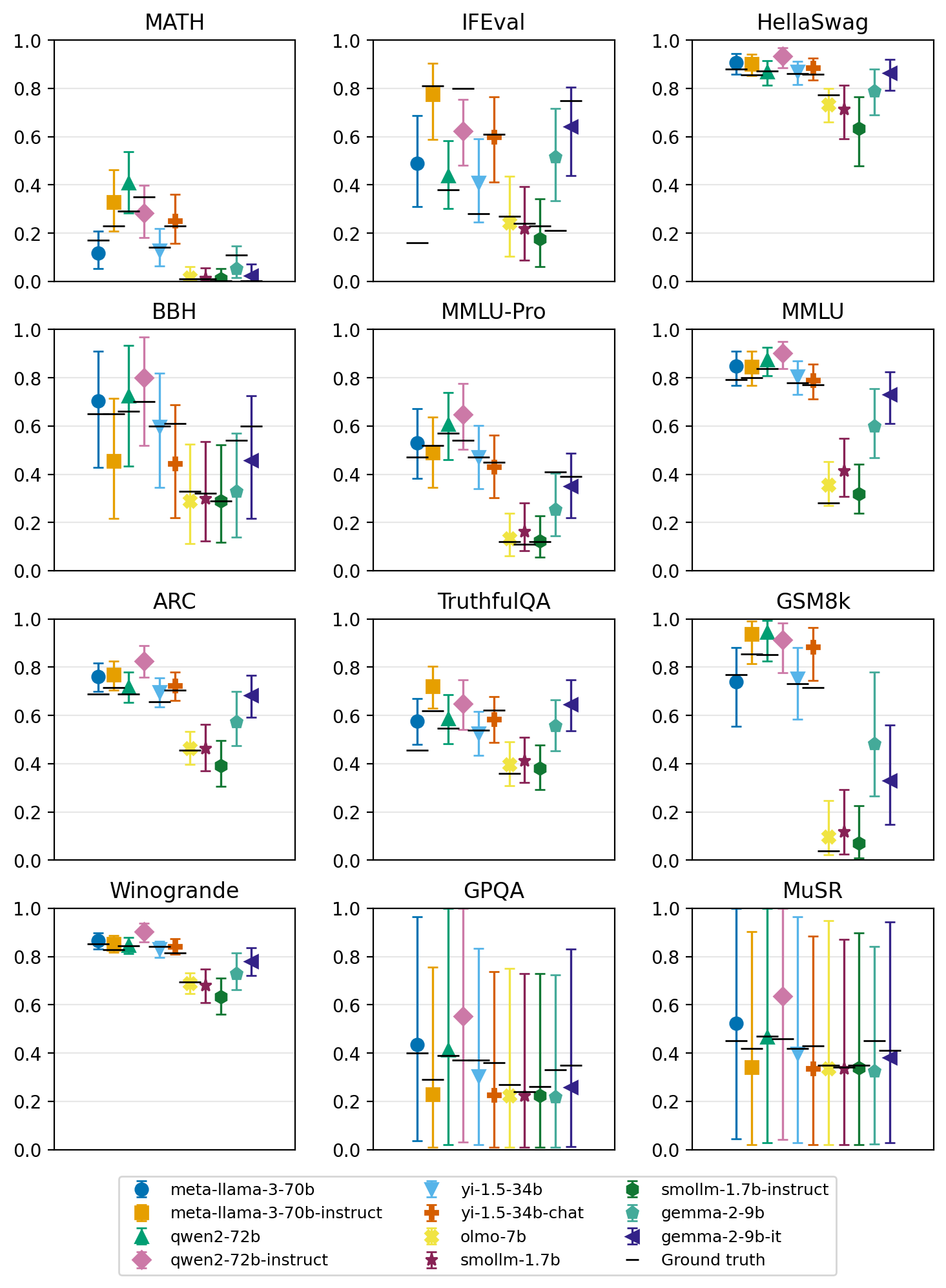}
    \caption{$95\%$ prediction intervals for eleven test LLMs.}
    \label{fig:pred_int}
\end{figure}

We apply the framework from Section~\ref{sec:pred_int} to construct 95\% prediction intervals using combined data from two Open LLM Leaderboards. Our test set includes Qwen-2-72B, Yi-1.5-34B, Yi-1.5-34B-chat, Meta-llama-3-70B, Meta-llama-3-70B-instruct, Olmo-7B, Smollm-1.7b, Smollm-1.7b-instruct, Gemma-2-9b, and Gemma-2-9b-it models withheld from fitting but with smaller variants present in the training data\footnote{{ For each test family, we fit a different scaling law and omit all models derived/related to the test model (with the same size) at training time. For example, when building a scaling law to predict results for LLaMa-3-Instruct-70b, we omit LLaMa-3-70b (and all derived models) from the training set.}}. As before, we fit our scaling model with $K=4$. Figure~\ref{fig:pred_int} summarizes the results\footnote{{ For some test LLMs, we had no access to their ground truth score for some of the benchmarks; we plot their predictions regardless.}}. Across benchmarks, the method produces reliable predictions: nearly all intervals contain the observed scores,  demonstrating that our scaling model's capability of capturing the underlying scaling of the LLM latent abilities across different benchmarks. A key advantage of our approach is that it delivers prediction intervals directly. Prior work has reported predicted benchmark scores as compute scales, but these are point predictions lacking any quantification of uncertainty \citep{hoffmann2022training,polo2024sloth, bhagia2024establishing, montgomery2025predictingtaskperformancecontextaware}. Interval width varies substantially by benchmark, which is itself informative. Most benchmarks yield tight intervals, with GPQA and MuSR standing out as exceptions. Notably, performance on these two benchmarks does not scale cleanly with parameter count. { In Appendix~E.3 online supplementary materials, we show that the width of the prediction intervals is driven primarily by the intrinsic uncertainty in the conditional 
distribution of $Y^{(l)}_{ij}$, rather than by estimation error, and we propose a strategy to obtain narrower intervals.}

\subsection{Optimal Scaling of Skills}\label{sec_data_optimal}
In practice, researchers may be interested in question related to optimal scalings: given a fixed computational budget measured in FLOPs, how can we optimally allocate this budget to maximize performance in a specific skill? Prior analyses of this nature have predominantly targeted validation loss optimization (e.g., by \citet{kaplan2020scalinglawsneurallanguage,hoffmann2022training}), but skill-specific optimization represents a novel area. In this section, we derive compute-optimal scaling laws tailored explicitly to skills following the recipe provided by \citet{polo2024sloth}. Our objective is clear: for a given language model family and a chosen skill, identify the optimal model configuration that maximizes performance under the constraint of a fixed computation budget $6st = c$. Consider a model family $l$ associated with skill $k$, characterized by:
$$
\theta^{(l)}_{k} = \alpha_{ik} + \beta_{k,0}\log(s) + \beta_{k,1}\log(t) + \beta_{k,2}\log(s)\log(t).
$$
Introducing the substitutions $u = \log(s)$, $v = \log(t)$, and $\bar{c} = \log(c) - \log(6)$, the optimization problem can be reformulated as:
$$
\max_{u,v} \alpha_{l,k} + \beta_{k,0}u + \beta_{k,1}v + \beta_{k,2}uv \quad \text{s.t.} \quad u + v = \bar{c}.
$$
This simplifies further into a univariate optimization problem in $u$:
$$
\max_{u} g_{ik}(u), \quad \text{with} \quad g_{ik}(u) \triangleq -\beta_{k,2} u^2 + (\beta_{k,0} -\beta_{k,1}+ \beta_{k,2}\bar{c})u + (\alpha_{l,k} + \beta_{k,1}\bar{c}).
$$
To ensure that our compute-optimal scaling procedure does not yield unrealistic predictions, we constrain the solution within the range of values for $u$ and $v$ observed in our training data. Specifically, we define constraints $u \in [\underline{u}, \overline{u}]$ and $v = \bar{c} - u \in [\underline{v}, \overline{v}]$, with the boundaries $\underline{u}, \overline{u}, \underline{v}, \overline{v}$ determined by quantiles from the training data. Combining these conditions provides the feasible set:
$$
u \in \mathcal{U} \triangleq [\max(\bar{c} - \overline{v}, \underline{u}), \min(\bar{c} - \underline{v}, \overline{u})]
$$
Thus, our optimization task becomes $\max_{u \in \mathcal{U}} g_{ik}(u)$, which can be efficiently solved. In particular, if $\beta_{k,2} > 0$, then $g_{ik}(u)$ is concave, and the optimal solution is either at the parabola's vertex (provided it lies within $\mathcal{U}$) or at one of the interval endpoints, $\max(\bar{c}-\overline{v},\underline{u})$ or $\min(\bar{c}-\underline{v},\overline{u})$. In practice, we replace parameters (like $\beta$) by their estimates.

Our results for our four skills (represented by MATH, IFEval, HellaSwag, and BBH) over a varying compute budget are presented in Table \ref{tab:opt_full}. { Here, we apply caps at 180B parameters and 15T training tokens to avoid extrapolating beyond the empirical range of the scaling model. An optimum at one of these boundaries indicates a boundary solution within the supported data range.} We see that the optimal point(s) for MATH are more token-heavy, while for QA benchmarks such as HellaSwag and BBH the optimal point(s) are parameter-heavy. 
{Our results align with \citet{roberts2025compute}, where they show knowledge-oriented QA benchmarks tend to favor more parameter-heavy allocations, whereas reasoning-oriented tasks may benefit more from additional training tokens. Our framework recovers this pattern within a likelihood-based latent-variable model and further provides formal uncertainty quantification for the corresponding scaling effects. This also provides a downstream-performance analogue of the compute-optimal scaling analysis in \citet{hoffmann2022training}, with latent skill performance replacing pre-training loss as the optimization objective. In summary, our results provide empirical support for skill-dependent compute-optimal scaling and show that our framework can recover interpretable differences across downstream abilities in a heterogeneous multi-family benchmark setting.} 

\begin{table}[h]
\centering
\footnotesize
\resizebox{\textwidth}{!}{%
\begin{tabular}{c|cc|cc|cc|cc}
\toprule
\textbf{FLOPs (1E19)} & \multicolumn{2}{c|}{\textbf{MATH}} & \multicolumn{2}{c|}{\textbf{IFEval}} & \multicolumn{2}{c|}{\textbf{HellaSwag}} & \multicolumn{2}{c}{\textbf{BBH}} \\
 & \textbf{Params (B)} & \textbf{Tokens (T)} & \textbf{Params (B)} & \textbf{Tokens (T)} & \textbf{Params (B)} & \textbf{Tokens (T)} & \textbf{Params (B)} & \textbf{Tokens (T)} \\
\midrule
6.30 & 0.07 & 0.15 & 0.07 & 0.15 & 0.07 & 0.15 & 0.07 & 0.15 \\
25.15 & 0.07 & 0.60 & 0.13 & 0.32 & 0.28 & 0.15 & 0.28 & 0.15 \\
100.37 & 0.07 & 2.39 & 0.26 & 0.63 & 1.12 & 0.15 & 1.12 & 0.15 \\
400.62 & 0.07 & 9.54 & 0.53 & 1.27 & 4.45 & 0.15 & 4.45 & 0.15 \\
1599.05 & 0.18 & 15.00 & 1.06 & 2.52 & 17.77 & 0.15 & 17.77 & 0.15 \\
6382.53 & 0.71 & 15.00 & 2.10 & 5.05 & 70.92 & 0.15 & 70.92 & 0.15 \\
25475.55 & 2.83 & 15.00 & 4.22 & 10.06 & 180.00 & 0.24 & 180.00 & 0.24 \\
101684.33 & 11.30 & 15.00 & 11.30 & 15.00 & 180.00 & 0.94 & 180.00 & 0.94 \\
405867.73 & 45.10 & 15.00 & 45.10 & 15.00 & 180.00 & 3.76 & 180.00 & 3.76 \\
1620000.00 & 180.00 & 15.00 & 180.00 & 15.00 & 180.00 & 15.00 & 180.00 & 15.00 \\
\bottomrule
\end{tabular}
}
\caption{Optimal allocation across skills for various compute budgets.}
\label{tab:opt_full}
\end{table}

\section{Conclusion and Future Work}\label{sec_conclude}

This work introduces a latent-variable scaling framework that captures how modern LLMs develop multiple distinct abilities as compute and other design choices change. Unlike classical scaling laws, which focus on perplexity or pre-training loss and assume homogeneous model families, our approach explicitly models family-specific latent abilities, benchmark-specific loadings, and covariate-driven scaling in a single likelihood-based setup with formal guarantees. Empirically, we find that a small number of latent skills is enough to explain performance across a wide range of benchmarks, and that these skills align naturally with interpretable abilities such as mathematics, instruction following, common-sense reasoning, and logical/linguistic reasoning. The estimated loadings, family abilities, and regression coefficients together provide a coherent picture of how compute translates into structured capability gains.


Our work opens several directions for future studies. A first direction is to move from benchmark-level averages to item-level modeling in a high-dimensional regime, which would sharpen skill estimates and make the latent structure more tightly grounded in the underlying questions. {A second direction is to enrich the covariates beyond $(\log(s),\log(t),\log(s)\log(t))$ by exploring alternative transformations of $s$ and $t$, nonlinear effects, and richer interactions, while also incorporating additional factors of interest such as data mixture, synthetic data usage, training pipeline choices, and pre-training and post-training procedures, including instruction tuning and RLHF, to better understand how these factors shape specific skills.
Another important direction is to allow family-level heterogeneity for the slopes of model covariates. For example, base models and post-trained variants may exhibit different relative gains from additional parameters or training tokens.}
Finally, our work provides an important foundation for more complex settings such as multimodal benchmarks and non-text modalities, where similar questions arise about how different capabilities co-evolve with compute. It lays the groundwork for tools to plan compute, design benchmarks, and understand how modeling decisions shape the emergence of capabilities in large-scale foundation models.


\section{Significance Statement} { 
Predicting how AI language models improve with scale is essential for guiding costly development decisions. Existing formulas treat all models alike and focus on a single performance measure, overlooking the diversity of modern model families and evaluation benchmarks. We propose a statistical framework in which each model family possesses hidden, interpretable skills such as mathematical reasoning, instruction following, and common sense reasoning, all of which grow predictably with model size and training data. Unlike prior work, our framework provides formal statistical guarantees and quantifies uncertainty around predictions. Applying it to hundreds of language models across twelve benchmarks reveals that different skills benefit from different compute tradeoffs, offering practical, evidence-based guidance for designing more efficient AI systems.
}

\bibliographystyle{agsm} 
\bibliography{bibliography}

@article{ferrari2004beta,
  title={Beta regression for modelling rates and proportions},
  author={Ferrari, Silvia and Cribari-Neto, Francisco},
  journal={Journal of applied statistics},
  volume={31},
  number={7},
  pages={799--815},
  year={2004},
  publisher={Taylor \& Francis}
}

@article{ferrari2011improved,
  title={Improved likelihood inference in beta regression},
  author={Ferrari, Silvia LP and Pinheiro, Eliane C},
  journal={Journal of Statistical Computation and Simulation},
  volume={81},
  number={4},
  pages={431--443},
  year={2011},
  publisher={Taylor \& Francis}
}

@article{bing2020adaptive,
  title={Adaptive estimation in structured factor models with applications to overlapping clustering},
  author={Bing, Xin and Bunea, Florentina and Ning, Yang and Wegkamp, Marten},
  journal={The Annals of Statistics},
  pages={2055–2081},
  volumn={48},
  year={2020}
}

@article{ouyang2024statistical,
  title={Statistical inference for covariate-adjusted and interpretable generalized factor model with application to testing fairness},
  author={Ouyang, Jing and Cui, Chengyu and Tan, Kean Ming and Xu, Gongjun},
  journal={Annals of Applied Statistics, to appear},
  year={2025}
}

@article{cui2025identifiability,
  title={Identifiability and Inference for Generalized Latent Factor Models},
  author={Cui, Chengyu and Xu, Gongjun},
  journal={arXiv preprint arXiv:2508.05866},
  year={2025}
}

@article{choshenhitchhiker,
  title={A Hitchhiker's Guide to Scaling Law Estimation},
  author={Choshen, Leshem and Zhang, Yang and Andreas, Jacob},
  journal={arXiv preprint arXiv:2410.11840},
  year={2024}
}

@book{bartholomew2011latent,
  title={Latent variable models and factor analysis: A unified approach},
  author={Bartholomew, David J and Knott, Martin and Moustaki, Irini},
  year={2011},
  publisher={John Wiley \& Sons}
}

@article{paszke2019pytorch,
  title={Pytorch: An imperative style, high-performance deep learning library},
  author={Paszke, Adam and Gross, Sam and Massa, Francisco and Lerer, Adam and Bradbury, James and Chanan, Gregory and Killeen, Trevor and Lin, Zeming and Gimelshein, Natalia and Antiga, Luca and others},
  journal={Advances in Neural Information Processing Systems},
  volume={32},
  year={2019}
}

@article{chen2025item,
  title={Item response theory—A statistical framework for educational and psychological measurement},
  author={Chen, Yunxiao and Li, Xiaoou and Liu, Jingchen and Ying, Zhiliang},
  journal={Statistical Science},
  volume={40},
  number={2},
  pages={167--194},
  year={2025},
  publisher={Institute of Mathematical Statistics}
}

@inproceedings{Sabour2024EmoBenchET,
  title={Emobench: Evaluating the emotional intelligence of large language models},
  author={Sabour, Sahand and Liu, Siyang and Zhang, Zheyuan and Liu, June and Zhou, Jinfeng and Sunaryo, Alvionna and Lee, Tatia and Mihalcea, Rada and Huang, Minlie},
  booktitle={Proceedings of the 62nd Annual Meeting of the Association for Computational Linguistics (Volume 1: Long Papers)},
  pages={5986--6004},
  year={2024}
}

@article{browne2001overview,
  title={An overview of analytic rotation in exploratory factor analysis},
  author={Browne, Michael W},
  journal={Multivariate Behavioral Research},
  volume={36},
  number={1},
  pages={111--150},
  year={2001},
  publisher={Taylor \& Francis}
}

@article{Rohe2020VintageFA,
  title={Vintage Factor Analysis with Varimax Performs Statistical Inference},
  author={Karl Rohe and Muzhe Zeng},
  journal={Journal of the Royal Statistical Society Series B: Statistical Methodology},
  volume={85},
  number={4},
  pages={1037-1060},
  year={2023}
}

@article{demszky2023using,
  title={Using large language models in psychology},
  author={Demszky, Dorottya and Yang, Diyi and Yeager, David S and Bryan, Christopher J and Clapper, Margarett and Chandhok, Susannah and Eichstaedt, Johannes C and Hecht, Cameron and Jamieson, Jeremy and Johnson, Meghann and others},
  journal={Nature Reviews Psychology},
  volume={2},
  number={11},
  pages={688--701},
  year={2023},
  publisher={Nature Publishing Group US New York}
}

@inproceedings{maclaurin2015autograd,
  title={Autograd: Effortless gradients in numpy},
  author={Maclaurin, Dougal and Duvenaud, David and Adams, Ryan P},
  booktitle={ICML 2015 AutoML workshop},
  volume={238},
  number={5},
  year={2015},
  organization={CNRS}
}

@article{mcculloch1997maximum,
  title={Maximum likelihood algorithms for generalized linear mixed models},
  author={McCulloch, Charles E},
  journal={Journal of the American statistical Association},
  volume={92},
  number={437},
  pages={162--170},
  year={1997},
  publisher={Taylor \& Francis}
}

@article{zhang2020improved,
  title={An improved stochastic EM algorithm for large-scale full-information item factor analysis},
  author={Zhang, Siliang and Chen, Yunxiao and Liu, Yang},
  journal={British Journal of Mathematical and Statistical Psychology},
  volume={73},
  number={1},
  pages={44--71},
  year={2020},
  publisher={Wiley Online Library}
}

@article{bai2012statistical,
author = {Jushan Bai and Kunpeng Li},
title = {{Statistical analysis of factor models of high dimension}},
volume = {40},
journal = {The Annals of Statistics},
number = {1},
publisher = {Institute of Mathematical Statistics},
pages = {436 -- 465},

  year={2012}
}

@article{bai2003inferential,
  title={Inferential theory for factor models of large dimensions},
  author={Bai, Jushan},
  journal={Econometrica},
  volume={71},
  number={1},
  pages={135--171},
  year={2003},
  publisher={Wiley Online Library}
}

@article{newey1994large,
  title={Large sample estimation and hypothesis testing},
  author={Newey, Whitney K and McFadden, Daniel},
  journal={Handbook of econometrics},
  volume={4},
  pages={2111--2245},
  year={1994},
  publisher={Elsevier}
}

@article{amemiya1985advanced,
  title={Advanced econometrics},
  author={Amemiya, Takeshi},
  journal={Harvard University Press},
  volume={2},
  pages={153--161},
  year={1985}
}

@article{kingma2014adam,
  title={Adam: A method for stochastic optimization},
  author={Kingma, Diederik P and Ba, Jimmy},
  journal={arXiv preprint arXiv:1412.6980},
  year={2014}
}

@Inbook{akaike1998information,
author="Akaike, Hirotogu",
title="Information Theory and an Extension of the Maximum Likelihood Principle",
bookTitle="Selected Papers of Hirotugu Akaike",
year="1998",
publisher="Springer New York"
}

@misc{beeching2023open,
  title={Open llm leaderboard},
  author={Beeching, Edward and Fourrier, Cl{\'e}mentine and Habib, Nathan and Han, Sheon and Lambert, Nathan and Rajani, Nazneen and Sanseviero, Omar and Tunstall, Lewis and Wolf, Thomas},
  year={2023}
}

@article{chen2023statistical,
  title={Statistical inference for noisy incomplete binary matrix},
  author={Chen, Yunxiao and Li, Chengcheng and Ouyang, Jing and Xu, Gongjun},
  journal={Journal of Machine Learning Research},
  volume={24},
  number={95},
  pages={1--66},
  year={2023}
}

@article{royall1970finite,
  title={On finite population sampling theory under certain linear regression models},
  author={Royall, Richard M},
  journal={Biometrika},
  volume={57},
  number={2},
  pages={377--387},
  year={1970},
  publisher={Oxford University Press}
}

@article{Liu2023RotationTS,
  title={Rotation to Sparse Loadings Using ${L}^p$ Losses and Related Inference Problems},
  author={Xinyi Liu and Gabriel Wallin and Yunxiao Chen and Irini Moustaki},
  journal={Psychometrika},
  year={2023},
  volume={88},
  pages={527-553}
}

@article{hendrickson1964promax,
  title={Promax: A quick method for rotation to oblique simple structure},
  author={Hendrickson, Alan E and White, Paul Owen},
  journal={British Journal of Statistical Psychology},
  volume={17},
  number={1},
  pages={65--70},
  year={1964},
  publisher={Wiley Online Library}
}

@book{mulaik2009foundations,
  title={Foundations of factor analysis},
  author={Mulaik, Stanley A},
  year={2009},
  publisher={CRC press}
}

@book{fabrigar2012exploratory,
  title={Exploratory factor analysis},
  author={Fabrigar, Leandre R and Wegener, Duane T},
  year={2012},
  publisher={Oxford University Press}
}

@book{ValliantDorfmanRoyall2000,
  author    = {Valliant, Richard and Dorfman, Alan H. and Royall, Richard M.},
  title     = {Finite Population Sampling and Inference: A Prediction Approach},
  publisher = {Wiley},
  year      = {2000}
}

@article{little2004model,
  title={To model or not to model? Competing modes of inference for finite population sampling},
  author={Little, Roderick J},
  journal={Journal of the American Statistical Association},
  volume={99},
  number={466},
  pages={546--556},
  year={2004},
  publisher={Taylor \& Francis}
}

@article{B1981,
  title={Marginal maximum likelihood estimation of item parameters: Application of an {EM} algorithm},
  author={Bock, R Darrell and Aitkin, Murray},
  journal={Psychometrika},
  volume={46},
  number={4},
  pages={443--459},
  year={1981},
  publisher={Springer-Verlag}
}

@book{andersonintroduction,
  title={An introduction to multivariate statistical analysis},
  author={Anderson, Theodore Wilbur},
   year={2003},
  publisher={Wiley New York}
}

@article{fan2017sufficient,
  title={Sufficient forecasting using factor models},
  author={Fan, Jianqing and Xue, Lingzhou and Yao, Jiawei},
  journal={Journal of Econometrics},
  volume={201},
  number={2},
  pages={292--306},
  year={2017},
  publisher={Elsevier}
}

@article{chen2019joint,
  title={Joint maximum likelihood estimation for high-dimensional exploratory item factor analysis},
  author={Chen, Yunxiao and Li, Xiaoou and Zhang, Siliang},
  journal={Psychometrika},
  volume={84},
  pages={124--146},
  year={2019},
  publisher={Springer}
}

@book{skrondal2004generalized,
  title={Generalized latent variable modeling: Multilevel, longitudinal, and structural equation models},
  author={Skrondal, Anders and Rabe-Hesketh, Sophia},
  year={2004},
  publisher={Chapman and Hall/CRC}
}

@article{muthen2004latent,
  title={Latent variable analysis},
  author={Muth{\'e}n, Bengt},
  journal={The Sage Handbook of Quantitative Methodology for the Social Sciences},
  volume={345},
  number={368},
  pages={106--109},
  year={2004},
  publisher={Los Angeles, CA; Thousand Oaks}
}

@article{bollen2002latent,
  title={Latent variables in psychology and the social sciences},
  author={Bollen, Kenneth A},
  journal={Annual Review of Psychology},
  volume={53},
  number={1},
  pages={605--634},
  year={2002},
  publisher={Annual Reviews 4139 El Camino Way, PO Box 10139, Palo Alto, CA 94303-0139, USA}
}

@article{sun2012multiple,
  title={Multiple hypothesis testing adjusted for latent variables, with an application to the {AGEMAP} gene expression data},
  author={Sun, Yunting and Zhang, Nancy R and Owen, Art B},
journal={The Annals of Applied Statistics},
  volumne={6},
  number={4},
  pages={1664--1688},
  year={2012},
  publisher={JSTOR}
}

@article{shen2009integrative,
  title={Integrative clustering of multiple genomic data types using a joint latent variable model with application to breast and lung cancer subtype analysis},
  author={Shen, Ronglai and Olshen, Adam B and Ladanyi, Marc},
  journal={Bioinformatics},
  volume={25},
  number={22},
  pages={2906--2912},
  year={2009},
  publisher={Oxford University Press}
}

@article{zhang2020imputed,
  title={Imputed factor regression for high-dimensional block-wise missing data},
  author={Zhang, Yanqing and Tang, Niansheng and Qu, Annie},
  journal={Statistica Sinica},
  volume={30},
  number={2},
  pages={631--651},
  year={2020},
  publisher={JSTOR}
}

@article{chen2022determining,
  title={Determining the number of factors in high-dimensional generalized latent factor models},
  author={Chen, Yunxiao and Li, Xiaoou},
  journal={Biometrika},
  volume={109},
  number={3},
  pages={769--782},
  year={2022},
  publisher={Oxford University Press}
}

@article{cui2024variational,
  title={Variational estimation for multidimensional generalized partial credit model},
  author={Cui, Chengyu and Wang, Chun and Xu, Gongjun},
  journal={Psychometrika},
  volume={89},
  number={3},
  pages={929--957},
  year={2024},
  publisher={Cambridge University Press \& Assessment}
}

@article{cho2021gaussian,
  title={Gaussian variational estimation for multidimensional item response theory},
  author={Cho, April E and Wang, Chun and Zhang, Xue and Xu, Gongjun},
  journal={British Journal of Mathematical and Statistical Psychology},
  volume={74},
  pages={52--85},
  year={2021},
  publisher={Wiley Online Library}
}

@article{C2010,
  title={{High-dimensional exploratory item factor analysis by a Metropolis--Hastings Robbins--Monro algorithm}},
  author={Cai, Li},
  journal={Psychometrika},
  volume={75},
  number={1},
  pages={33--57},
  year={2010},
  publisher={Springer}
}

@article{lewandowski2009generating,
  title={Generating random correlation matrices based on vines and extended onion method},
  author={Lewandowski, Daniel and Kurowicka, Dorota and Joe, Harry},
  journal={Journal of Multivariate Analysis},
  volume={100},
  number={9},
  pages={1989--2001},
  year={2009},
  publisher={Elsevier}
}

@article{wolfinger1993generalized,
  title={Generalized linear mixed models a pseudo-likelihood approach},
  author={Wolfinger, Russ and O'connell, Michael},
  journal={Journal of statistical Computation and Simulation},
  volume={48},
  number={3-4},
  pages={233--243},
  year={1993},
  publisher={Taylor \& Francis}
}

@article{sinha2021practitioner,
  title={Practitioner’s guide to latent class analysis: Methodological considerations and common pitfalls},
  author={Sinha, Pratik and Calfee, Carolyn S and Delucchi, Kevin L},
  journal={Critical Care Medicine},
  volume={49},
  number={1},
  pages={e63--e79},
  year={2021},
  publisher={LWW}
}

@article{bai2018consistency,
  title={Consistency of {AIC} and {BIC} in estimating the number of significant components in high-dimensional principal component analysis},
  author={Bai, Zhidong and Choi, Kwok Pui and Fujikoshi, Yasunori},
  journal={The Annals of Statistics},
  volume={46},
  number={3},
  pages={1050--1076},
  year={2018},
  publisher={JSTOR}
}

@article{polo2024sloth,
  title={Sloth: scaling laws for llm skills to predict multi-benchmark performance across families},
  author={Maia Polo, Felipe and Somerstep, Seamus and Choshen, Leshem and Sun, Yuekai and Yurochkin, Mikhail},
  journal={Advances in Neural Information Processing Systems},
  volume={38},
  pages={39511--39556},
  year={2026}
}

@article{ilic2023unveiling,
  title={Unveiling the general intelligence factor in language models: A psychometric approach},
  author={Ili{\'c}, David},
  journal={arXiv preprint arXiv:2310.11616},
  year={2023}
}

@article{hoffmann2022training,
  title={Training compute-optimal large language models},
  author={Hoffmann, Jordan and Borgeaud, Sebastian and Mensch, Arthur and Buchatskaya, Elena and Cai, Trevor and Rutherford, Eliza and Casas, Diego de Las and Hendricks, Lisa Anne and Welbl, Johannes and Clark, Aidan and others},
  journal={arXiv preprint arXiv:2203.15556},
  year={2022}
}

@article{polo2024efficient,
  title={Efficient multi-prompt evaluation of LLMs},
  author={Maia Polo, Felipe and Xu, Ronald and Weber, Lucas and Silva, M{\'\i}rian and Bhardwaj, Onkar and Choshen, Leshem and de Oliveira, Allysson Flavio Melo and Sun, Yuekai and Yurochkin, Mikhail},
  journal={arXiv preprint arXiv:2405.17202},
  year={2024}
}

@article{sprague2023musr,
  title={Musr: Testing the limits of chain-of-thought with multistep soft reasoning},
  author={Sprague, Zayne and Ye, Xi and Bostrom, Kaj and Chaudhuri, Swarat and Durrett, Greg},
  journal={arXiv preprint arXiv:2310.16049},
  year={2023}
}

@article{polo2024tinybenchmarks,
  title={tinyBenchmarks: evaluating LLMs with fewer examples},
  author={Maia Polo, Felipe and Weber, Lucas and Choshen, Leshem and Sun, Yuekai and Xu, Gongjun and Yurochkin, Mikhail},
  journal={arXiv preprint arXiv:2402.14992},
  year={2024}
}

@article{lin2021truthfulqa,
  title={Truthful{QA}: Measuring how models mimic human falsehoods},
  author={Lin, Stephanie and Hilton, Jacob and Evans, Owain},
  journal={arXiv preprint arXiv:2109.07958},
  year={2021}
}

@article{young2024yi,
  title={Yi: Open foundation models by 01. ai},
  author={Young, Alex and Chen, Bei and Li, Chao and Huang, Chengen and Zhang, Ge and Zhang, Guanwei and Li, Heng and Zhu, Jiangcheng and Chen, Jianqun and Chang, Jing and others},
  journal={arXiv preprint arXiv:2403.04652},
  year={2024}
}

@article{yang2024qwen2,
  title={Qwen2 technical report},
  author={Yang, An and Yang, Baosong and Hui, Binyuan and Zheng, Bo and Yu, Bowen and Zhou, Chang and Li, Chengpeng and Li, Chengyuan and Liu, Dayiheng and Huang, Fei and others},
  journal={arXiv preprint arXiv:2407.10671},
  year={2024}
}

@article{dubey2024llama,
  title={The {L}lama 3 herd of models},
  author={Dubey, Abhimanyu and Jauhri, Abhinav and Pandey, Abhinav and Kadian, Abhishek and Al-Dahle, Ahmad and Letman, Aiesha and Mathur, Akhil and Schelten, Alan and Yang, Amy and Fan, Angela and others},
  journal={arXiv preprint arXiv:2407.21783},
  year={2024}
}

@article{liang2022holistic,
  title={Holistic evaluation of language models},
  author={Liang, Percy and Bommasani, Rishi and Lee, Tony and Tsipras, Dimitris and Soylu, Dilara and Yasunaga, Michihiro and Zhang, Yian and Narayanan, Deepak and Wu, Yuhuai and Kumar, Ananya and others},
  journal={arXiv preprint arXiv:2211.09110},
  year={2022}
}

@article{wang2018glue,
  title={GLUE: A multi-task benchmark and analysis platform for natural language understanding},
  author={Wang, Alex and Singh, Amanpreet and Michael, Julian and Hill, Felix and Levy, Omer and Bowman, Samuel R},
  journal={arXiv preprint arXiv:1804.07461},
  year={2018}
}

@article{burnell2023revealing,
  title={Revealing the structure of language model capabilities},
  author={Burnell, Ryan and Hao, Han and Conway, Andrew RA and Orallo, Jose Hernandez},
  journal={arXiv preprint arXiv:2306.10062},
  year={2023}
}

@article{kipnis2024texttt,
  title={Metabench - A Sparse Benchmark to Measure General Ability in Large Language Models},
  author={Kipnis, Alex and Voudouris, Konstantinos and Buschoff, Luca M Schulze and Schulz, Eric},
  journal={arXiv preprint arXiv:2407.12844},
  year={2024}
}

@article{owen2024predictable,
  title={How predictable is language model benchmark performance?},
  author={Owen, David},
  journal={arXiv preprint arXiv:2401.04757},
  year={2024}
}

@article{rosenfeld2019constructivepredictiongeneralizationerror,
  title={A constructive prediction of the generalization error across scales},
  author={Rosenfeld, Jonathan S and Rosenfeld, Amir and Belinkov, Yonatan and Shavit, Nir},
  journal={arXiv preprint arXiv:1909.12673},
  year={2019}
}

@article{kaplan2020scalinglawsneurallanguage,
  title={Scaling laws for neural language models},
  author={Kaplan, Jared and McCandlish, Sam and Henighan, Tom and Brown, Tom B and Chess, Benjamin and Child, Rewon and Gray, Scott and Radford, Alec and Wu, Jeffrey and Amodei, Dario},
  journal={arXiv preprint arXiv:2001.08361},
  year={2020}
}

@article{gadre2024languagemodelsscalereliably,
  title={Language models scale reliably with over-training and on downstream tasks},
  author={Gadre, Samir Yitzhak and Smyrnis, Georgios and Shankar, Vaishaal and Gururangan, Suchin and Wortsman, Mitchell and Shao, Rulin and Mercat, Jean and Fang, Alex and Li, Jeffrey and Keh, Sedrick and others},
  journal={arXiv preprint arXiv:2403.08540},
  year={2024}
}

@misc{openllmldrboard,
      title={Open {LLM} leaderboard.}, 
      author={Edward Beeching and Clémentine Fourrier and Nathan Habib and Sheon Han and Nathan Lambert and Nazneen Rajani and Omar Sanseviero and Lewis Tunstall and Thomas Wolf},
      year={2023}
}

@misc{open-llm-leaderboard-v2,
  author = {Clémentine Fourrier and Nathan Habib and Alina Lozovskaya and Konrad Szafer and Thomas Wolf},
  title = {Open LLM Leaderboard v2},
  year = {2024}
}

@article{clark2018thinksolvedquestionanswering,
  title={Think you have solved question answering? try {ARC}, the {AI}2 reasoning challenge},
  author={Clark, Peter and Cowhey, Isaac and Etzioni, Oren and Khot, Tushar and Sabharwal, Ashish and Schoenick, Carissa and Tafjord, Oyvind},
  journal={arXiv preprint arXiv:1803.05457},
  year={2018}
}

@article{zellers2019hellaswagmachinereallyfinish,
  title={Hella{S}wag: Can a machine really finish your sentence?},
  author={Zellers, Rowan and Holtzman, Ari and Bisk, Yonatan and Farhadi, Ali and Choi, Yejin},
  journal={arXiv preprint arXiv:1905.07830},
  year={2019}
}

@article{cui2026beyond,
  title={Beyond Vintage Rotation: Bias-Free Sparse Representation Learning with Oracle Inference},
  author={Cui, Chengyu and Chen, Yunxiao and Ouyang, Jing and Xu, Gongjun},
  journal={arXiv preprint arXiv:2602.22590},
  year={2026}
}

@article{sakaguchi2019winograndeadversarialwinogradschema,
  title={Wino{G}rande: An adversarial {W}inograd schema challenge at scale},
  author={Sakaguchi, Keisuke and Bras, Ronan Le and Bhagavatula, Chandra and Choi, Yejin},
  journal={Communications of the ACM},
  volume={64},
  number={9},
  pages={99--106},
  year={2021},
  publisher={ACM New York, NY, USA}
}

@article{zhou2023instructionfollowingevaluationlargelanguage,
  title={Instruction-following evaluation for large language models},
  author={Zhou, Jeffrey and Lu, Tianjian and Mishra, Swaroop and Brahma, Siddhartha and Basu, Sujoy and Luan, Yi and Zhou, Denny and Hou, Le},
  journal={arXiv preprint arXiv:2311.07911},
  year={2023}
}

@article{cobbe2021trainingverifierssolvemath,
  title={Training verifiers to solve math word problems},
  author={Cobbe, Karl and Kosaraju, Vineet and Bavarian, Mohammad and Chen, Mark and Jun, Heewoo and Kaiser, Lukasz and Plappert, Matthias and Tworek, Jerry and Hilton, Jacob and Nakano, Reiichiro and others},
  journal={arXiv preprint arXiv:2110.14168},
  year={2021}
}

@article{hendrycks2021measuringmathematicalproblemsolving,
  title={Measuring mathematical problem solving with the {MATH} dataset},
  author={Hendrycks, Dan and Burns, Collin and Kadavath, Saurav and Arora, Akul and Basart, Steven and Tang, Eric and Song, Dawn and Steinhardt, Jacob},
  journal={arXiv preprint arXiv:2103.03874},
  year={2021}
}

@article{wang2024mmluprorobustchallengingmultitask,
  title={{MMLU}-{P}ro: A more robust and challenging multi-task language understanding benchmark},
  author={Wang, Yubo and Ma, Xueguang and Zhang, Ge and Ni, Yuansheng and Chandra, Abhranil and Guo, Shiguang and Ren, Weiming and Arulraj, Aaran and He, Xuan and Jiang, Ziyan and others},
  journal={Advances in Neural Information Processing Systems},
  volume={37},
  pages={95266--95290},
  year={2024}
}

@inproceedings{suzgun2022challengingbigbenchtaskschainofthought,
  title={Challenging big-bench tasks and whether chain-of-thought can solve them},
  author={Suzgun, Mirac and Scales, Nathan and Sch{\"a}rli, Nathanael and Gehrmann, Sebastian and Tay, Yi and Chung, Hyung Won and Chowdhery, Aakanksha and Le, Quoc and Chi, Ed and Zhou, Denny and others},
  booktitle={Findings of the Association for Computational Linguistics: ACL 2023},
  pages={13003--13051},
  year={2023}
}

@inproceedings{rein2023gpqagraduatelevelgoogleproofqa,
  title={{GPQA}: A graduate-level google-proof Q\&A benchmark},
  author={Rein, David and Hou, Betty Li and Stickland, Asa Cooper and Petty, Jackson and Pang, Richard Yuanzhe and Dirani, Julien and Michael, Julian and Bowman, Samuel R},
  booktitle={First Conference on Language Modeling},
  year={2024}
}

@article{hendrycks2020measuring,
  title={Measuring massive multitask language understanding},
  author={Hendrycks, Dan and Burns, Collin and Basart, Steven and Zou, Andy and Mazeika, Mantas and Song, Dawn and Steinhardt, Jacob},
  journal={arXiv preprint arXiv:2009.03300},
  year={2020}
}

@book{kumbhakar2003stochastic,
  title={Stochastic frontier analysis},
  author={Kumbhakar, Subal C and Lovell, CA Knox},
  year={2003},
  publisher={Cambridge university press}
}

@article{ruan2024observational,
  title={Observational scaling laws and the predictability of langauge model performance},
  author={Ruan, Yangjun and Maddison, Chris J and Hashimoto, Tatsunori B},
  journal={Advances in Neural Information Processing Systems},
  volume={37},
  pages={15841--15892},
  year={2024}
}

@book{reckase2009,
  title={Multidimensional Item Response Theory},
  author={M.D. Reckase},
  year={2009},
  publisher={Springer New York, NY}
}

@article{hastings1970monte,
  title={Monte Carlo sampling methods using Markov chains and their applications},
  author={W. K. Hastings},
  year={1970},
  journal={Biometrika},
  volume={57},
  number={1},
  pages={97–109},
  publisher={Oxford University Press}
}

@inproceedings{roberts2025compute,
  title={Compute optimal scaling of skills: Knowledge vs reasoning},
  author={Roberts, Nicholas and Chatterji, Niladri S and Narang, Sharan and Lewis, Mike and Hupkes, Dieuwke},
  booktitle={Findings of the Association for Computational Linguistics: ACL 2025},
  pages={13295--13316},
  year={2025}
}

@article{bhagia2024establishing,
  title={Establishing task scaling laws via compute-efficient model ladders},
  author={Bhagia, Akshita and Liu, Jiacheng and Wettig, Alexander and Heineman, David and Tafjord, Oyvind and Jha, Ananya Harsh and Soldaini, Luca and Smith, Noah A and Groeneveld, Dirk and Koh, Pang Wei and others},
  journal={arXiv preprint arXiv:2412.04403},
  year={2024}
}

@article{montgomery2025predictingtaskperformancecontextaware,
  title={Predicting Task Performance with Context-aware Scaling Laws},
  author={Montgomery, Kyle and Park, David and Tu, Jianhong and Bendersky, Michael and Gunel, Beliz and Song, Dawn and Wang, Chenguang},
  journal={arXiv preprint arXiv:2510.14919},
  year={2025}
}

@inproceedings{song2025dynamicsinstructionfinetuningchinese,
  title={Dynamics of Instruction Fine-Tuning for Chinese Large Language Models},
  author={Song, Chiyu and Zhou, Zhanchao and Yan, Jianhao and Fei, Yuejiao and Lan, Zhenzhong and Zhang, Yue},
  booktitle={Proceedings of the 31st International Conference on Computational Linguistics},
  pages={10345--10366},
  year={2025}
}

@inproceedings{zhang2024unveilingimpactcodingdata,
  title={Unveiling the impact of coding data instruction fine-tuning on large language models reasoning},
  author={Zhang, Xinlu and Chen, Zhiyu Zoey and Ye, Xi and Yang, Xianjun and Chen, Lichang and Wang, William Yang and Petzold, Linda Ruth},
  booktitle={Proceedings of the AAAI Conference on Artificial Intelligence},
  volume={39},
  number={24},
  pages={25949--25957},
  year={2025}
}



\newpage 
\begin{center}
    {\Large \bf Supplement to ``A Latent Variable Framework for \\Scaling Laws in Large Language Models''}
\end{center}
\appendix

\section{Related Work}\label{sec:related_work}

\subsection*{Latent Variable Models}

Latent variable modeling has long been a popular statistical framework for capturing complex dependency structures among multiple observed variables \citep{skrondal2004generalized,reckase2009,bartholomew2011latent}. Its flexibility and interpretability make it particularly suitable for analyzing high-dimensional, complex-structured data. Latent variable methods have found widespread applications across diverse fields, such as economics~\citep{bai2003inferential,fan2017sufficient}, public health~\citep{shen2009integrative,sun2012multiple}, and social sciences~\citep{bollen2002latent,muthen2004latent}.

There has been much empirical evidence that the LLM performance across different families and benchmarks can also be effectively characterized by low-dimensional latent variables/factors. For example, \citet{ilic2023unveiling} extracts a general intelligence factor (or “g-factor”) from the Open LLM Leaderboard \citep{openllmldrboard} and GLUE \citep{wang2018glue} via factor analysis, further demonstrating that this factor positively correlates with LLM size. Similarly, \citet{burnell2023revealing} utilizes HELM \citep{liang2022holistic} data to show that LLM intelligence can be decomposed into three interrelated factors, each positively correlated with LLM size, though without incorporating training data or LLM family information or proposing a formal scaling law. In another approach, \citet{kipnis2024texttt} apply a unidimensional item response theory model to six Open LLM Leaderboard benchmarks, finding that the primary factor is strongly correlated with the overall (grand) score of the LLMs. Building on this insight, \citet{polo2024tinybenchmarks,polo2024efficient} demonstrate that extracting low-dimensional latent skills can significantly enhance evaluation efficiency, achieving up to a 140-fold reduction in computing requirements. Another related work \citep{polo2024sloth} introduces scaling laws for low-dimensional LLM skills; however, their formulation lacks basic statistical properties, including identifiability and consistency. This limits their impact in terms of uncertainty quantification and interpretability. 

\subsection*{Scaling Laws in Machine Learning}

Scaling laws have become a central topic in modern machine learning, particularly in the context of training large neural networks on datasets containing billions or even trillions of data points, offering valuable guidelines for decisions about optimal model size and dataset scale~\citep{choshenhitchhiker}.
A particularly prominent application of scaling laws arises in the development of large language models, which are typically trained on massive corpora containing billions or trillions of tokens. Two seminal works in this area are \citet{kaplan2020scalinglawsneurallanguage} and \citet{hoffmann2022training}, which formalize empirical relationships between compute, model size, dataset size, and performance, as described in \eqref{eq_scaling_base}. 

Building on the classical scaling laws, which primarily focus on language model perplexity or pre-training loss, recent studies have substantially enriched this framework by extending it to accurately predict performance on downstream benchmarks. Recent works \citep{owen2024predictable,gadre2024languagemodelsscalereliably,ruan2024observational,polo2024sloth} have developed scaling laws tailored explicitly for benchmark performance, recognizing that loss metrics do not always directly correlate with downstream task capabilities. 
Among recent work, \citet{owen2024predictable},
\citet{ruan2024observational}, and \citet{polo2024sloth} are the most closely related to ours. Specifically, \citet{owen2024predictable} presents a method for evaluating predictability through simple statistical models by fitting simple functional forms against this estimated loss. { Two key limitations of their approach are that it does not account for: (i) family-specific heterogeneity, which reduces model complexity but may compromise predictive accuracy by overlooking systematic differences across LLM families, and (ii) the correlation structure among benchmark scores, limiting interpretability because predictions are not grounded in underlying model capabilities.} In contrast, \citet{ruan2024observational} and \citet{polo2024sloth} incorporate family-wise heterogeneity when modeling LLM performance and account for low-rank correlation structure between scores, but their methods { either (i) rely on orthogonal designs that are difficult to satisfy when latent LLM capabilities exhibit complex dependence structures, (ii) are not intended to extrapolate to hypothetical/future LLMs, or (iii) lack rigorous statistical guarantees. Our work aims to fill these gaps in previous work by introducing a scaling law that accommodates family-specific heterogeneity and captures the correlation structure among benchmark scores through latent capabilities. Our approach provides a principled and interpretable description of LLM performance while preserving the ability to generalize beyond observed models. In addition, our framework is supported by rigorous statistical guarantees, enabling parameter inference and both accurate prediction and uncertainty quantification.
}

\section{Proof of Theorems \ref{thm:consistency}--\ref{thm:ci}}
In this section, we present proof details of the theoretical results given in Section \ref{sec_method}.
Section \ref{sec:asumps} states technical assumptions.
Sections \ref{sec:consistency.proof}--\ref{sec:noramlity.proof} present the proofs of Theorems \ref{thm:consistency}--\ref{thm:ci}, respectively.

\subsection{Assumptions}\label{sec:asumps}

In this section, we present a set of technical conditions required to establish Theorems~\ref{thm:consistency}--\ref{thm:ci}.
\begin{assumption}\label{asump:consistency}
    The following sub-assumptions hold:
    \begin{enumerate}[label=(\alph*), ref=(\alph*)]
        \item \label{asump:consistency:b} $\xi^*$ is an interior point of 
        $\Xi(M)$.
        \item \label{asump:consistency:c}For $\forall~\xi \in \Xi(M)$ and $\forall~l$, if~$\mathbb{P}_{Y^{(l)} \sim f_l(\cdot\mid \xi^*)}\{f_l(Y^{(l)}\mid \xi^*) = f_l(Y^{(l)}\mid \xi) \} = 1$, then $\xi = \xi^*$.
        \item \label{asump:consistency:d} There exists a constant $\Delta_0$ such that for any $l$, $\mathbb{E}_{Y^{(l)} \sim f_l(\cdot \mid \xi^*)}\{\sup_{\xi \in \Xi(M)} \log^2f_l(Y^{(l)} \mid \xi) \} \leq \Delta_0$.
        
    \end{enumerate}
\end{assumption}
Here, assumptions~\ref{asump:consistency:b}--\ref{asump:consistency:c} of~\ref{asump:consistency} impose standard and mild conditions in the theory of $M$-estimation and can be verified under the setting of Section~\ref{subsec_setup}.
Moreover, Assumption~\ref{asump:consistency:d} of ~\ref{asump:consistency} provides a sufficient condition for the uniform law of large numbers to hold for the empirical loss function $\mathcal{L}(Y \mid \xi)$.

\begin{assumption}\label{asump:normality}
    The following sub-assumptions hold:
    \begin{enumerate}[label=(\alph*)]
        \item \label{asump:normality:a} Let $I(\xi_f) = \EE\{\sum_{l =1}^N\frac{\partial \log f_l(Y^{(l)}\mid \xi)}{\partial \xi_f } \frac{\partial \log f_l(Y^{(l)}\mid \xi)}{\partial \xi_f^\T }\} /N$, $v = \lambda_{\min}\{I(\xi_f^*)\}>0$, where we use $\lambda_{\min}(A)$ to denote the minimal eigenvalue of matrix $A$. 
        
        \item \label{asump:normality:b} There exists positive constant $\delta$ such that $\sup_{l \in [N]} \EE\left\{\left\| \left. \frac{\partial \log f_l(Y^{(l)}\mid \xi)}{\partial \xi_f} \right|_{\xi_f = \xi_f^*} \right\|_2^{2+\delta} \right\} < \infty$.
        
        \item \label{asump:normality:c} Let $\|\cdot\|_{\max}$ be the maximum absolute entry of a matrix. There exists a finite constant $\Delta_0$ and a neighborhood $\mathcal{N}$ of the true model parameter $\xi^*$ such that $\forall~l \in [N]$, $\EE\left\{\sup_{\xi \in \mathcal{N}} \left\| \frac{\partial^2 \log f_l(Y^{(l)} \mid \xi)}{\partial \xi_f \partial \xi_f^\T} \right\|_{\max}^2 \right\} \leq \Delta_0$.
    \end{enumerate}
\end{assumption}

Assumption~\ref{asump:normality} outlines standard regularity conditions required for establishing Theorem~\ref{thm:ci}. Assumption~\ref{asump:normality:a} of \ref{asump:normality} guarantees that the asymptotic variance matrix of $\hat{\xi}_f$ is well conditioned and positive definite. Assumption~\ref{asump:normality:b} of \ref{asump:normality} provides the condition for the central limit theorem to be applied to the score function. Assumption~\ref{asump:normality:c} of \ref{asump:normality} ensures the uniform convergence of the empirical Hessian matrix associated with~\eqref{eq_mmle} in a neighborhood around $\xi^*$.

\subsection{Proof for Theorem \ref{thm:consistency}}\label{sec:consistency.proof}
\begin{proof}
To prove the consistency of $\hat{\xi}$, we apply Theorem 2.1 of \cite{newey1994large}.
The proof steps consist of four major components:
\begin{itemize}
    \item[(i)] $\mathbb{E}_{Y^{(l)} \sim f_l(\cdot \mid \xi^*)}\left\{\mathcal{L}(Y\mid \xi) \right\}$ is uniquely maximized at $\xi^*$ over $\Xi(M)$.
    \item[(ii)] $\Xi(M)$ is compact.
    \item[(iii)] $\mathbb{E}_{Y^{(l)} \sim f_l(\cdot \mid \xi^*)}\left\{\mathcal{L}(Y\mid \xi) \right\}$ is continuous.
    \item[(iv)] $\mathcal{L}(Y\mid \xi) \overset{p}{\to}\mathbb{E}_{Y^{(l)} \sim f_l(\cdot \mid \xi^*)}\left\{\mathcal{L}(Y\mid \xi) \right\}$ uniformly over $\Xi(M)$.
\end{itemize}

For condition (i),
let $\lambda(\cdot)$ denote Lebesgue measure in Euclidean space $\mathbb{R}^{n_l\times J}$, an application of Jensen's inequality yields
\begin{align*}
    \mathbb{E}_{Y^{(l)} \sim f_l(\cdot \mid \xi^*)}\left\{\frac{1}{N}\sum_{l=1}^N\log \frac{f_l\left(Y^{(l)} \mid \xi \right)}{f_l\left(Y^{(l)} \mid \xi^* \right)} \right\} & \leq \frac{1}{N}\sum_{l=1}^N\log\left[\mathbb{E}_{Y^{(l)} \sim f_l(\cdot \mid \xi^*)}\left\{ \frac{f_l\left(Y^{(l)} \mid \xi \right)}{f_l\left(Y^{(l)} \mid \xi^* \right)} \right\}\right]\\
    &=\frac{1}{N}\sum_{l=1}^N\log \int_{y^{(l)} \in [0,1]^{ n_l \times J }}f_l(y^{(l)}\mid \xi)d\lambda(y^{(l)})\\
    &= \frac{1}{N}\sum_{l =1}^N\log 1\\
    &=0.
\end{align*}
Based on Assumption \ref{asump:consistency}, $\xi^*$ uniquely maximizes $\mathbb{E}_{Y^{(l)} \sim f_l(\cdot \mid \xi^*)}\left\{\mathcal{L}(Y\mid \xi) \right\}$,
condition (i) is satisfied.
Conditions (ii) and (iii) are clearly satisfied by Assumption \ref{asump:consistency} and our scaling model specification.
Condition (iv) requires the uniform law of large numbers to hold for $\mathcal{L}(Y\mid \xi)$ over $\xi \in \Xi(M)$.
Given Assumption \ref{asump:consistency}, we can apply Theorem 4.2.1 of \cite{amemiya1985advanced}, and hence condition (iv) holds.
\end{proof}
\subsection{Proof for Theorem \ref{thm:ci}}\label{sec:noramlity.proof}
\begin{proof}
The proof is based on Theorem 3.1 of \cite{newey1994large}.
By Theorem~\ref{thm:consistency}, $\hat{\xi}$ is the solution to \eqref{eq_mmle} and satisfies $\hat{\xi}\overset{p}{\to} \xi^*$.
Let $\xi_f$ denote the vector of free parameters in the main manuscript. Since $\xi^*$ is an interior point of $\Xi(M)$,  we can apply a Taylor expansion around $\xi_f^*$:
\[
\underbrace{\frac{\partial \mathcal{L}(Y\mid \xi)}{\partial \xi_f}\bigg|_{\xi_f = \hat{\xi}_f}}_{ = {0}} - \frac{\partial \mathcal{L}(Y\mid \xi)}{\partial \xi_f}\bigg|_{\xi_f = \xi_f^*} = \frac{\partial^2 \mathcal{L}(Y\mid \xi)}{\partial \xi_f \partial \xi_f^\T}\bigg |_{\xi_f = \tilde{\xi}_f}\left( \hat{\xi}_f - \xi_f^*\right),
\]
where $\tilde{\xi}_f$ is a vector that lies between $\xi_f^*$ and $\hat{\xi}_f$.
We start by studying the asymptotic distribution of the second term in the above equation.
We have
\begin{align*}
    &\EE\left\{\frac{\partial \mathcal{L}(Y\mid \xi)}{\partial \xi_f}\bigg|_{\xi_f = \xi_f^*} \right\}\\ 
    &= \EE\left\{\frac{1}{N}\sum_{l =1}^N\frac{\partial\log f_l(Y^{(l)}\mid \xi)}{\partial \xi_f}\bigg |_{\xi_f = \xi_f^*} \right\}\\
    &=\frac{1}{N}\sum_{l =1}^N\EE\left\{ \frac{1}{f_l(Y^{(l)}\mid \xi^*)}\frac{\partial f_l(Y^{(l)}\mid \xi)}{\partial \xi_f}\bigg |_{\xi_f = \xi_f^*}\right\}\\
    &=\frac{1}{N}\sum_{l =1}^N \int_{y^{(l)} \in [0,1]^{ n_l \times  J }}\frac{1}{f_l(y^{(l)}\mid \xi^*)}\frac{\partial f_l(y^{(l)}\mid \xi)}{\partial \xi_f}\bigg |_{\xi_f = \xi_f^*} \times f_l(y^{(l)}\mid \xi^*)d\lambda(y^{(l)})\\
    &=\frac{1}{N}\sum_{l =1}^N \int_{y^{(l)} \in [0,1]^{ n_l \times  J }}\frac{\partial f_l(y^{(l)}\mid \xi)}{\partial \xi_f}\bigg |_{\xi_f = \xi_f^*} d\lambda(y^{(l)})\\
    &=\frac{1}{N}\sum_{l =1}^N \frac{\partial \int_{y^{(l)} \in [0,1]^{ n_l \times  J }}f_l(y^{(l)}\mid \xi)d\lambda(y^{(l)})}{\partial \xi_f}\bigg |_{\xi_f = \xi_f^*}\\
    &= \frac{1}{N}\sum_{l =1}^N\frac{\partial 1}{\partial \xi_f}\bigg |_{\xi_f = \xi_f^*}\\
    &= {0}.
\end{align*}
Moreover, $\{Y^{(l)} \}_{l = 1}^{N}$ is a sequence of independent random variables.
From the derivation of Fisher information matrix and Assumption \ref{asump:normality}, we know that
\begin{align*}
  & \mathrm{var}\left\{\frac{\partial \mathcal{L}(Y\mid \xi)}{\partial \xi_f}\bigg |_{\xi_f = \xi_f^*} \right\}\\
  &= \EE\left[\left\{\frac{\partial \mathcal{L}(Y\mid \xi)}{\partial \xi_f}\bigg|_{\xi_f = \xi_f^*} \right\}\left\{\frac{\partial \mathcal{L}(Y\mid \xi)}{\partial \xi_f}\bigg|_{\xi_f = \xi_f^*} \right\}^\T \right] \\
  &= \frac{1}{N^2}\sum_{l =1}^{N}\EE\left\{\frac{\partial \log f_l(Y^{(l)}\mid \xi)}{\partial \xi_f} \frac{\partial \log f_l(Y^{(l)}\mid \xi)}{\partial \xi_f^\T} \bigg |_{\xi_f = \xi_f^*}\right\}\\ &= \frac{1}{N}I(\xi_f^*).
\end{align*}
We verify the multivariate Lindeberg-Feller condition to establish the central limit theorem of the sequence
\[
{
\frac{1}{\sqrt{N}}\sum_{l=1}^{N}
\left.
\frac{\partial \log f_l(Y^{(l)}\mid \xi)}
{\partial \xi_f}
\right|_{\xi_f=\xi_f^*}.}
\]
Let $g_l(Y^{(l)}) = \frac{\partial \log f_l(Y^{(l)} \mid \xi)}{\partial \xi_f}\mid_{\xi_f = \xi_f^*}$.
We set $C_0 = \max[\sup_{l =1}^{N}\EE\{\lVert \partial \log f_l(Y^{(l)}\mid \xi)/\partial \xi_f \mid_{\xi_f = \xi_f^*} \rVert_2^{2+\delta}\},\sup_{l =1}^{N}\EE\{\lVert \partial \log f_l(Y^{(l)}\mid \xi)/\partial \xi_f \mid_{\xi_f = \xi_f^*} \rVert_2^{2}\}]$ from Assumption \ref{asump:normality}.

For any $\epsilon>0$, by Markov's inequality and H{\"o}lder's inequality, we have
\begin{align*}
    &\lim_{N\rightarrow \infty}\frac{1}{Nv^2}\sum_{l =1}^{N}\EE\left[\norm{g_l(Y^{(l)})}^2_2\mathbf{1}\left\{\norm{g_l(Y^{(l)})}_2^2\geq \epsilon Nv^2 \right\}\right]\\
    & \leq \lim_{N\rightarrow \infty}\frac{1}{Nv^2}\sum_{l =1}^{N} \EE^{\frac{2}{2+\delta}}\left\{\norm{g_l(Y^{(l)})}_2^{2+\delta} \right\}\mathbb{P}^{\frac{\delta}{2+\delta}}\left\{ \norm{g_l(Y^{(l)})}_2^2 \geq \epsilon Nv^2\right\}\\
    & \leq \lim_{N\rightarrow \infty}\frac{1}{Nv^2}\sum_{l =1}^{N} \frac{C_0}{\left(\epsilon Nv^2 \right)^{\frac{\delta}{2+\delta}}}\\
    & \leq \lim_{N\rightarrow \infty}\frac{C_0}{Nv^2}\frac{N^{\frac{2}{2+\delta}}}{(\epsilon v^2)^\frac{\delta}{2+\delta}}\\
    &= \lim_{N\rightarrow \infty}\frac{C_0}{N^{\frac{\delta}{2+\delta}}v^{\frac{4+4\delta}{2+\delta}}\epsilon^{\frac{\delta}{2+\delta}}} = 0.
\end{align*}
Hence, the Lindeberg-Feller condition holds and we have
\[
\frac{1}{\sqrt{N}}\sum_{l =1}^{N}\frac{\partial \log f_l(Y^{(l)}\mid \xi)}{\partial \xi_f}\bigg |_{\xi_f = \xi_f^*} \overset{d}{\longrightarrow}\mathcal{N}\{0, I(\xi_f^*) \}.
\]

Finally, under Assumption \ref{asump:normality}, the uniform law of large numbers holds for the quantity $N^{-1}\sum_{l =1}^{N}\partial^2\log f_l(Y^{(l)}\mid \xi)/\partial \xi_f\partial \xi_f^\T$ over a small neighborhood of $\xi_f^*$. Given the consistency of $\hat{\xi}_f$ and the continuity of $\partial^2\log f_l(Y^{(l)}\mid \xi)/\partial \xi_f\partial \xi_f^\T$, we have
\[
\frac{1}{N}\sum_{l =1}^{N}\frac{\partial^2 \log f_l(Y^{(l)}\mid \xi)}{\partial \xi_f \partial \xi_f^\T}\bigg |_{\xi_f = \tilde{\xi}_f} \overset{p}{\longrightarrow} -I(\xi_f^*).
\]
Hence, all conditions in Theorem 3.1 of \cite{newey1994large} are satisfied, and we have that the result in Theorem \ref{thm:ci} holds by setting $\Psi = I(\xi_f^*)^{-1}$.
\end{proof}

\section{Derivation for the Fisher Information Matrix}\label{appendix.phi}

In this section, we derive the explicit form of the fisher information matrix $I(\xi_f^*)$, which has the form:
\[
I(\xi_f^*) = -\EE\left\{\frac{1}{N}\sum_{l =1}^{N}\frac{\partial^2 \log f_l(Y^{(l)}\mid \xi)}{\partial \xi_f \partial\xi_f^\T}\bigg |_{ \xi = \xi^*}\right\} ,
\]
where $f_l(Y^{(l)}\mid \xi)$ is defined for each $l\in [N]$ as
\[
f_l(Y^{(l)}\mid \xi) = \int_{\mathbb{R}^K}\prod_{i\in [n_l]}\prod_{j \in [J]} p\left\{Y_{ij}^{(l)}\mid \eta_{ij}^{(l)},\phi_j\right\}\pi (\alpha_l|\Sigma)d\alpha_l,
\]
Define
\[
P(Y^{(l)}\mid \xi, \alpha_l) = \prod_{i\in [n_l]}\prod_{j \in [J]} p\left\{Y_{ij}^{(l)}\mid \eta_{ij}^{(l)},\phi_j\right\},
\]
and $Q(Y^{(l)}, \alpha_l\mid \xi) = P(Y^{(l)}\mid \xi, \alpha_l)\pi(\alpha_l\mid \Sigma)$, observe that
\[
{
\frac{\partial  Q(Y^{(l)}, \alpha_l\mid \xi)}{\partial \xi_f} = \frac{\partial \log Q(Y^{(l)}, \alpha_l\mid\xi)}{\partial \xi_f}Q(Y^{(l)}, \alpha_l\mid \xi).}
\]
Let $\tilde{\pi}(\alpha_l\mid Y^{(l)}, \xi) = \{P(Y^{(l)}\mid \xi, \alpha_l)\pi(\alpha_l\mid \Sigma)/f_l(Y^{(l)}\mid \xi) \}$ be the posterior distribution of $\alpha_l$, under the assumption of interchangeability between integration and differentiation, we have
\begin{align*}
    \frac{\partial \log f_l}{\partial \xi_f} &= \frac{1}{f_l}\frac{\partial \int_{\mathbb{R}^K}Q(Y^{(l)}, \alpha_l \mid \xi)d\alpha_l}{\partial \xi_f}\\
    &=\frac{1}{f_l}\int_{\mathbb{R}^K}\frac{\partial  Q(Y^{(l)}, \alpha_l\mid \xi)}{\partial \xi_f}d\alpha_l\\
     &= \frac{1}{f_l}\int_{\mathbb{R}^K}\frac{\partial \log Q(Y^{(l)}, \alpha_l\mid \xi)}{\partial \xi_f}Q(Y^{(l)}, \alpha_l\mid \xi)d\alpha_l\\
    &= \frac{1}{f_l}\int_{\mathbb{R}^K}\frac{\partial \log Q(Y^{(l)}, \alpha_l\mid \xi)}{\partial \xi_f}P(Y^{(l)}\mid \xi, \alpha_l)\pi(\alpha_l\mid \Sigma)d\alpha_l\\
    &= \int_{\mathbb{R}^K}\frac{\partial \log Q(Y^{(l)},\alpha_l\mid \xi)}{\partial \xi_f} \frac{P(Y^{(l)}\mid \xi, \alpha_l)\pi(\alpha_l\mid \Sigma)}{f_l(Y^{(l)}\mid \xi)}d\alpha_l\\
    &= \int_{\mathbb{R}^K}\frac{\partial \log Q(Y^{(l)},\alpha_l\mid \xi)}{\partial \xi_f}\tilde{\pi}(\alpha_l \mid \xi, Y^{(l)})d\alpha_l.
\end{align*}
Moreover,
\begin{align*}
    &\frac{\partial^2 \log f_l}{\partial \xi_f \partial \xi_f^\T}\notag\\ &= \int_{\mathbb{R}^K}\left\{\frac{\partial^2 \log Q(Y^{(l)}, \alpha_l\mid \xi)}{\partial \xi_f \partial \xi_f^\T}\tilde{\pi}(\alpha_l\mid \xi, Y^{(l)}) + \frac{\partial \log Q(Y^{(l)}, \alpha_l\mid \xi)}{\partial \xi_f}\frac{\partial \tilde{\pi}(\alpha_l\mid\xi, Y^{(l)})}{\partial \xi_f^\T}\right\}d\alpha_l\notag\\
    &= \int_{\mathbb{R}^K}\left\{\frac{\partial^2 \log Q(Y^{(l)}, \alpha_l\mid \xi)}{\partial \xi_f \partial \xi_f^\T} + \frac{\partial \log Q(Y^{(l)}, \alpha_l\mid \xi)}{\partial \xi_f}\frac{\partial \log\tilde{\pi}(\alpha_l\mid \xi, Y^{(l)})}{\partial \xi_f^\T}\right\}\tilde{\pi}(\alpha_l\mid \xi, Y^{(l)})d\alpha_l\notag\\
&=\int_{\mathbb{R}^k}\left\{ \frac{\partial^2 \log Q(Y^{(l)}, \alpha_l\mid \xi)}{\partial \xi_f \partial \xi_f^\T} +\frac{\partial \log Q(Y^{(l)},\alpha_l\mid\xi)}{\partial \xi_f}\frac{\partial \log Q(Y^{(l)}, \alpha_l\mid\xi)}{\partial \xi_f^\T} \right\}\\
&\times \tilde{\pi}(\alpha_l\mid \xi, Y^{(l)})d\alpha_l\\
&-\bigg[ \left\{ \int_{\mathbb{R}^K}\frac{\partial \log Q(Y^{(l)},\alpha_l\mid \xi)}{\partial \xi_f}\tilde{\pi}(\alpha_l \mid \xi, Y^{(l)})d\alpha_l\right\}\\
&\times\left\{ \int_{\mathbb{R}^K}\frac{\partial \log Q(Y^{(l)},\alpha_l\mid \xi)}{\partial\xi_f^\T}\tilde{\pi}(\alpha_l \mid \xi, Y^{(l)})d\alpha_l\right\}\bigg]\notag.
\end{align*}

Next, we derive $\partial \log Q(Y^{(l)}, \alpha_l\mid \xi)/\partial \xi$ and $\partial^2 \log Q(Y^{(l)}, \alpha_l)/\partial \xi_f \partial \xi_f^\T$. Here, we write the numerator as $\log Q$ for simplicity.
Recall the definition of free loadings in Section~\ref{sec_theory} such that $\lambda_v = (\bar{\lambda}_{1}^\T, \dots, \bar{\lambda}_{J}^\T)^\T$ and $\beta_v = (\beta_1^\T,\beta_2^\T,\beta_3^\T)^\T$. Then
\begin{align}\label{QGrad}
\frac{\partial \log Q}{\partial \xi_f} = \left(
    \frac{\partial \log Q}{\partial \lambda_v},
     \frac{\partial \log Q}{\partial \beta_v},
     \frac{\partial \log Q}{\partial b},
     \frac{\partial \log Q}{\partial \phi},
     \frac{\partial \log Q}{\partial \mathrm{vech}(\Sigma)}
\right)^T,
\end{align}
\begin{align}\label{QHessian}
    &\frac{\partial^2 \log Q}{\partial \xi_f \partial\xi_f^\T}=\notag\\
    & \begin{pmatrix}
    \frac{\partial^2 \log Q}{\partial \lambda_v\partial\lambda_v^\T} &\frac{\partial^2 \log Q}{\partial \lambda_v\partial\beta_v^\T} &\frac{\partial^2 \log Q}{\partial \lambda_v\partial b^\T}  & \frac{\partial^2 \log Q}{\partial \lambda_v\partial \phi^\T} & \frac{\partial^2 \log Q}{\partial \lambda_v\partial \mathrm{vech}^\T(\Sigma)}\\
    \frac{\partial^2\log Q}{\partial \beta_v \partial \lambda_v^\T}&  \frac{\partial^2\log Q}{\partial \beta_v \partial \beta_v^\T}&  \frac{\partial^2\log Q}{\partial \beta_v \partial  b^\T}&\frac{\partial^2\log Q}{\partial \beta_v \partial\phi^\T} & \frac{\partial^2\log Q}{\partial \beta_v \partial \mathrm{vech}^\T(\Sigma)}\\
    \frac{\partial^2\log Q}{\partial b \partial \lambda_v^\T} & \frac{\partial^2\log Q}{\partial b \partial \beta_v^\T}&  \frac{\partial^2\log Q}{\partial b \partial b^\T}&\frac{\partial^2\log Q}{\partial b \partial\phi^\T} & \frac{\partial^2\log Q}{\partial b \partial \mathrm{vech}^\T(\Sigma)}\\
    \frac{\partial^2\log Q}{\partial \phi \partial \lambda_v^\T}&\frac{\partial^2\log Q}{\partial \phi \partial \beta_v^\T}& \frac{\partial^2\log Q}{\partial \phi \partial b^\T}&\frac{\partial^2\log Q}{\partial \phi \partial\phi^\T} & \frac{\partial^2\log Q}{\partial \phi \partial \mathrm{vech}^\T(\Sigma)}\\
    \frac{\partial^2\log Q}{\partial \mathrm{vech}(\Sigma) \partial \lambda_v^\T} & \frac{\partial^2\log Q}{\partial \mathrm{vech}(\Sigma) \partial \beta_v^\T} & \frac{\partial^2\log Q}{\partial \mathrm{vech}(\Sigma) \partial b^\T}&\frac{\partial^2\log Q}{\partial \mathrm{vech}(\Sigma) \partial\phi^\T} & \frac{\partial^2\log Q}{\partial \mathrm{vech}(\Sigma) \partial \mathrm{vech}^\T(\Sigma)}\\
\end{pmatrix}.
\end{align}

Next, we calculate the elements in the above equations explicitly.
For any $l$, we start by making some definitions for the following intermediate quantities.
Here we omit the superscript $(l)$ for simplicity. We start with introducing some intermediate functions.
{Let $\sigma(x) = 1/( 1+e^{-x})$, $\psi(x) = \Gamma'(x)/\Gamma(x)$, $\psi'(x) = \{\Gamma''(x)\Gamma(x) - \Gamma^{'2}(x) \}/\Gamma^2(x)$, $\mu'(\eta_{ij})
=
(1-\gamma_j)\sigma(\eta_{ij})\{1-\sigma(\eta_{ij})\}$, and $\mu''(\eta_{ij})
=
(1-\gamma_j)\sigma(\eta_{ij})\{1-\sigma(\eta_{ij})\}
\{1-2\sigma(\eta_{ij})\}$.
We have
\begin{enumerate}
\item 
\begin{align*}
&G^\eta_{ij}\\ &= \frac{\partial \log p\left(Y_{ij}\mid \eta_{ij},\phi_j\right)}{\partial \eta_{ij}}\\
& = \bigg[ -\psi\{\mu(\eta_{ij}){\phi_j}\}  + \psi[\{1 - \mu(\eta_{ij}) \}{\phi_j} ]+ \log \frac{Y_{ij}}{1 - Y_{ij}} \bigg]\phi_j \mu'(\eta_{ij}).
\end{align*}
\item \begin{align*}
G^{\phi_j}_{ij} & = \frac{\partial \log p\left(Y_{ij}\mid \eta_{ij},\phi_j\right)}{\partial {\phi_j}}\\
&=\psi(\phi_j) - \psi\{ \mu(\eta_{ij}){\phi_j}\}\mu(\eta_{ij})- \psi[\{1 - \mu(\eta_{ij})\}{\phi_j}]\{1 - \mu(\eta_{ij}) \}\\
&+\mu(\eta_{ij})\log Y_{ij}+ \{1 - \mu(\eta_{ij}) \}\log(1 - Y_{ij}).
    \end{align*}
    \item 
    \begin{align*}
J^{\eta\eta}_{ij}
&=
\frac{\partial^2 \log p\left(Y_{ij}\mid \eta_{ij},\phi_j\right)}
{\partial \eta_{ij}^2}\\
&=
-\phi_j^2
\left[
\psi'\{\phi_j\mu(\eta_{ij})\}
+
\psi'[\phi_j\{1-\mu(\eta_{ij})\}]
\right]
\{\mu'(\eta_{ij})\}^2\\
&\quad+
\bigg[
-\phi_j\psi\{\mu(\eta_{ij})\phi_j \}
+\phi_j \psi[\{1 - \mu(\eta_{ij}) \}\phi_j]
+
\phi_j\log\frac{Y_{ij}}{1-Y_{ij}}
\bigg]
\mu''(\eta_{ij}),
\end{align*}
\item 
\begin{align*}
J^{\eta{\phi_j}}_{ij}
&= \frac{\partial^2\log p\left(Y_{ij}\mid \eta_{ij},\phi_j\right)}
{\partial \eta_{ij} \partial \phi_j}\\
&=
\bigg[
-\psi\{\mu(\eta_{ij})\phi_j\}
-\phi_j\mu(\eta_{ij})\psi'\{\mu(\eta_{ij})\phi_j\}\\
&\quad
+\psi[\{1-\mu(\eta_{ij})\}\phi_j]
+\phi_j\{1-\mu(\eta_{ij})\}
\psi'[\{1-\mu(\eta_{ij})\}\phi_j]\\
&\quad
+\log\frac{Y_{ij}}{1-Y_{ij}}
\bigg]\mu'(\eta_{ij}),
\end{align*}
\item 
\begin{align*}
J^{{\phi_j}{\phi_j}}_{ij}
&=
\frac{\partial^2\log p\left(Y_{ij}\mid \eta_{ij},\phi_j\right)}
{\partial \phi_j^2}\\
&= \psi'(\phi_j) - \mu^2(\eta_{ij})\psi'\{\mu(\eta_{ij})\phi_j \} - \{1 - \mu(\eta_{ij}) \}^2 \psi'[\{ 1 - \mu(\eta_{ij})\}\phi_j].
\end{align*}
\end{enumerate}}
Based on the definitions above, we calculate the key quantities involved in \eqref{QGrad}--\eqref{QHessian}.
For each $j \in [J]$ and $\bar{\lambda}_j \in \mathbb{R}^{K}$ or $\mathbb{R}$, we have
\begin{align*}
    &\frac{\partial \log Q(Y, \alpha_l \mid \xi)}{\partial \bar{\lambda}_j}\\
    &= \frac{\sum_{i = 1}^{n_l}\sum_{r =1}^{J}\log p\left(Y_{ir}\mid \eta_{ir},\phi_r\right)+ \log \pi (\alpha_l\mid \Sigma)}{\partial \bar{\lambda}_j}
    =\frac{\sum_{i = 1}^{n_l}\log p\left(Y_{ij}\mid \eta_{ij},\phi_j\right)}{\partial \bar{\lambda}_j}\\
    &=\sum_{i = 1}^{n_l}G^{\eta}_{ij}\bar{\theta}_{ij}(s,t).
\end{align*}
Here, $\bar{\theta}_{ij}(s,t) = \theta_{i,k}(s,t)$ for $j \in \mathcal{S}_k,~k \in [K]$, and $\bar{\theta}_{ij}(s,t)= \theta_i(s,t)$ otherwise.
For each $p\in \{1, 2,3 \}$, let $\beta_p^\T \in \mathbb{R}^K$ be the $p$-th row of the $3\times K$ matrix $\beta$.
We have
\begin{align*}
     \frac{\partial \log Q(Y, \alpha_l \mid \xi)}{\partial \beta_p}
    = \frac{\sum_{i = 1}^{n_l}\sum_{j =1}^{J} \log p\left(Y_{ij}\mid \eta_{ij},\phi_j\right)}{\partial \beta_p}
    =\sum_{i = 1}^{n_l}\sum_{j =1}^{J} G^\eta_{ij}\frac{\partial \eta_{ij}}{\partial \beta_p}.
\end{align*}
Note that $\eta_{ij} = \lambda_j^\T(\alpha_l + \beta^\T x_i) + b_j$, we have
\begin{align*}
    \frac{\partial \eta_{ij}}{\partial \beta_p} = \frac{\lambda_j^\T (\beta_1x_{i1} + \beta_2x_{i2} +\beta_3x_{i3}+\alpha_l ) }{\partial\beta_p} = \lambda_{j}x_{ip}.
\end{align*}
We therefore obtain the following.
\[
\frac{\partial \log Q(Y, \alpha_l \mid \xi)}{\partial \beta_p} = \sum_{i = 1}^{n_l}\sum_{j =1}^{J} G^\eta_{ij}\lambda_jx_{ip}.
\]
Next, for each $j \in [J]$, the quantity $b_j$ is a scalar, we have
\begin{align*}
&\frac{\partial \log Q(Y, \alpha_l\mid \xi)}{\partial b_j}
= \frac{\sum_{i =1}^{n_l}\log p\left(Y_{ij}\mid \eta_{ij},\phi_j\right)}{\partial b_j}
 =\sum_{i = 1}^{n_l}G^{\eta}_{ij}.\\
 &\frac{\partial \log Q(Y, \alpha_l\mid \xi)}{\partial b^\T} = \left\{\sum_{i = 1}^{n_l}G^{\eta}_{i1}, \ldots, \sum_{i = 1}^{n_l}G^{\eta}_{iJ}\right\}^\T.
\end{align*}
Similarly, the quantity $\phi_j$ is a scalar, and we have
\begin{align*}
    &\frac{\partial \log Q(Y, \alpha_l \mid \xi)}{\partial \phi_j}
    =\frac{\sum_{i =1}^{n_l}\log p\left(Y_{ij}\mid \eta_{ij},\phi_j\right)}{\partial \phi_j}
    =\sum_{i = 1}^{n_l}G^{\phi_j
}_{ij}.\\
& \frac{\partial \log Q(Y, \alpha_l\mid \xi)}{\partial \phi^\T} = \left\{\sum_{i = 1}^{n_l}G^{\phi_1
}_{i1}, \ldots,\sum_{i = 1}^{n_l}G^{\phi_J
}_{iJ} \right\}^\T
\end{align*}
Finally, let $\sigma_{k_1k_2}$ be the $(k_1,k_2)$-th entry of the matrix $\Sigma$ such that $1 \leq k_1< k_2 \leq K$.
We have
\begin{align*}
    \frac{\partial \log Q(Y, \alpha_l\mid \xi)}{\partial \sigma_{k_1k_2}} &= \frac{\partial \log \pi (\alpha_l\mid \Sigma)}{\partial \sigma_{k_1 k_2}}\\
    &= \frac{\partial\left\{-\frac{K}{2}\log 2\pi - \frac{1}{2}\log |\Sigma| - \frac{1}{2}\alpha_l^\T \Sigma^{-1}\alpha_l \right\}}{\partial \sigma_{k_1, k_2}} \\
    &=-\left\{\Sigma^{-1} - \Sigma^{-1}\alpha_l\alpha_l^\T\Sigma^{-1} \right\}_{k_1, k_2}.
\end{align*}
And
\begin{align*}
    &\frac{\partial \log Q(Y, \alpha_l\mid \xi)}{\partial \mathrm{vech}^\T(\Sigma)}\\&=\left\{\frac{\partial \log Q(Y, \alpha_l\mid \xi)}{\partial \sigma_{21}}, \ldots, \frac{\partial \log Q(Y, \alpha_l\mid \xi)}{\partial \sigma_{K1}}, \ldots, \frac{\partial \log Q(Y, \alpha_l\mid \xi)}{\partial \sigma_{K,K-1}} \right\}^\T.
\end{align*}
Now we compute the Hessian components.
For each $j,k\in [J]$ and $\bar{\lambda}_j, \bar{\lambda}_k \in \mathbb{R}^{K}$ or $\mathbb{R}$, we have
\begin{align*}
    \frac{\partial^2 \log Q(Y, \alpha_l \mid \xi)}{\partial \bar{\lambda}_j\partial\bar{\lambda}_j^\T} = \sum_{i = 1}^{n_l}J^{\eta\eta}_{ij}\bar{\theta}_{ij}(s,t)\bar{\theta}_i^\T(s,t).
\end{align*}
If $j\neq k$, we have
\begin{align*}
    \frac{\partial^2 \log Q(Y, \alpha_l, \mid \xi)}{\partial \bar{\lambda}_j \partial \bar{\lambda}_k^\T} = {0}.
\end{align*}
For each $p \in \{1,2,3 \},~\beta_p \in \mathbb{R}^K$, and some $j$ such that $\bar{\lambda}_j \in \mathbb{R}^k$, we have
\begin{align*}
    &\frac{\partial^2 \log Q(Y, \alpha_l\mid \xi)}{\partial \bar{\lambda}_j \partial \beta_p^\T}=\sum_{i = 1}^{n_l}J^{\eta\eta}_{ij}x_{ip}\theta_i(s,t)\bar{\lambda}_j^\T+\sum_{i = 1}^{n_l}G^\eta_{ij}x_{ip}{I}_K.
\end{align*}
Here, we only consider the unconstrained $\lambda_j$ for simplicity. For constrained case, 
when $\bar{\lambda}_j \in \mathbb{R}$, such a quantity can be computed similarly taking the corresponding entries in the above equation.
For each $j,k \in [J]$, $\bar{\lambda}_j \in \mathbb{R}^K$ or $\mathbb{R}$ and $b_k \in \mathbb{R}$, when $j = k$, we have
\begin{align*}
    &\frac{\partial^2 \log Q(Y.\alpha_l\mid \xi)}{\partial \bar{\lambda}_j\partial b_j}
    =\sum_{i = 1}^{n_l}J^{\eta\eta}_{ij}\bar{\theta}_{ij}(s,t).
\end{align*}
When $j\neq k$,
\begin{align*}
    &\frac{\partial^2 \log Q(Y.\alpha_l\mid \xi)}{\partial \bar{\lambda}_j\partial b_k} = {0}.
\end{align*}
For each $j \in [J]$, $\bar{\lambda}_j \in \mathbb{R}^K$ or $\mathbb{R}$ and $\phi_j \in \mathbb{R}$, we have
\begin{align*}
    &\frac{\partial^2 \log Q(Y,\alpha_l\mid \xi)}{\partial \bar{\lambda}_j\partial \phi_j} = \sum_{i = 1}^{n_l}J^{\eta\phi_j}_{ij}\bar{\theta}_{ij}(s,t).
\end{align*}
When $j \neq k$
\[
\frac{\partial^2 \log Q(Y,\alpha_l\mid \xi)}{\partial \bar{\lambda}_j\partial \phi_k} = 0.
\]
Moreover
\begin{align*}
    \frac{\partial^2 \log Q(Y, \alpha_l\mid \xi)}{\partial \bar{\lambda}_j \partial \mathrm{vech}(\Sigma)^\T} = {0}.
\end{align*}
For each $p_1, p_2 \in \{1,2,3 \}$, $\beta_{p_1}, \beta_{p_2}\in \mathbb{R}^K$, we have
\begin{align*}
    &\frac{\partial^2 \log Q(Y, \alpha_l\mid \xi)}{\partial \beta_{p_1} \partial \beta_{p_2}^\T}
     =\sum_{j =1}^{J}\sum_{i = 1}^{n_l}J^{\eta\eta}_{ij}\lambda_j\lambda_j^\T x_{ip_1}x_{ip_2}.
\end{align*}
For each $p\in \{1,2,3 \}$ and each $j \in [J]$, we have
\begin{align*}
    &\frac{\partial^2 \log Q(Y, \alpha_l\mid \xi)}{\partial \beta_{p} \partial b_j}=\sum_{i = 1}^{n_l}J^{\eta\eta}_{ij}\lambda_jx_{ip}.
\end{align*}
Moreover, for each $j \in [J]$
\begin{align*}
    &\frac{\partial^2 \log Q(Y, \alpha_l\mid \xi)}{\partial \beta_{p} \partial \phi_j}=\sum_{i = 1}^{n_l}J^{\eta\phi_j}_{ij}\lambda_jx_{ip}.
\end{align*}
And
\begin{align*}
    &\frac{\partial^2 \log Q(Y, \alpha_l\mid \xi)}{\partial \beta_{p} \partial \mathrm{vech}^\T(\Sigma)} = {0}.
\end{align*}
For any $j, k \in [J]$, we have
\begin{align*}
    &\frac{\partial^2 \log Q(Y, \alpha_l\mid \xi)}{\partial b_j^2}= \sum_{i = 1}^{n_l}J^{\eta\eta}_{ij}.
\end{align*}
If $j \neq k$, we have
\[
\frac{\partial^2 \log Q(Y, \alpha_l\mid \xi)}{\partial b_j \partial b_k} =0.
\]
Moreover
\begin{align*}
    &\frac{\partial^2 \log Q(Y, \alpha_l\mid \xi)}{\partial b_j \partial \phi_j}=\sum_{i = 1}^{n_l}J^{\eta\phi_j}_{ij}.
\end{align*}
If $j \neq k$
\[
\frac{\partial^2 \log Q(Y, \alpha_l\mid \xi)}{\partial b_j \partial \phi_k} = 0.
\]
And
\begin{align*}
   &\frac{\partial^2 \log Q(Y, \alpha_l\mid \xi)}{\partial b_j \partial \mathrm{vech}^\T(\Sigma)}  = {0}.
\end{align*}
For $\phi_j \in \mathbb{R}$ and $j \in [J]$, we have
\begin{align*}
    & \frac{\partial^2 \log Q(Y, \alpha_l \mid 
    \xi)}{\partial \phi^2_j} =\sum_{i = 1}^{n_l}J^{\phi_j\phi_j}_{ij}.
\end{align*}
Additionally, 
\begin{align*}
    &\frac{\partial^2 \log Q(Y, \alpha_l\mid \xi)}{\partial \phi \partial \mathrm{vech}^\T(\Sigma)}  = {0}.
\end{align*}
Finally, for $1\leq k_1, k_2, k_3, k_4\leq K$, $k_1<k_2$, $k_3<k_4$, we have
\begin{align*}
    &\frac{\partial^2 \log Q(Y, \alpha_l\mid \xi)}{\partial\sigma_{k_1k_2}\partial \sigma_{k_3, k_4}}\\
    &=\frac{1}{2}\left\{\Sigma^{-1}\otimes \Sigma^{-1} - \left(\Sigma^{-1}\alpha_l\alpha_l^\T\Sigma^{-1}\right) \otimes \Sigma^{-1} - \Sigma^{-1}\otimes\left(\Sigma^{-1}\alpha_l\alpha_l^\T\Sigma^{-1}\right) \right\}_{(k_1,k_2),(k_3, k_4)}.
\end{align*}
By using above calculations, one can obtain $I(\xi_f^*)$ by taking the expectation with respect to $\{Y^{(l)}\}_{l = 1}^N$.
And the covariance matrix $\Psi$ in Theorem~\ref{thm:ci} in the main paper is defined as $I(\xi_f^*)^{-1}$.


\section{Metropolis--Hastings Sampling for the Approximated Posterior of $\alpha_l$}\label{append:mcmc}

In this section, we present Algorithm \ref{alg:mcmc}, which gives the steps to sample from the posterior distribution of $\alpha_l$. The algorithm requires the specification of $\alpha_l^{(0)}$ and $\Sigma_{\text{prop}}$, which we set to the mean and covariance of the Laplace approximation of the posterior distribution of $\alpha_l$, respectively.

\renewcommand{\thealgorithm}{S\arabic{algorithm}}
\begin{algorithm}[H]
\caption{Metropolis--Hastings Sampling for the Approximated Posterior of $\alpha_l$}
\label{alg:mcmc}
\begin{algorithmic}[1]
\Require \\
\begin{itemize}
    \item Data: responses $Y^{(l)}$, covariates $\{x_i\}_{i\in I_l}$;
    \item Model parameter estimates: $\hat{\Lambda}$, $\hat{\beta}$, $\hat{b}$, $\hat{\phi}$, $\hat{\Sigma}$
    \item Hyperparameters: Burn-in $B$ (default $100$), Thinning $T$ (default $10$), Total iterations $N$ (default $5000$)
    \item Initial state: $\alpha_l^{(0)}$
    \item Proposal covariance: $\Sigma_{\text{prop}}$.
\end{itemize}
\For{$t=1$ to $N$}
    \State \textbf{Propose} $\alpha_l^{*} \sim N\Big(\alpha_l^{(t-1)}, \Sigma_{\text{prop}}\Big)$
    \State \textbf{Compute}:
    \[
    q(\alpha_l^{*}) = \prod_{i\in [n_l]}\prod_{j \in [J]} f\Big(Y_{ij} : \mu\Big\{ (\hat{\lambda}_j)^\intercal \Big(\alpha_l^{*}+(\hat{\beta})^\T x_i\Big) + \hat{b}_j\Big\},\hat{\phi}\Big) \pi\Big(\alpha_l^{*}|\hat{\Sigma}\Big)
    \]
    \State Similarly compute $q(\alpha_l^{(t-1)})$.
    \State \textbf{Calculate} the log acceptance ratio:
    \[
     r = \log q(\alpha_l^{*})  - \log q(\alpha_l^{(t-1)}) 
    \]
    \State \textbf{Draw} $u \sim \text{Uniform}(0,1)$.
    \If{$\log u \le r$}
        \State Set $\alpha_l^{(t)} \gets \alpha_l^{*}$.
    \Else
        \State Set $\alpha_l^{(t)} \gets \alpha_l^{(t-1)}$.
    \EndIf
\EndFor
\State \textbf{Discard} the first $B$ samples as burn-in.
\State \textbf{Thin} the remaining samples by retaining every $T$th iteration.
\State \Return The thinned sample set as approximate draws from $f(\alpha_l|Y^{(l)},\xi)$.
\end{algorithmic}
\end{algorithm}

\section{Additional Experiment Results}
In this section, we present additional results for the experiments in Section~\ref{sec_experiment}. Section~\ref{append:llms} provides the details of the used LLMs in the experiment. Section~\ref{supp_sec_aic} presents the AIC curve and the results under the setting of latent dimension $K=6$.
\subsection{LLMs Used}\label{append:llms}

The following table summarizes the LLMs used in the experiments in Section~\ref{sec_experiment}, including the model name, the model family, and the leaderboard on which each model appears.

{\scriptsize
\begin{longtable}{lllrr}
\toprule
 & model & family & leaderboard-1 & leaderboard-2 \\
\midrule
0 & athena-gemma-2-2b-it & athena-gemma-2-it & False & True \\
1 & bio-medical-llama-3-8b & bio-medical-llama-3 & False & True \\
2 & bloom & bloom & True & False \\
3 & bloom-1b1 & bloom & True & True \\
4 & bloom-3b & bloom & True & True \\
5 & bloom-560m & bloom & True & True \\
6 & bloom-7b1 & bloom & True & True \\
7 & blossom-v5.1-34b & blossom-v5.1 & True & True \\
8 & blossom-v5.1-9b & blossom-v5.1 & False & True \\
9 & braincog-8b-0.1-instruct & braincog-0.1-instruct & False & True \\
10 & calme-2.1-qwen2-72b & calme-2.1-qwen2 & False & True \\
11 & calme-2.1-qwen2-7b & calme-2.1-qwen2 & False & True \\
12 & calme-2.2-llama3-70b & calme-2.2-llama3 & False & True \\
13 & calme-2.2-qwen2-72b & calme-2.2-qwen2 & False & True \\
14 & calme-2.2-qwen2-7b & calme-2.2-qwen2 & False & True \\
15 & calme-2.3-llama3-70b & calme-2.3-llama3 & False & True \\
16 & calme-2.3-qwen2-72b & calme-2.3-qwen2 & False & True \\
17 & calme-2.3-qwen2-7b & calme-2.3-qwen2 & False & True \\
18 & calme-2.4-llama3-70b & calme-2.4-llama3 & False & True \\
19 & calme-2.4-qwen2-7b & calme-2.4-qwen2 & False & True \\
20 & calme-2.5-qwen2-7b & calme-2.5-qwen2 & False & True \\
21 & calme-2.6-qwen2-7b & calme-2.6-qwen2 & False & True \\
22 & calme-2.7-qwen2-7b & calme-2.7-qwen2 & False & True \\
23 & codegen-16b-nl & codegen-nl & True & False \\
24 & codegen-6b-nl & codegen-nl & True & False \\
25 & codellama-13b & codellama & True & False \\
26 & codellama-34b & codellama & True & False \\
27 & codellama-70b & codellama & True & False \\
28 & codellama-70b-instruct & codellama-instruct & True & False \\
29 & codellama-7b & codellama & True & False \\
30 & configurable-hermes-2-pro-llam & configurable-hermes-2-pro-llam & True & True \\
31 & cosmosage-v3 & cosmosage-v3 & False & True \\
32 & deepseek-coder-1.3b-base & deepseek-coder-base & True & False \\
33 & deepseek-coder-33b-base & deepseek-coder-base & True & False \\
34 & deepseek-coder-6.7b-base & deepseek-coder-base & True & False \\
35 & delirium-v1 & delirium-v1 & False & True \\
36 & dolly-v2-12b & dolly-v2 & True & True \\
37 & dolly-v2-3b & dolly-v2 & False & True \\
38 & dolly-v2-7b & dolly-v2 & True & True \\
39 & dolphin-2.9-llama3-8b & dolphin-2.9-llama3 & True & True \\
40 & dolphin-2.9.1-llama-3-70b & dolphin-2.9.1-llama-3 & False & True \\
41 & dolphin-2.9.1-yi-1.5-34b & dolphin-2.9.1-yi-1.5 & True & True \\
42 & dolphin-2.9.1-yi-1.5-9b & dolphin-2.9.1-yi-1.5 & True & True \\
43 & dolphin-2.9.2-qwen2-72b & dolphin-2.9.2-qwen2 & False & True \\
44 & dolphin-2.9.2-qwen2-7b & dolphin-2.9.2-qwen2 & False & True \\
45 & dolphin-2.9.4-gemma2-2b & dolphin-2.9.4-gemma2 & False & True \\
46 & einstein-v6.1-llama3-8b & einstein-v6.1-llama3 & True & True \\
47 & einstein-v7-qwen2-7b & einstein-v7-qwen2 & False & True \\
48 & falcon-180b & falcon & True & False \\
49 & falcon-40b & falcon & True & True \\
50 & falcon-40b-instruct & falcon-instruct & False & True \\
51 & falcon-7b & falcon & True & True \\
52 & falcon-7b-instruct & falcon-instruct & True & True \\
53 & falcon-rw-1b & falcon-rw & True & False \\
54 & fietje-2 & fietje-2 & False & True \\
55 & fietje-2-chat & fietje-2-chat & False & True \\
56 & fietje-2-instruct & fietje-2-instruct & False & True \\
57 & gemma-2-2b & gemma-2 & False & True \\
58 & gemma-2-2b-it & gemma-2-it & False & True \\
59 & gemma-2-2b-jpn-it & gemma-2-jpn-it & False & True \\
60 & gemma-2-2b-jpn-it-abliterated- & gemma-2-jpn-it-abliterated-17 & False & True \\
61 & gemma-2-2b-jpn-it-abliterated- & gemma-2-jpn-it-abliterated-17- & False & True \\
62 & gemma-2-2b-jpn-it-abliterated- & gemma-2-jpn-it-abliterated-17- & False & True \\
63 & gemma-2-2b-jpn-it-abliterated- & gemma-2-jpn-it-abliterated-18 & False & True \\
64 & gemma-2-2b-jpn-it-abliterated- & gemma-2-jpn-it-abliterated-18- & False & True \\
65 & gemma-2-2b-jpn-it-abliterated- & gemma-2-jpn-it-abliterated-24 & False & True \\
66 & gemma-2-2b-opus-instruct & gemma-2-opus-instruct & False & True \\
67 & gemma-2-2b-orpo-jpn-it-abliter & gemma-2-orpo-jpn-it-abliterate & False & True \\
68 & gemma-2-2b-orpo-jpn-it-abliter & gemma-2-orpo-jpn-it-abliterate & False & True \\
69 & gemma-2-2b-stheno-filtered & gemma-2-stheno-filtered & False & True \\
70 & gemma-2-9b & gemma-2 & False & True \\
71 & gemma-2-9b-it & gemma-2-it & False & True \\
72 & gemma-2-9b-it-dpo & gemma-2-it-dpo & False & True \\
73 & gemma-2-9b-it-simpo & gemma-2-it-simpo & False & True \\
74 & gemma-2-9b-it-wpo-hb & gemma-2-it-wpo-hb & False & True \\
75 & gemma-2-9b-moth & gemma-2-moth & False & True \\
76 & gemma-2b & gemma & True & True \\
77 & gemma-2b-it & gemma-it & True & True \\
78 & gemma-2b-orpo & gemma-orpo & True & True \\
79 & gemma-7b & gemma & True & True \\
80 & gemma-7b-it & gemma-it & True & True \\
81 & gemma2-9b-it-psy10k-mental-hea & gemma2-it-psy10k-mental-health & False & True \\
82 & gemma2-9b-it-simpo-infinity-pr & gemma2-it-simpo-infinity-prefe & False & True \\
83 & gemma2-9b-it-train6 & gemma2-it-train6 & False & True \\
84 & gpt-j-6b & gpt-j-neo-neox & True & False \\
85 & gpt-neo-1.3b & gpt-j-neo-neox & True & True \\
86 & gpt-neo-125m & gpt-j-neo-neox & True & False \\
87 & gpt-neo-2.7b & gpt-j-neo-neox & True & True \\
88 & gpt-neox-20b & gpt-j-neo-neox & True & False \\
89 & henbane-7b-attempt2 & henbane-attempt2 & False & True \\
90 & hermes-2-pro-llama-3-8b & hermes-2-pro-llama-3 & True & True \\
91 & hermes-2-theta-llama-3-8b & hermes-2-theta-llama-3 & True & True \\
92 & higgs-llama-3-70b & higgs-llama-3 & False & True \\
93 & humanish-llama3-8b-instruct & humanish-llama3-instruct & False & True \\
94 & internlm2-20b & internlm2 & True & False \\
95 & internlm2-7b & internlm2 & True & False \\
96 & l3-pneuma-8b & l3-pneuma & False & True \\
97 & l3-pneuma-8b & l3-pneuma & False & True \\
98 & lambda-gemma-2-9b-dpo & lambda-gemma-2-dpo & False & True \\
99 & leniachat-gemma-2b-v0 & leniachat-gemma-v0 & False & True \\
100 & leniachat-qwen2-1.5b-v0 & leniachat-qwen2-v0 & False & True \\
101 & llama-13b & llama & True & True \\
102 & llama-2-13b & llama-2 & True & True \\
103 & llama-2-13b-chat & llama-2-chat & True & True \\
104 & llama-2-70b & llama-2 & True & True \\
105 & llama-2-70b-chat & llama-2-chat & True & True \\
106 & llama-2-7b & llama-2 & True & True \\
107 & llama-2-7b-chat & llama-2-chat & True & True \\
108 & llama-3-6.3b-v0.1 & llama-3-v0.1 & False & True \\
109 & llama-3-70b-instruct-v0.1 & llama-3-instruct-v0.1 & False & True \\
110 & llama-3-8b-instruct & llama-3-instruct & False & True \\
111 & llama-3-8b-instruct-gapo-v2-be & llama-3-instruct-gapo-v2-bert- & False & True \\
112 & llama-3-8b-instruct-gapo-v2-be & llama-3-instruct-gapo-v2-bert- & False & True \\
113 & llama-3-8b-instruct-gapo-v2-be & llama-3-instruct-gapo-v2-bert- & False & True \\
114 & llama-3-8b-instruct-gapo-v2-ro & llama-3-instruct-gapo-v2-rouge & False & True \\
115 & llama-3-8b-instruct-gapo-v2-ro & llama-3-instruct-gapo-v2-rouge & False & True \\
116 & llama-3-8b-instruct-gapo-v2-ro & llama-3-instruct-gapo-v2-rouge & False & True \\
117 & llama-3-8b-instruct-v0.10 & llama-3-instruct-v0.10 & False & True \\
118 & llama-3-8b-instruct-v0.8 & llama-3-instruct-v0.8 & True & True \\
119 & llama-3-8b-instruct-v0.9 & llama-3-instruct-v0.9 & True & True \\
120 & llama-3-8b-ita & llama-3-ita & True & True \\
121 & llama-3-8b-magpie-align-sft-v0 & llama-3-magpie-align-sft-v0.1 & False & True \\
122 & llama-3-8b-magpie-align-sft-v0 & llama-3-magpie-align-sft-v0.3 & False & True \\
123 & llama-3-8b-magpie-align-v0.1 & llama-3-magpie-align-v0.1 & False & True \\
124 & llama-3-8b-magpie-align-v0.3 & llama-3-magpie-align-v0.3 & False & True \\
125 & llama-3-groq-8b-tool-use & llama-3-groq-tool-use & False & True \\
126 & llama-3-instruct-8b-simpo & llama-3-instruct-simpo & False & True \\
127 & llama-3-sauerkrautlm-70b-instr & llama-3-sauerkrautlm-instruct & False & True \\
128 & llama-3-sauerkrautlm-8b-instru & llama-3-sauerkrautlm-instruct & True & True \\
129 & llama-3.1-distilled & llama-3.1-distilled & False & True \\
130 & llama-30b & llama & True & False \\
131 & llama-65b & llama & True & True \\
132 & llama-7b & llama & True & True \\
133 & llama3-openbiollm-70b & llama3-openbiollm & False & True \\
134 & llamantino-3-anita-8b-inst-dpo & llamantino-3-anita-inst-dpo-it & True & True \\
135 & luminia-13b-v3 & luminia-v3 & True & True \\
136 & magnum-72b-v1 & magnum-v1 & False & True \\
137 & magnum-v1-72b & magnum-v1 & False & True \\
138 & magnum-v2-72b & magnum-v2 & False & True \\
139 & magnum-v3-9b-customgemma2 & magnum-v3-customgemma2 & False & True \\
140 & meta-llama-3-70b & meta-llama-3 & True & True \\
141 & meta-llama-3-70b-instruct & meta-llama-3-instruct & True & True \\
142 & meta-llama-3-8b & meta-llama-3 & True & True \\
143 & meta-llama-3-8b-instruct & meta-llama-3-instruct & True & True \\
144 & meta-llama-3-8bee & meta-llama-3-ee & True & True \\
145 & mpt-30b & mpt & True & False \\
146 & mpt-30b-chat & mpt-chat & True & False \\
147 & mpt-30b-instruct & mpt-instruct & True & False \\
148 & mpt-7b & mpt & True & False \\
149 & mpt-7b-chat & mpt-chat & True & False \\
150 & mpt-7b-instruct & mpt-instruct & True & False \\
151 & n3n-delirium-v1-1030-0227 & n3n-delirium-v1-1030-0227 & False & True \\
152 & n3n-gemma-2-9b-it-20241029-153 & n3n-gemma-2-it-20241029-1532 & False & True \\
153 & n3n-gemma-2-9b-it-20241110-202 & n3n-gemma-2-it-20241110-2026 & False & True \\
154 & nepali-llm & nepali-llm & False & True \\
155 & olmo-1b & olmo & True & True \\
156 & olmo-7b & olmo & True & True \\
157 & open-llama-13b & open-llama- & True & False \\
158 & open-llama-3b & open-llama- & True & False \\
159 & open-llama-3b-v2 & open-llama--v2 & True & False \\
160 & open-llama-7b & open-llama- & True & False \\
161 & open-llama-7b-v2 & open-llama--v2 & True & False \\
162 & openchat-3.6-8b-20240522 & openchat-3.6-20240522 & True & True \\
163 & openhermes-13b & openhermes & True & True \\
164 & openhermes-7b & openhermes & True & True \\
165 & openlongcot-base-gemma2-2b & openlongcot-base-gemma2 & False & True \\
166 & opt-1.3b & opt & True & True \\
167 & opt-125m & opt & True & False \\
168 & opt-13b & opt & True & False \\
169 & opt-2.7b & opt & True & False \\
170 & opt-30b & opt & True & True \\
171 & opt-350m & opt & True & False \\
172 & opt-6.7b & opt & True & False \\
173 & opt-66b & opt & True & False \\
174 & orca-2-13b & orca-2 & True & True \\
175 & orca-2-7b & orca-2 & True & True \\
176 & orca-mini-v3-13b & orca-mini-v3- & True & True \\
177 & orca-mini-v3-70b & orca-mini-v3- & False & True \\
178 & orca-mini-v3-7b & orca-mini-v3- & True & True \\
179 & orca-mini-v7-72b & orca-mini-v7- & False & True \\
180 & orca-mini-v7-7b & orca-mini-v7- & False & True \\
181 & orpollama-3-8b & orpollama-3 & True & True \\
182 & pantheon-rp-1.0-8b-llama-3 & pantheon-rp-1.0-llama-3 & True & True \\
183 & phi-1-5 & phi-1-5 & True & False \\
184 & phi-1-5-instruct-v0.1 & phi-1-5-instruct-v0.1 & False & True \\
185 & phi-2 & phi-2 & True & False \\
186 & phi-2-instruct-apo & phi-2-instruct-apo & False & True \\
187 & phi-2-instruct-v0.1 & phi-2-instruct-v0.1 & False & True \\
188 & pythia-1.4b & pythia & True & False \\
189 & pythia-12b & pythia & True & True \\
190 & pythia-160m & pythia & True & True \\
191 & pythia-1b & pythia & True & False \\
192 & pythia-2.8b & pythia & True & True \\
193 & pythia-410m & pythia & True & True \\
194 & pythia-6.9b & pythia & True & True \\
195 & pythia-70m & pythia & True & False \\
196 & quantized-open-llama-3b-v2 & quantized-open-llama-v2 & False & True \\
197 & qwen-14b & qwen & True & False \\
198 & qwen-72b & qwen & True & False \\
199 & qwen-7b & qwen & True & False \\
200 & qwen-las-v0.1 & qwen-las-v0.1 & True & True \\
201 & qwen1.5-0.5b & qwen1.5 & True & True \\
202 & qwen1.5-0.5b-chat & qwen1.5-chat & True & True \\
203 & qwen1.5-1.8b & qwen1.5 & True & True \\
204 & qwen1.5-1.8b-chat & qwen1.5-chat & True & True \\
205 & qwen1.5-14b & qwen1.5 & True & True \\
206 & qwen1.5-14b-chat & qwen1.5-chat & True & True \\
207 & qwen1.5-32b & qwen1.5 & True & True \\
208 & qwen1.5-32b-chat & qwen1.5-chat & True & True \\
209 & qwen1.5-4b & qwen1.5 & True & True \\
210 & qwen1.5-4b-chat & qwen1.5-chat & True & True \\
211 & qwen1.5-72b & qwen1.5 & True & False \\
212 & qwen1.5-72b-chat & qwen1.5-chat & True & False \\
213 & qwen1.5-7b & qwen1.5 & True & True \\
214 & qwen1.5-7b-chat & qwen1.5-chat & True & True \\
215 & qwen1.5-7b-chat-sa-v0.1 & qwen1.5-chat-sa-v0.1 & True & True \\
216 & qwen2-0.5b & qwen2 & True & True \\
217 & qwen2-0.5b-abyme & qwen2-abyme & False & True \\
218 & qwen2-0.5b-instruct & qwen2-instruct & False & True \\
219 & qwen2-1.5b & qwen2 & True & True \\
220 & qwen2-1.5b-instruct & qwen2-instruct & False & True \\
221 & qwen2-72b & qwen2 & True & True \\
222 & qwen2-72b-instruct & qwen2-instruct & False & True \\
223 & qwen2-7b & qwen2 & True & True \\
224 & qwen2-7b-instruct & qwen2-instruct & False & True \\
225 & qwen2-7b-instruct-v0.1 & qwen2-instruct-v0.1 & False & True \\
226 & qwen2-7b-instruct-v0.8 & qwen2-instruct-v0.8 & False & True \\
227 & recurrentgemma-2b & recurrentgemma & True & True \\
228 & recurrentgemma-2b-it & recurrentgemma-it & True & True \\
229 & redpajama-incite-base-3b-v1 & redpajama-incite-base-v1 & True & False \\
230 & redpajama-incite-base-7b-v0.1 & redpajama-incite-base-v0.1 & True & False \\
231 & rwkv-4-14b-pile & rwkv-4-pile & True & False \\
232 & rwkv-4-169m-pile & rwkv-4-pile & True & False \\
233 & rwkv-4-1b5-pile & rwkv-4-pile & True & False \\
234 & rwkv-4-3b-pile & rwkv-4-pile & True & False \\
235 & rwkv-4-430m-pile & rwkv-4-pile & True & False \\
236 & rwkv-4-7b-pile & rwkv-4-pile & True & False \\
237 & samantha-qwen-2-7b & samantha-qwen-2 & False & True \\
238 & sauerkrautlm-gemma-2b & sauerkrautlm-gemma & True & True \\
239 & sauerkrautlm-gemma-7b & sauerkrautlm-gemma & True & True \\
240 & selm-llama-3-8b-instruct-iter- & selm-llama-3-instruct-iter-3 & False & True \\
241 & smollm-1.7b & smollm & False & True \\
242 & smollm-1.7b-instruct & smollm-instruct & False & True \\
243 & smollm-1.7b-instruct-ifeval & smollm-instruct-ifeval & False & True \\
244 & smollm-135m & smollm & False & True \\
245 & smollm-135m-instruct & smollm-instruct & False & True \\
246 & smollm-360m & smollm & False & True \\
247 & smollm-360m-instruct & smollm-instruct & False & True \\
248 & stablelm-2-1-6b & stablelm-2-1- & True & False \\
249 & stablelm-2-1-6b-chat & stablelm-2-1--chat & True & False \\
250 & stablelm-base-alpha-3b & stablelm-base-alpha & True & False \\
251 & stablelm-base-alpha-7b & stablelm-base-alpha & True & False \\
252 & stablelm-base-alpha-7b-v2 & stablelm-base-alpha-v2 & True & False \\
253 & starcoder2-15b & starcoder2 & True & True \\
254 & starcoder2-3b & starcoder2 & True & True \\
255 & starcoder2-7b & starcoder2 & True & True \\
256 & starcoderbase & starcoderbase & True & False \\
257 & starcoderbase-1b & starcoderbase & True & False \\
258 & starcoderbase-3b & starcoderbase & True & False \\
259 & starcoderbase-7b & starcoderbase & True & False \\
260 & suzume-llama-3-8b-multilingual & suzume-llama-3-multilingual & True & True \\
261 & suzume-llama-3-8b-multilingual & suzume-llama-3-multilingual-or & False & True \\
262 & suzume-llama-3-8b-multilingual & suzume-llama-3-multilingual-or & False & True \\
263 & suzume-llama-3-8b-multilingual & suzume-llama-3-multilingual-or & False & True \\
264 & suzume-llama-3-8b-multilingual & suzume-llama-3-multilingual-or & False & True \\
265 & wizardlm-13b-v1.0 & wizardlm-v1.0 & False & True \\
266 & wizardlm-70b-v1.0 & wizardlm-v1.0 & False & True \\
267 & xglm-1.7b & xglm & True & False \\
268 & xglm-4.5b & xglm & True & False \\
269 & xglm-564m & xglm & True & False \\
270 & xglm-7.5b & xglm & True & False \\
271 & yi-1.5-34b & yi-1.5 & True & True \\
272 & yi-1.5-34b-chat & yi-1.5-chat & True & True \\
273 & yi-1.5-6b & yi-1.5 & True & True \\
274 & yi-1.5-6b-chat & yi-1.5-chat & True & True \\
275 & yi-1.5-9b & yi-1.5 & True & True \\
276 & yi-1.5-9b-chat & yi-1.5-chat & True & True \\
277 & yi-1.5-9b-chat-abliterated & yi-1.5-chat-abliterated & False & True \\
278 & yi-34b & yi & True & True \\
279 & yi-34b-200k & yi-200k & True & False \\
280 & yi-34b-chat & yi-chat & True & False \\
281 & yi-6b & yi & True & True \\
282 & yi-6b-200k & yi-200k & True & False \\
283 & yi-6b-chat & yi-chat & False & True \\
284 & yi-9b & yi & True & True \\
285 & zephyr-7b-gemma-v0.1 & zephyr-gemma-v0.1 & True & True \\
\bottomrule
\end{longtable}}

\subsection{AIC curve and K=6 Results}\label{supp_sec_aic}

\begin{figure}[H]
    \centering
    \begin{subfigure}{0.48\textwidth}
        \centering
        \includegraphics[width=\linewidth]{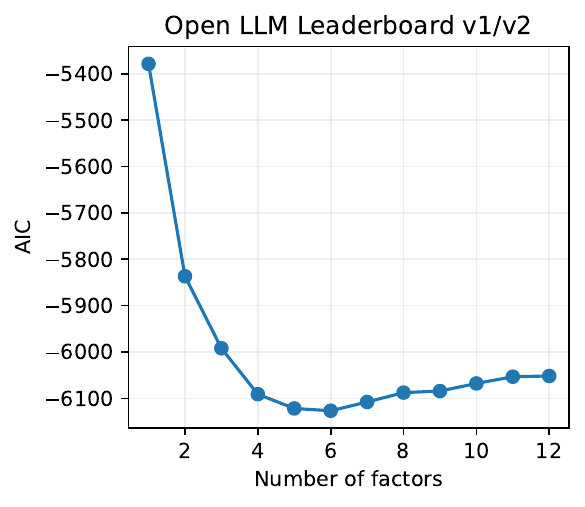}
        \caption{AIC for different $K$'s.}
        \label{fig:aic_sub}
    \end{subfigure}
    \hfill
    \begin{subfigure}{0.41\textwidth}
        \centering
        \includegraphics[width=\linewidth]{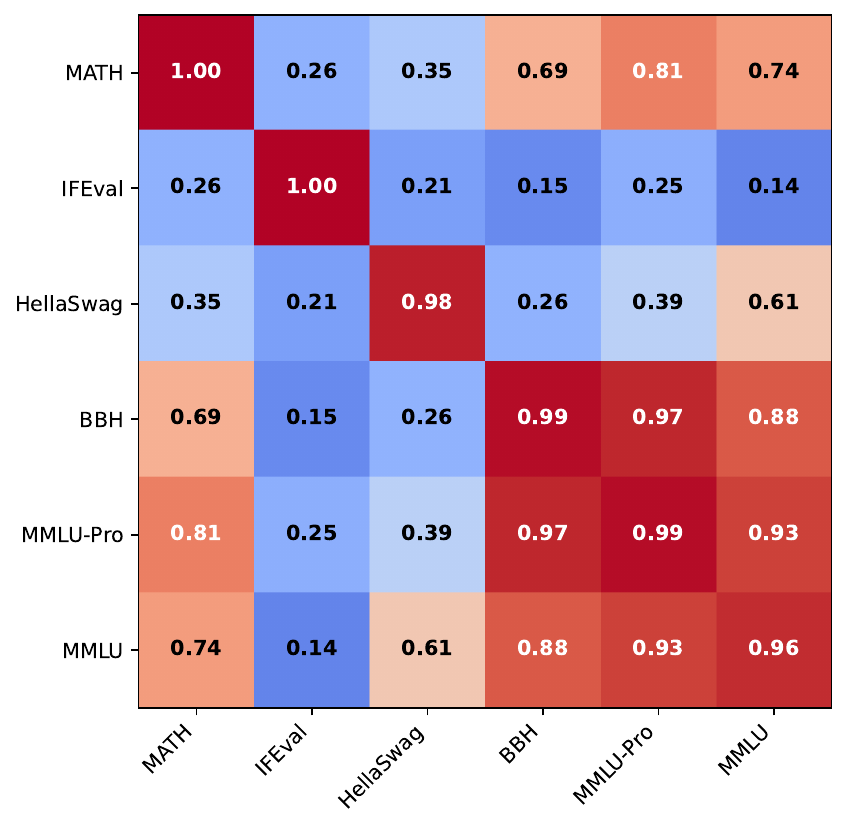}
        \caption{Correlation under $K=6$.}
        \label{fig:correlation_full}
    \end{subfigure}
    
     \caption{Model selection and estimated correlation structure. Left: AIC curve. Right: correlation of $\alpha$ under $K = 6$.}
    \label{fig:aic}
\end{figure}

\begin{figure}
    \centering
    \includegraphics[width=1\linewidth]{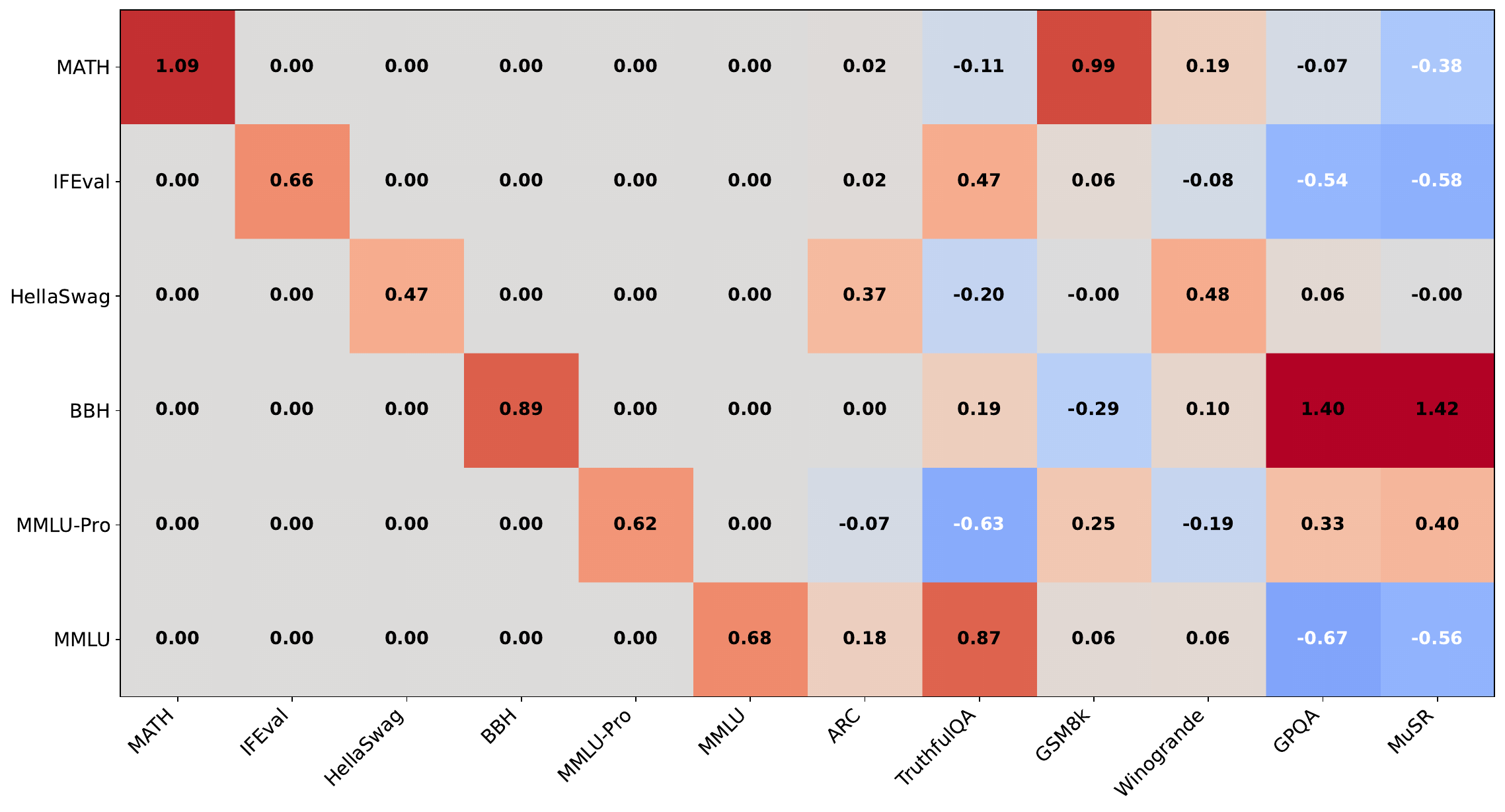}
    \caption{Estimated loadings for the $K=6$ model.}
    \label{fig:loadings_full}
\end{figure}

This section reports the AIC results for the experiment in Section~\ref{sec_experiment}. Figure~\ref{fig:aic} displays the AIC values for models with $K \in \{1,2,\dots 12\}$ and Figure~\ref{fig:correlation_full} shows the estimated correlation matrix of the latent variable $\alpha$ for $K=6$. We also present the estimated loadings for $K=6$ in Figure~\ref{fig:loadings_full}, which corresponds to the model attaining the smallest AIC. However, as indicated by the strong correlations in Figure~\ref{fig:correlation_full}, three of the components are highly correlated, suggesting that the six-dimensional solution may be over-specified and that a lower-dimensional model could provide a more parsimonious and interpretable representation. Given that the AIC value for $K=4$ is very close to that for $K=6$, we set the latent dimension to $K=4$ and report the corresponding estimation results in the main text.

\subsection{Prediction intervals}\label{supp_int}

\begin{figure}[H]
    \centering
    \includegraphics[width=1\linewidth]{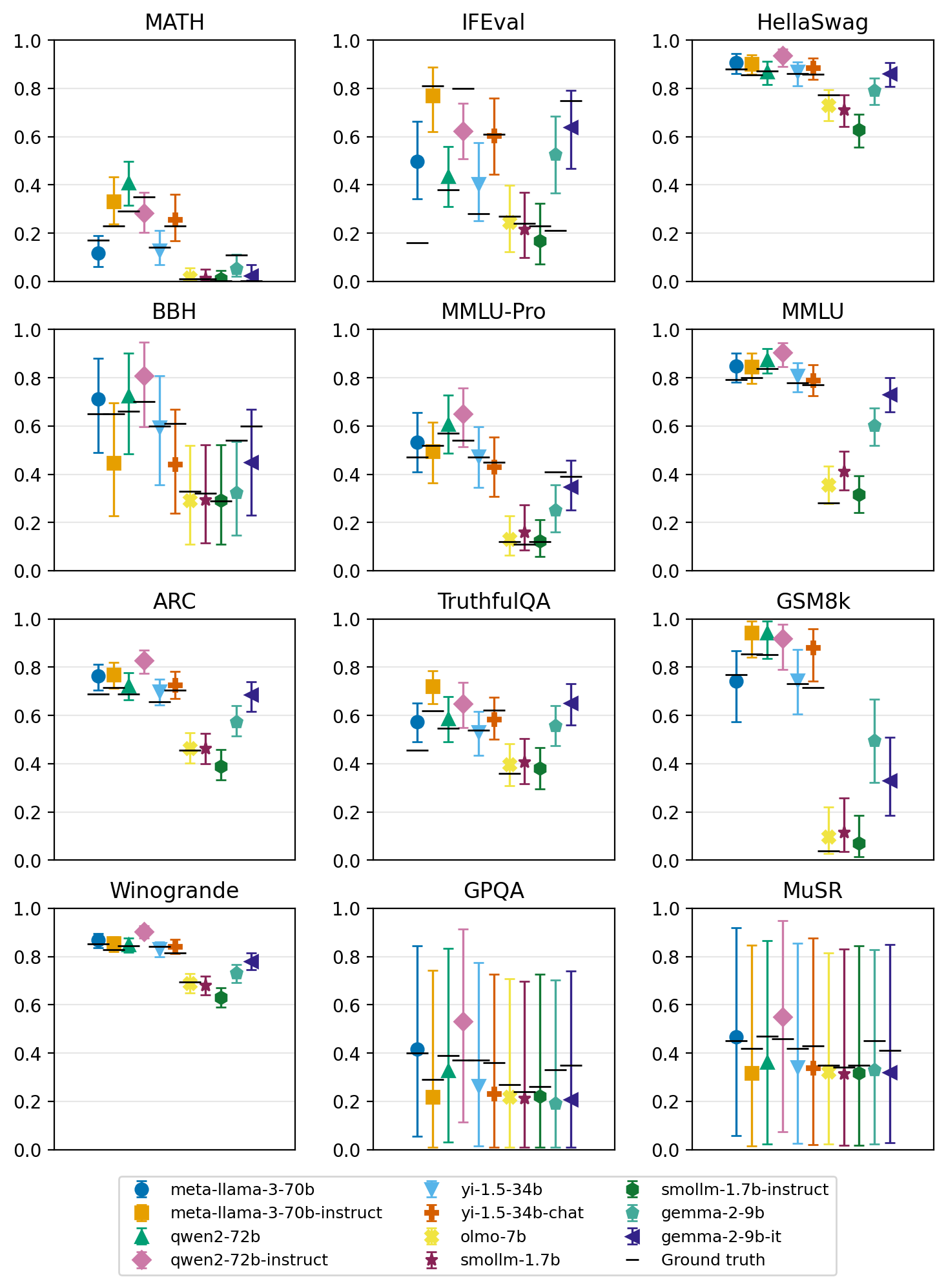}
    \caption{{Isolating uncertainty in $Y^{(l)}_{ij}$.}}
    \label{fig:pred_int_supp1}
\end{figure}

\begin{figure}[H]
    \centering
    \includegraphics[width=1\linewidth]{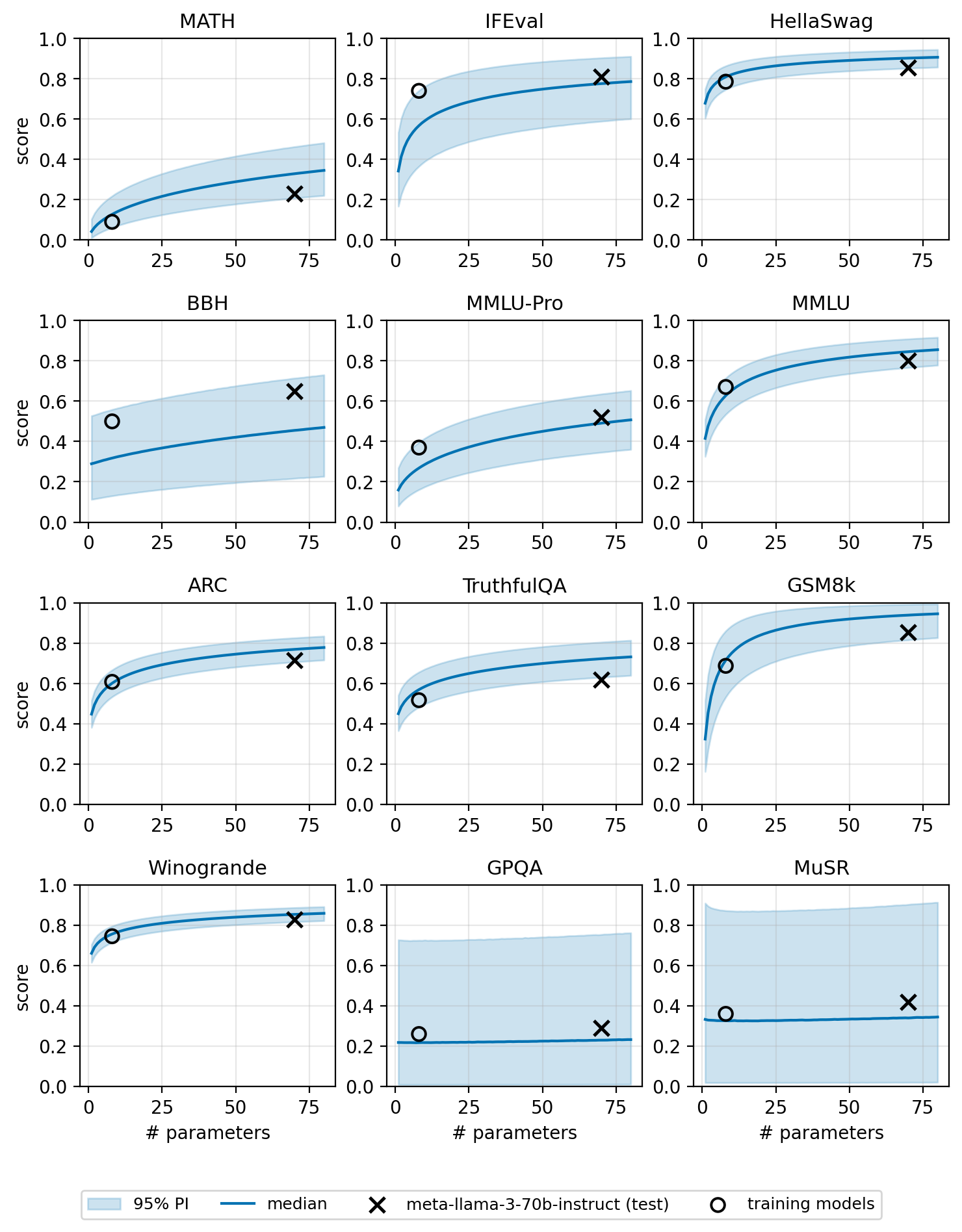}
    \caption{{Scaling behavior for LLaMa-3-Instruct while maintaining token count fixed.}}
    \label{fig:pred_int_supp2}
\end{figure}

\begin{figure}[H]
    \centering
    \includegraphics[width=1\linewidth]{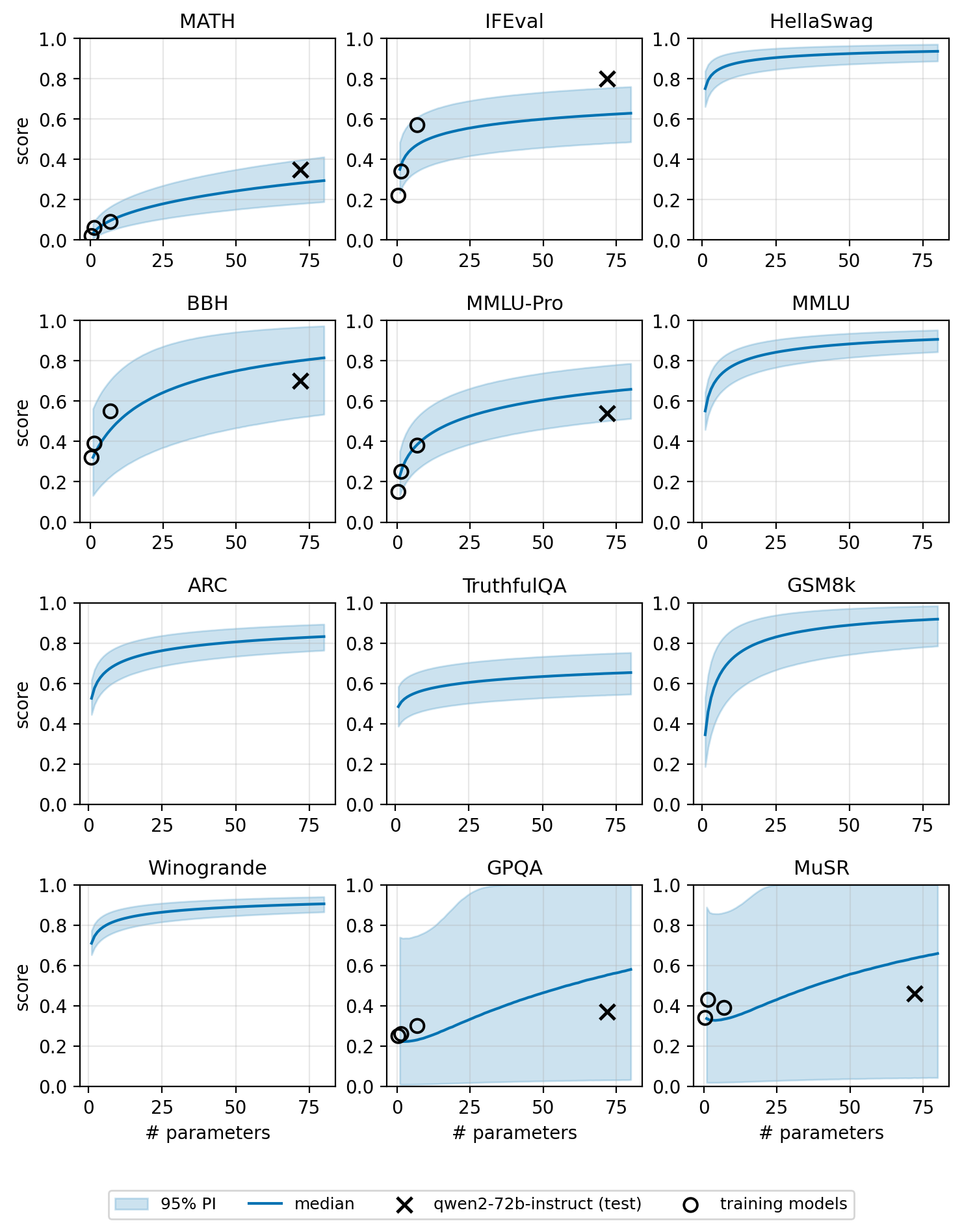}
    \caption{{Scaling behavior for Qwen-2-Instruct while maintaining token count fixed.}}
    \label{fig:pred_int_supp3}
\end{figure}

\begin{figure}[H]
    \centering
    \includegraphics[width=1\linewidth]{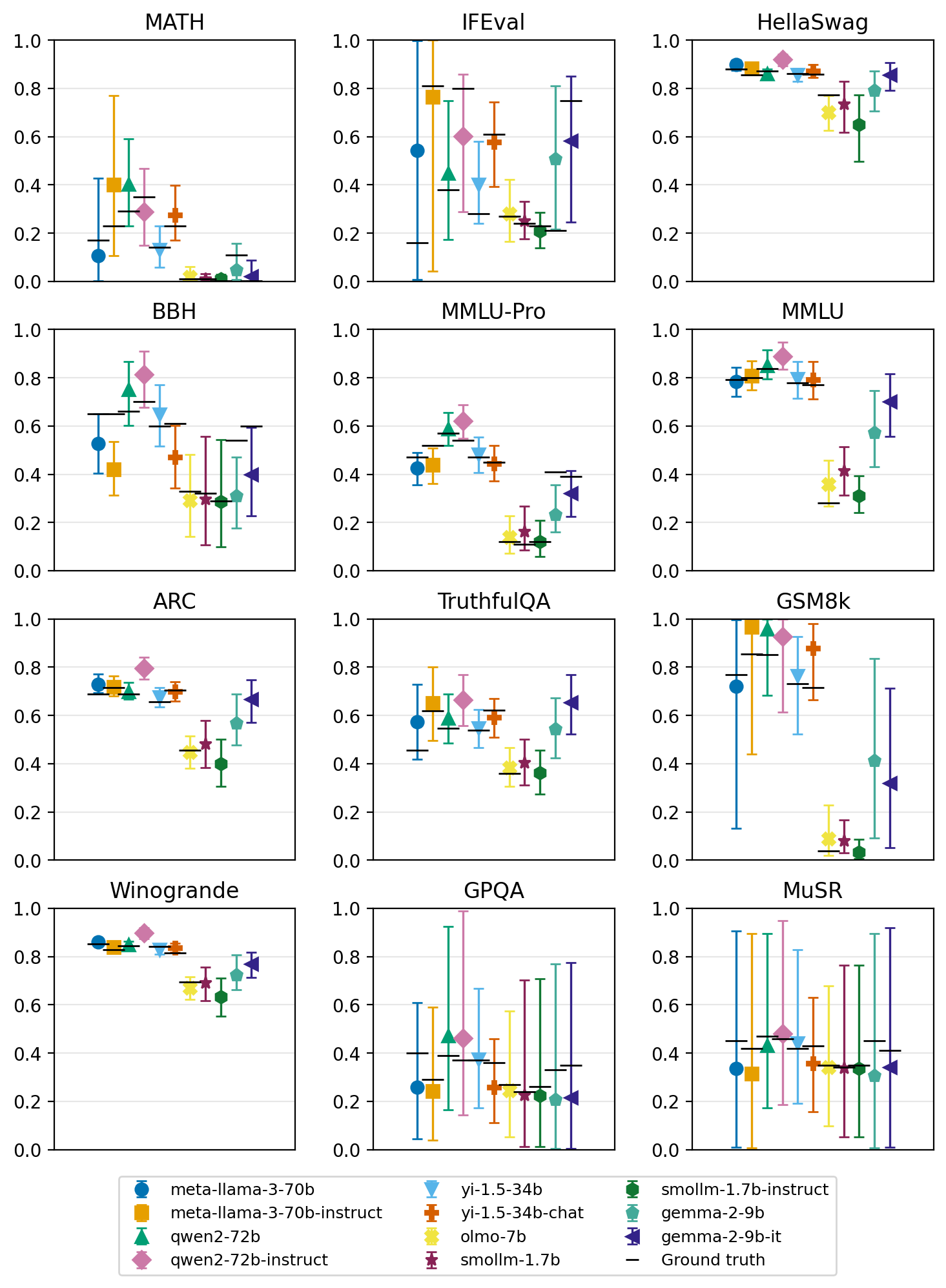}
    \caption{{Expanding on Figure \ref{fig:pred_int} by considering variable scale parameters $\phi_j$.}}
    \label{fig:pred_int_supp4}
\end{figure}

{ This section extends the results regarding the prediction intervals presented in Figure~\ref{fig:pred_int} of the main text.
In Figure~\ref{fig:pred_int_supp1}, we decompose the total uncertainty in the response variable into two components: uncertainty stemming from statistical estimation error, and uncertainty inherent to the response distribution. To isolate the latter, we fix point estimates for $(\alpha_l, \xi)$ by averaging over the samples $(\alpha_l^{(t)}, \hat{\xi}^{(t)})$ (Algorithm \ref{alg:pred_int}), yielding $(\bar{\alpha}_l, \bar{\xi})$, 
and then draw from the conditional distribution of $Y_{ij}^{(l)}$ given $x_{i}^{(l)}$ and $(\bar{\alpha}_l, \bar{\xi})$. Under this construction, all variability in Figure~\ref{fig:pred_int_supp1} reflects intrinsic response uncertainty alone. The figure confirms that the bulk of the interval width seen in Figure~\ref{fig:pred_int} is inherent to the response distribution and does not arise from statistical estimation.

In Figures~\ref{fig:pred_int_supp2} and~\ref{fig:pred_int_supp3}, we detail the results of Figure~\ref{fig:pred_int} for some model families, focusing on LLaMA and Qwen separately. For each family, we examine the scaling behavior with respect to model size while holding the token count fixed. In these plots, we use Algorithm \ref{alg:pred_int} for interval computation.

Finally, in Figure \ref{fig:pred_int_supp4}, we explore a strategy for obtaining narrower prediction intervals. Since our previous analyses indicated that most of the uncertainty arises from the conditional distribution of $Y_{ij}^{(l)}$, we parameterize the scale parameters $\phi_j$ as affine functions\footnote{As in \citet{ferrari2011improved}.} of $x_i^{(l)}$. Specifically, we assume
\[
\phi_j(x_i^{(l)}) = \tilde{\beta}^{(j)\top}(1, x_i^{(l)}).
\]
Comparing Figure \ref{fig:pred_int} with Figure \ref{fig:pred_int_supp4}, we observe that this approach can indeed reduce the width of the prediction intervals in some cases, while also producing wider intervals when warranted by the data. As before, we use Algorithm \ref{alg:pred_int} to compute the intervals, extending the estimators to incorporate the additional estimators for parameters $\tilde{\beta}^{(j)}$.}

\subsection{Predictive analysis}\label{supp_add_compare}

{
In this section, we investigate how our scaling law compares with existing baselines in terms of predictive performance. We consider the same test models used in Section \ref{sec:intervals}, but train a single scaling law across all models while excluding the test models and their corresponding LLM families from the training set.

We evaluate baselines similar to those considered in \citet{polo2024sloth}: (i) the simple scaling law proposed by \citet{owen2024predictable}, which does not incorporate family information (``FLOPs''), (ii) an extension of the previous baseline that decomposes FLOPs into model size and number of training tokens, including their interaction term (``Size/Tokens/Interaction''), (iii) versions of the first two baselines augmented with family-specific intercepts, and (iv) the two variants of Sloth introduced in \citet{polo2024sloth} (with and without a trainable activation function).

From Table \ref{tab:pred_results}, we observe that our method achieves mean absolute errors (MAE) comparable to the Sloth variants. This is particularly encouraging because, unlike Sloth, our approach relies on stronger distributional assumptions on the response variables, specifically the assumption of Beta-distributed outcomes. Although our method slightly underperforms the ``Family/FLOPs'' baseline, it offers a more interpretable framework by modeling performance in terms of fundamental skills, whereas the competing scaling laws are designed purely for predictive accuracy.
}

\begin{sidewaystable}
\centering
\resizebox{\textwidth}{!}{
\begin{tabular}{l| c c c c c c c c c c c c |c}
\toprule
 & MATH & IFEval & HellaSwag & BBH & MMLU-Pro & MMLU & ARC & TruthfulQA & GSM8k & Winogrande & GPQA & MuSR & All \\
\midrule
FLOPs & $\boldsymbol{0.079_{\pm 0.023}}$ & $0.197_{\pm 0.066}$ & $0.025_{\pm 0.008}$ & $0.237_{\pm 0.042}$ & $\boldsymbol{0.085_{\pm 0.017}}$ & $0.091_{\pm 0.019}$ & $\boldsymbol{0.050_{\pm 0.008}}$ & $0.131_{\pm 0.041}$ & $0.113_{\pm 0.025}$ & $0.027_{\pm 0.006}$ & $0.227_{\pm 0.029}$ & $0.256_{\pm 0.034}$ & $0.142_{\pm 0.013}$ \\
\midrule
Size/Tokens/Interaction & $\boldsymbol{0.079_{\pm 0.023}}$ & $0.192_{\pm 0.061}$ & $0.035_{\pm 0.010}$ & $0.118_{\pm 0.025}$ & $\boldsymbol{0.076_{\pm 0.015}}$ & $0.104_{\pm 0.024}$ & $0.062_{\pm 0.014}$ & $0.122_{\pm 0.034}$ & $0.174_{\pm 0.047}$ & $0.031_{\pm 0.007}$ & $0.129_{\pm 0.021}$ & $0.135_{\pm 0.023}$ & $0.110_{\pm 0.010}$ \\
\midrule
Family/FLOPs & $\boldsymbol{0.082_{\pm 0.031}}$ & $\boldsymbol{0.064_{\pm 0.010}}$ & $\boldsymbol{0.018_{\pm 0.004}}$ & $\boldsymbol{0.062_{\pm 0.010}}$ & $\boldsymbol{0.066_{\pm 0.007}}$ & $\boldsymbol{0.026_{\pm 0.005}}$ & $\boldsymbol{0.022_{\pm 0.006}}$ & $\boldsymbol{0.041_{\pm 0.017}}$ & $\boldsymbol{0.051_{\pm 0.019}}$ & $\boldsymbol{0.009_{\pm 0.002}}$ & $\boldsymbol{0.041_{\pm 0.008}}$ & $\boldsymbol{0.043_{\pm 0.008}}$ & $\boldsymbol{0.048_{\pm 0.005}}$ \\
\midrule
Family/Size/Tokens/Interaction & $0.106_{\pm 0.037}$ & $\boldsymbol{0.071_{\pm 0.015}}$ & $0.025_{\pm 0.005}$ & $\boldsymbol{0.040_{\pm 0.008}}$ & $0.093_{\pm 0.018}$ & $\boldsymbol{0.058_{\pm 0.012}}$ & $0.055_{\pm 0.009}$ & $0.123_{\pm 0.047}$ & $\boldsymbol{0.076_{\pm 0.018}}$ & $0.020_{\pm 0.005}$ & $\boldsymbol{0.028_{\pm 0.004}}$ & $\boldsymbol{0.032_{\pm 0.008}}$ & $\boldsymbol{0.061_{\pm 0.007}}$ \\
\midrule
Simple Sloth (d=2) & $0.110_{\pm 0.028}$ & $0.151_{\pm 0.036}$ & $\boldsymbol{0.021_{\pm 0.002}}$ & $0.073_{\pm 0.024}$ & $0.112_{\pm 0.023}$ & $0.082_{\pm 0.014}$ & $\boldsymbol{0.055_{\pm 0.013}}$ & $\boldsymbol{0.058_{\pm 0.019}}$ & $0.101_{\pm 0.020}$ & $0.021_{\pm 0.005}$ & $0.160_{\pm 0.046}$ & $\boldsymbol{0.028_{\pm 0.006}}$ & $0.088_{\pm 0.009}$ \\
\midrule
Simple Sloth (d=3) & $0.143_{\pm 0.054}$ & $0.134_{\pm 0.032}$ & $0.025_{\pm 0.004}$ & $\boldsymbol{0.057_{\pm 0.010}}$ & $\boldsymbol{0.092_{\pm 0.023}}$ & $0.073_{\pm 0.014}$ & $0.057_{\pm 0.013}$ & $\boldsymbol{0.061_{\pm 0.020}}$ & $0.107_{\pm 0.020}$ & $\boldsymbol{0.016_{\pm 0.004}}$ & $\boldsymbol{0.066_{\pm 0.020}}$ & $\boldsymbol{0.023_{\pm 0.007}}$ & $\boldsymbol{0.076_{\pm 0.009}}$ \\
\midrule
Simple Sloth (d=4) & $\boldsymbol{0.068_{\pm 0.024}}$ & $\boldsymbol{0.103_{\pm 0.031}}$ & $0.028_{\pm 0.005}$ & $0.083_{\pm 0.021}$ & $0.093_{\pm 0.022}$ & $\boldsymbol{0.049_{\pm 0.010}}$ & $0.063_{\pm 0.012}$ & $\boldsymbol{0.083_{\pm 0.031}}$ & $\boldsymbol{0.071_{\pm 0.018}}$ & $\boldsymbol{0.010_{\pm 0.003}}$ & $0.095_{\pm 0.024}$ & $0.098_{\pm 0.027}$ & $0.076_{\pm 0.007}$ \\
\midrule
Simple Sloth (d=5) & $0.086_{\pm 0.032}$ & $\boldsymbol{0.077_{\pm 0.017}}$ & $\boldsymbol{0.020_{\pm 0.004}}$ & $0.083_{\pm 0.019}$ & $0.109_{\pm 0.028}$ & $\boldsymbol{0.069_{\pm 0.015}}$ & $0.065_{\pm 0.013}$ & $0.091_{\pm 0.033}$ & $\boldsymbol{0.066_{\pm 0.019}}$ & $\boldsymbol{0.008_{\pm 0.003}}$ & $\boldsymbol{0.062_{\pm 0.018}}$ & $\boldsymbol{0.061_{\pm 0.011}}$ & $\boldsymbol{0.070_{\pm 0.007}}$ \\
\midrule
Sloth (d=2) & $0.385_{\pm 0.094}$ & $0.170_{\pm 0.063}$ & $\boldsymbol{0.017_{\pm 0.005}}$ & $\boldsymbol{0.068_{\pm 0.026}}$ & $0.137_{\pm 0.038}$ & $\boldsymbol{0.025_{\pm 0.007}}$ & $\boldsymbol{0.029_{\pm 0.006}}$ & $\boldsymbol{0.065_{\pm 0.022}}$ & $0.100_{\pm 0.024}$ & $0.021_{\pm 0.006}$ & $0.103_{\pm 0.025}$ & $0.080_{\pm 0.027}$ & $0.117_{\pm 0.017}$ \\
\midrule
Sloth (d=3) & $0.098_{\pm 0.035}$ & $\boldsymbol{0.125_{\pm 0.033}}$ & $\boldsymbol{0.008_{\pm 0.004}}$ & $\boldsymbol{0.020_{\pm 0.004}}$ & $\boldsymbol{0.050_{\pm 0.011}}$ & $\boldsymbol{0.023_{\pm 0.004}}$ & $\boldsymbol{0.019_{\pm 0.006}}$ & $\boldsymbol{0.086_{\pm 0.028}}$ & $\boldsymbol{0.061_{\pm 0.018}}$ & $0.029_{\pm 0.005}$ & $\boldsymbol{0.070_{\pm 0.020}}$ & $0.064_{\pm 0.027}$ & $\boldsymbol{0.059_{\pm 0.008}}$ \\
\midrule
Sloth (d=4) & $0.157_{\pm 0.052}$ & $0.145_{\pm 0.041}$ & $\boldsymbol{0.014_{\pm 0.003}}$ & $\boldsymbol{0.043_{\pm 0.012}}$ & $0.102_{\pm 0.029}$ & $\boldsymbol{0.060_{\pm 0.009}}$ & $\boldsymbol{0.028_{\pm 0.005}}$ & $\boldsymbol{0.077_{\pm 0.025}}$ & $\boldsymbol{0.061_{\pm 0.017}}$ & $\boldsymbol{0.013_{\pm 0.003}}$ & $0.083_{\pm 0.017}$ & $0.100_{\pm 0.039}$ & $0.083_{\pm 0.010}$ \\
\midrule
Sloth (d=5) & $0.271_{\pm 0.095}$ & $\boldsymbol{0.071_{\pm 0.019}}$ & $\boldsymbol{0.007_{\pm 0.003}}$ & $\boldsymbol{0.031_{\pm 0.008}}$ & $\boldsymbol{0.025_{\pm 0.006}}$ & $\boldsymbol{0.022_{\pm 0.005}}$ & $\boldsymbol{0.026_{\pm 0.005}}$ & $0.130_{\pm 0.043}$ & $\boldsymbol{0.067_{\pm 0.019}}$ & $0.021_{\pm 0.005}$ & $\boldsymbol{0.044_{\pm 0.007}}$ & $\boldsymbol{0.032_{\pm 0.008}}$ & $\boldsymbol{0.067_{\pm 0.013}}$ \\
\midrule
Ours (d=2) & $\boldsymbol{0.044_{\pm 0.018}}$ & $0.183_{\pm 0.038}$ & $0.025_{\pm 0.004}$ & $0.092_{\pm 0.021}$ & $\boldsymbol{0.063_{\pm 0.011}}$ & $0.091_{\pm 0.013}$ & $0.065_{\pm 0.011}$ & $\boldsymbol{0.088_{\pm 0.025}}$ & $0.088_{\pm 0.012}$ & $\boldsymbol{0.016_{\pm 0.005}}$ & $0.081_{\pm 0.015}$ & $0.070_{\pm 0.013}$ & $0.079_{\pm 0.007}$ \\
\midrule
Ours (d=3) & $\boldsymbol{0.045_{\pm 0.018}}$ & $0.141_{\pm 0.040}$ & $0.025_{\pm 0.004}$ & $\boldsymbol{0.073_{\pm 0.024}}$ & $\boldsymbol{0.066_{\pm 0.012}}$ & $0.087_{\pm 0.016}$ & $0.063_{\pm 0.011}$ & $0.098_{\pm 0.029}$ & $0.098_{\pm 0.011}$ & $\boldsymbol{0.017_{\pm 0.004}}$ & $\boldsymbol{0.071_{\pm 0.013}}$ & $\boldsymbol{0.062_{\pm 0.012}}$ & $\boldsymbol{0.072_{\pm 0.007}}$ \\
\midrule
Ours (d=4) & $\boldsymbol{0.034_{\pm 0.010}}$ & $\boldsymbol{0.105_{\pm 0.027}}$ & $0.021_{\pm 0.005}$ & $0.114_{\pm 0.026}$ & $0.099_{\pm 0.023}$ & $0.095_{\pm 0.010}$ & $0.056_{\pm 0.010}$ & $0.099_{\pm 0.033}$ & $0.079_{\pm 0.014}$ & $\boldsymbol{0.012_{\pm 0.004}}$ & $\boldsymbol{0.055_{\pm 0.016}}$ & $\boldsymbol{0.054_{\pm 0.011}}$ & $\boldsymbol{0.071_{\pm 0.007}}$ \\
\midrule
Ours (d=5) & $\boldsymbol{0.067_{\pm 0.024}}$ & $\boldsymbol{0.102_{\pm 0.028}}$ & $\boldsymbol{0.020_{\pm 0.004}}$ & $0.153_{\pm 0.028}$ & $0.110_{\pm 0.027}$ & $0.072_{\pm 0.011}$ & $\boldsymbol{0.046_{\pm 0.009}}$ & $0.102_{\pm 0.035}$ & $\boldsymbol{0.065_{\pm 0.013}}$ & $0.020_{\pm 0.005}$ & $0.119_{\pm 0.036}$ & $0.085_{\pm 0.024}$ & $0.088_{\pm 0.009}$ \\
\bottomrule
\end{tabular}
}
\caption{{Mean absolute error ($\pm$ standard error) of different scaling laws.}}
\label{tab:pred_results}
\end{sidewaystable}

\subsection{Selecting Anchor Benchmarks}\label{supp_sec_anchor}
\begin{figure}[t]
    \centering
    \includegraphics[width=0.8\linewidth]{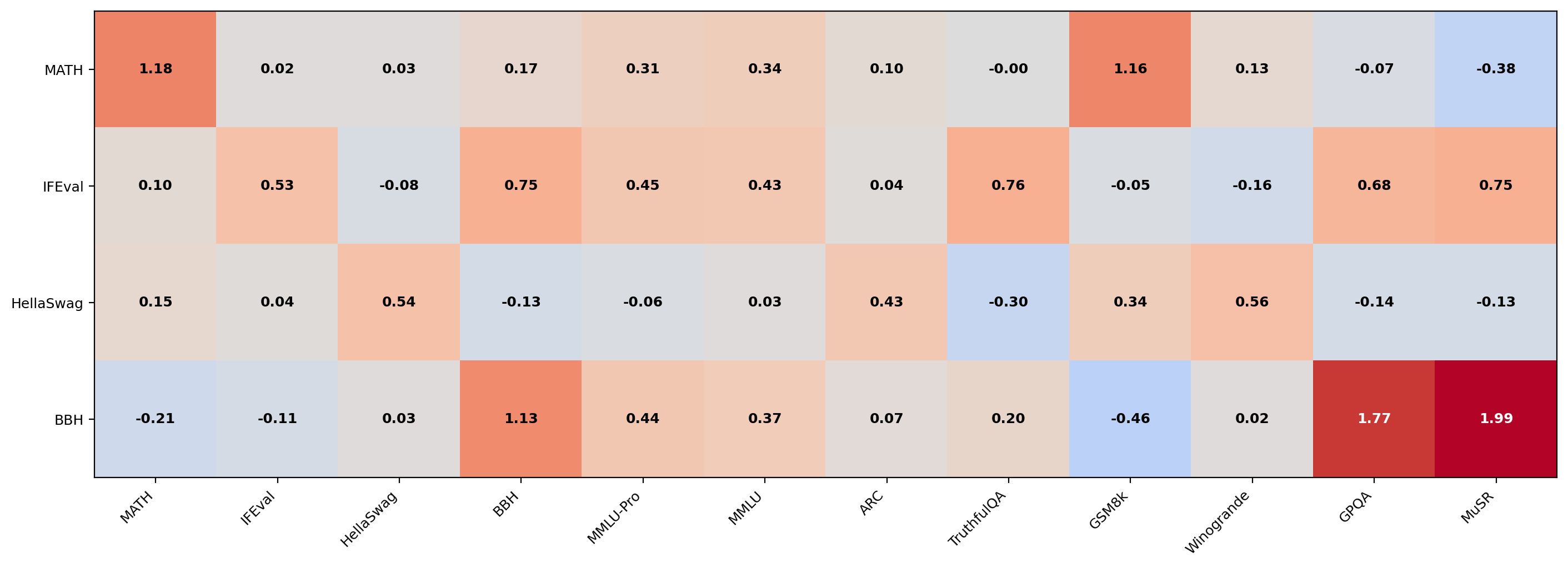}
    \caption{Estimated loadings after promax rotation.}
    \label{fig:loadings_promax}
\end{figure}
\begin{figure}[t]
    \centering
    \includegraphics[width=0.8\linewidth]{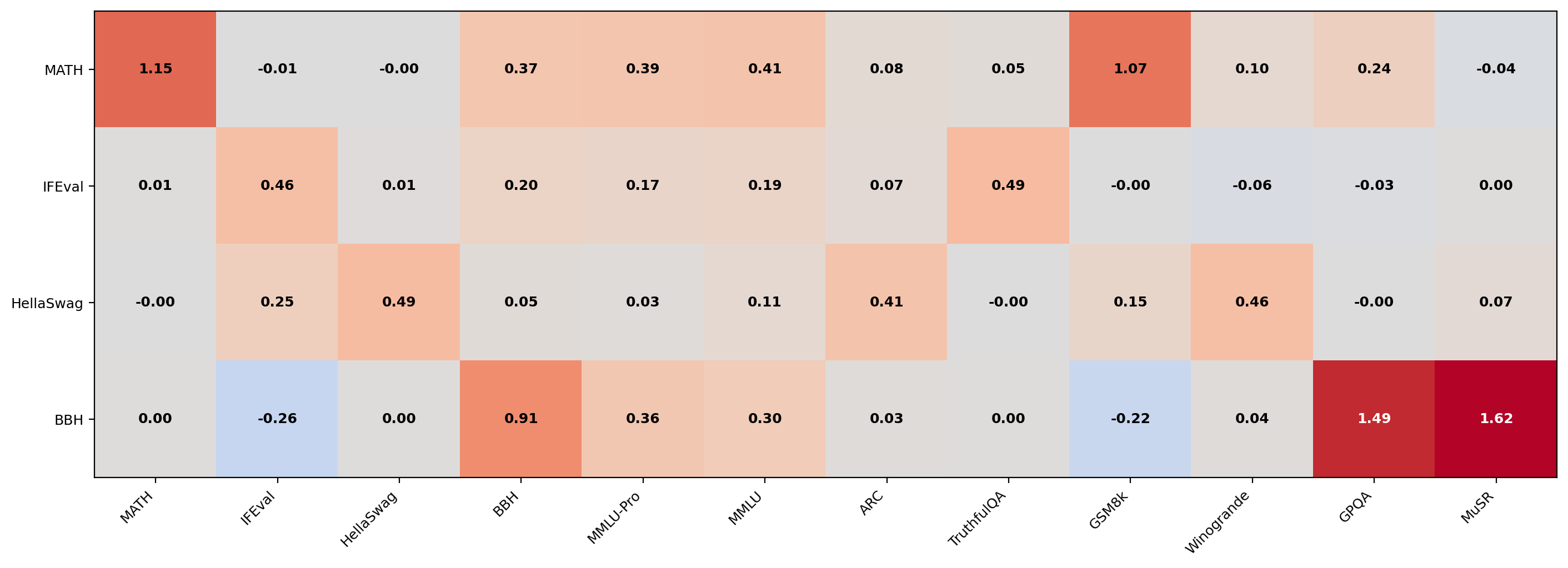}
    \caption{Estimated loadings after $\ell_1$ rotation.}
    \label{fig:loadings_l1}
\end{figure}
\begin{figure}[t]
    \centering
    \includegraphics[width=0.8\linewidth]{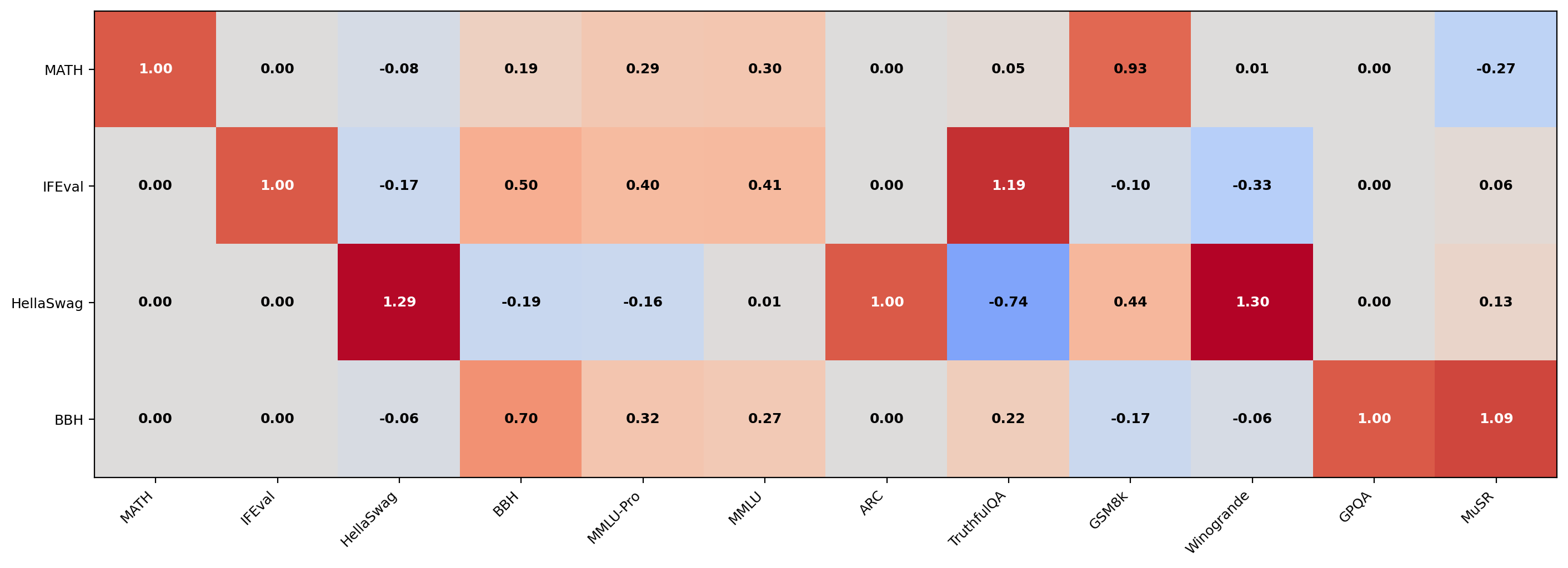}
    \caption{Estimated loadings with Math, IFEval, ARC, and GPQA  as the anchor benchmarks.}
    \label{fig:loadings_promax_guide}
\end{figure}
\begin{figure}[t]
    \centering
    \includegraphics[width=0.8\linewidth]{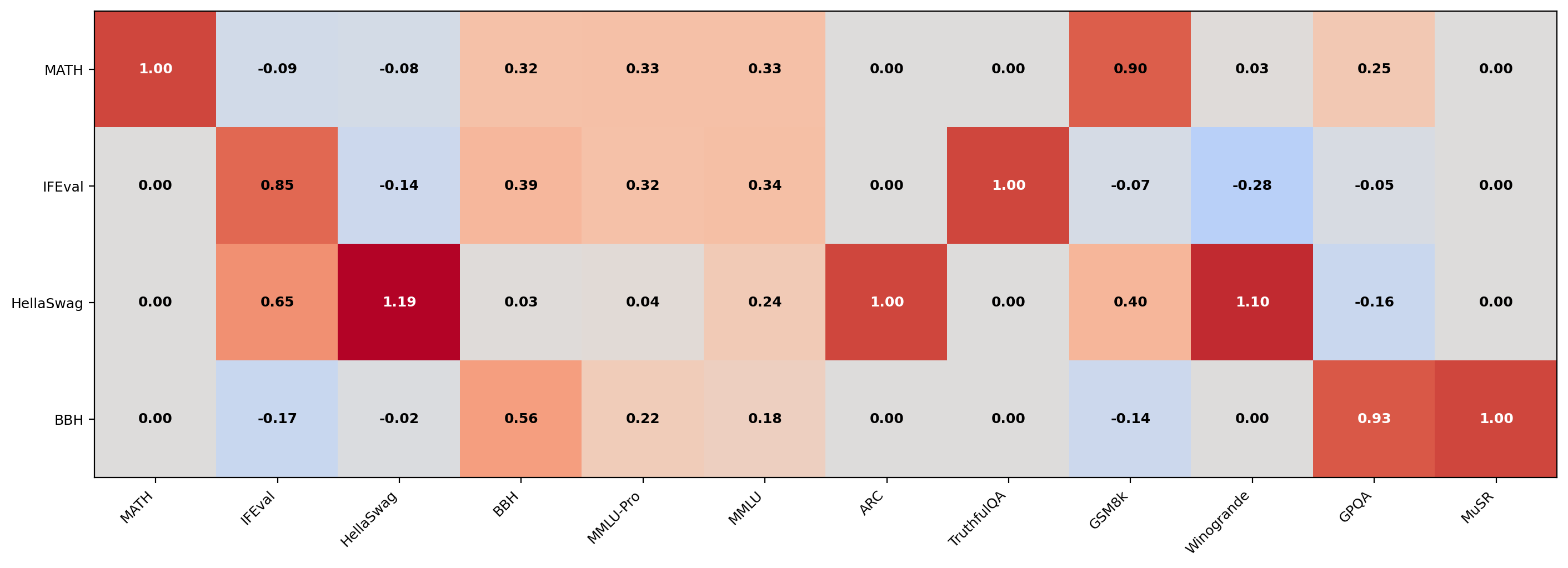}
    \caption{Estimated loadings with MATH, TruthfulQA, HellaSwag, and MuSR as the anchor benchmarks.}
    \label{fig:loadings_l1_guide}
\end{figure}

{ In Section~\ref{sec_experiment}, we choose MATH, IFEval, HellaSwag, and BBH as anchor benchmarks because they are designed to target distinct skills. To assess whether this choice is empirically appropriate, we adopt the exploratory factor analysis framework. Specifically, we use rotation methods that seek a transformation matrix $ G$ such that the rotated loading matrix $\hat\Lambda G$ minimizes a prescribed rotation criterion, typically designed to promote sparsity or simple structure in the resulting loadings. The rotated solution $\hat\Lambda\hat G$, where $\hat G$ denotes the corresponding minimizer, is then commonly used as a diagnostic guide for imposing constraints on the loading matrix under the standard EFA framework~\citep{mulaik2009foundations,fabrigar2012exploratory,cui2026beyond}.
Under this framework, we consider two widely used oblique rotation criteria that allow correlated latent variables: promax rotation~\citep{hendrickson1964promax} and $\ell_1$ rotation~\citep{Liu2023RotationTS}. The resulting rotated loading matrices are presented in Figures~\ref{fig:loadings_promax} and~\ref{fig:loadings_l1}, respectively. In both cases, the four selected benchmarks exhibit the desired anchor-like structure: each has a dominant loading on one latent dimension, with relatively small cross-loadings on the remaining dimensions. This pattern supports the use of MATH, IFEval, HellaSwag, and BBH as anchor benchmarks.

At the same time, the rotation results also suggest alternative choices of anchor benchmarks. Implied by the promax rotation result in Figure~\ref{fig:loadings_promax}, we consider an alternative specification using MATH, IFEval, ARC, and GPQA as anchor benchmarks. Under this specification, we refit the model and present the resulting loading matrix in Figure~\ref{fig:loadings_promax_guide}. Similarly, motivated by the $\ell_1$ rotation result in Figure~\ref{fig:loadings_l1}, we consider another alternative specification using MATH, TruthfulQA, HellaSwag, and MuSR as anchor benchmarks, and the resulting loading matrix is presented in Figure~\ref{fig:loadings_l1_guide}. The overall loading patterns remain similar to Figure~\ref{fig:loadings} under these alternative specifications, suggesting that the main conclusions are robust to reasonable changes in the anchor selection. Moreover, the inference for $\beta$ is not affected by these alternative choices of anchors.}

{
\section{Simulation Results}\label{supp_sec_simu}

In this section, we conduct numerical experiments to assess the performance of our proposed approach on synthetic data.
We start with introducing the data-generating process.
To sample the true parameters $\xi = \{\vvec^\T(\Lambda), \vvec^\T(\beta), b^\T, \phi^\T, \mathrm{vec}^\T(\Sigma)\}^\T$, we adopt the following scheme.
The regression coefficient matrix $\beta \in \mathbb{R}^{3 \times K}$ is generated elementwise according to
\[
\beta_{rk} \sim 0.2N(0,1),
\qquad r=1,2,3,\;\; k=1,\ldots,K.
\]
The intercept parameters are generated by
\[
b_j = -1 + 0.3Z_j,
\qquad Z_j \overset{\mathrm{i.i.d.}}{\sim} N(0,1),
\qquad j=1,\ldots,J,
\]
while the precision parameters and guessing parameters are independently generated from
\[
\phi_j \overset{\mathrm{i.i.d.}}{\sim} \mathrm{Unif}(5,10),
\qquad
\gamma_j \overset{\mathrm{i.i.d.}}{\sim} \mathrm{Unif}(0,0.05),
\qquad j=1,\ldots,J.
\]
To satisfy the identifiability condition, the loading matrix $\Lambda=(\lambda_{jk}) \in \mathbb{R}^{J\times K}$ is generated with a structured form. Specifically, the first $K$ rows form a diagonal matrix with diagonal entries independently sampled from $\mathrm{Unif}(0.5,1.0)$, namely,
\[
\Lambda_{1:K,\,1:K}
=
\mathrm{diag}(d_1,\ldots,d_K),
\qquad
d_k \overset{\mathrm{i.i.d.}}{\sim} \mathrm{Unif}(0.5,1.0).
\]
The remaining entries are independently generated according to
\[
\lambda_{jk} \sim 0.2N(0,1),
\qquad j=K+1,\ldots,J,\;\; k=1,\ldots,K.
\]
Finally, we set $\Sigma = I_K$.
Consider $N$ LLM families, with each family $l$ containing $n_l$ LLMs.
Recall we assume that each $i$-th LLM in the $l$-th family has a $K$-dimensional latent ability $\theta_i^{(l)}$, given as
\begin{equation}\label{eq:model.specification}
    \theta_i^{(l)} ~=\;\alpha_l ~+~ \beta^\intercal x_i^{(l)},
\end{equation}
where $x_i^{(l)}\in \RR^p$ is the observed feature vector of LLM $i$ in family $l$.
We generate both $x_i^{(l)}$ and $\alpha_l$ independently from a standard normal distribution.
For $J$ benchmark tests, we use a beta distribution to model 
the response of LLM $i$ in the $l$-th family to the benchmark $j$, denoted by $Y_{ij}^{(l)}\in[0,1]$.
 Given this $K$-dimensional latent ability $\theta_i^{(l)} = \alpha_l + \beta^\T x_i^{(l)}$, we generate $Y_{ij}^{(l)}$ from the Beta distribution as in the setup of Section~\ref{subsec_setup}.

We vary the number of LLM families $N$ from $100$ to $200$ and the number of benchmarks $J\in\{8, 10\}$. The latent dimension is fixed at $K=2$, and the covariate dimension is set to $p=3$ to align with the settings considered in the real data application section.
Moreover, based on our empirical analysis of LLM leaderboard data, we observe that each LLM family contains at least one model and at most eight models. Motivated by this observation, we independently generate the family sizes $n_l$ from the discrete uniform distribution on $\{1,\ldots,8\}$.

\begin{table}[htbp]
\caption{{Pointwise coverage probabilities for entries of $\beta$ under $K = 2$, $N \in \{ 100, 150,200\}$ and $J \in \{8,10 \}$. The entry of $(\mathrm{Row}r,\mathrm{Column}k)$ corresponds to the empirical coverage probability of $\beta_{rk}$.}}
\label{tab:beta_pointwise_coverage}
\begin{tabular}{llrrrrrr}
\toprule
 &  & \multicolumn{2}{c}{$N = 100$} & \multicolumn{2}{c}{$N = 150$} & \multicolumn{2}{c}{$N = 200$} \\
 &  & Column1 & Column2 & Column1 & Column2 & Column1 & Column2 \\
\midrule
$J = 8$  & Row1 & 1.000 & 0.985 & 0.990 & 0.975 & 0.990 & 0.980 \\
         & Row2 & 0.995 & 0.995 & 0.985 & 0.985 & 0.980 & 0.955 \\
         & Row3 & 0.995 & 0.975 & 0.985 & 0.990 & 0.985 & 0.975 \\
\midrule
$J = 10$ & Row1 & 0.969 & 0.969 & 0.985 & 0.974 & 0.960 & 0.940 \\
         & Row2 & 0.949 & 0.959 & 0.954 & 0.929 & 0.910 & 0.930 \\
         & Row3 & 1.000 & 0.980 & 0.985 & 0.974 & 0.975 & 0.950 \\
\bottomrule
\end{tabular}
\end{table}

Next, we compute empirical coverage probabilities to evaluate the finite-sample performance of the proposed inference procedure. The target parameters include the coefficient matrix $\beta$, the loading matrix $\Lambda$, and the intercept vector $b$. 
For each parameter, we obtain the corresponding estimator by solving Equation (6) in the main manuscript and construct elementwise $95\%$ confidence intervals based on Theorem 2. The empirical coverage probabilities are then computed over $200$ repetitions.

\begin{table}[htbp]
\caption{{Pointwise coverage probabilities for entries of $\Lambda$ under $K = 2$, $N \in \{ 100, 150,200\}$ and $J \in \{8,10 \}$. The entry of $(\mathrm{Row}j,\mathrm{Column}k)$ corresponds to the empirical coverage probability of $\lambda_{jk}$. The empty cells correspond to the entries in $\Lambda$ that were fixed at true value $0$ throughout the optimization routine.}}
\label{tab:lambda_pointwise_coverage}
\begin{tabular}{llrrrrrr}
\toprule
 &  & \multicolumn{2}{c}{$N = 100$} & \multicolumn{2}{c}{$N = 150$} & \multicolumn{2}{c}{$N = 200$} \\
 &  & Column1 & Column2 & Column1 & Column2 & Column1 & Column2 \\
\midrule
$J = 8$  & Row1 & 0.950 & --    & 0.920 & --    & 0.800 & --    \\
         & Row2 & --    & 0.985 & --    & 0.940 & --    & 0.940 \\
         & Row3 & 0.985 & 0.985 & 0.995 & 0.985 & 0.975 & 0.975 \\
         & Row4 & 1.000 & 0.990 & 1.000 & 0.990 & 0.995 & 0.985 \\
         & Row5 & 1.000 & 0.970 & 0.995 & 0.970 & 0.995 & 0.960 \\
         & Row6 & 1.000 & 1.000 & 1.000 & 0.995 & 0.990 & 1.000 \\
         & Row7 & 0.985 & 0.980 & 0.975 & 0.965 & 0.925 & 0.945 \\
         & Row8 & 1.000 & 1.000 & 1.000 & 0.985 & 1.000 & 0.975 \\
\midrule
$J = 10$ & Row1  & 0.939 & --    & 0.934 & --    & 0.874 & --    \\
         & Row2  & --    & 0.918 & --    & 0.852 & --    & 0.724 \\
         & Row3  & 0.980 & 0.980 & 0.949 & 0.944 & 0.910 & 0.925 \\
         & Row4  & 0.898 & 0.944 & 0.872 & 0.944 & 0.724 & 0.894 \\
         & Row5  & 0.990 & 0.990 & 0.985 & 0.969 & 0.985 & 0.950 \\
         & Row6  & 0.964 & 0.969 & 0.964 & 0.980 & 0.965 & 0.965 \\
         & Row7  & 0.964 & 0.959 & 0.939 & 0.929 & 0.854 & 0.910 \\
         & Row8  & 0.995 & 1.000 & 0.980 & 0.995 & 0.985 & 0.985 \\
         & Row9  & 0.918 & 0.898 & 0.908 & 0.832 & 0.879 & 0.749 \\
         & Row10 & 0.944 & 0.944 & 0.929 & 0.918 & 0.879 & 0.859 \\
\bottomrule
\end{tabular}
\end{table}

Tables~\ref{tab:beta_pointwise_coverage}--\ref{tab:b_pointwise_coverage} report the empirical pointwise coverage probabilities for $\beta\in \RR^{p\times K}$, $\Lambda\in \RR^{J\times K}$, and $b\in \RR^J$, respectively. The results in Table~\ref{tab:beta_pointwise_coverage} show that the empirical coverage probabilities are generally close to the nominal level of $95\%$ for both $J = 8$ and $J = 10$, even under moderate sample sizes with $N$ on the order of $10^2$. 
However, we also also observe that for some entries, especially when $N = 100$, the coverage probabilities are slightly above $95\%$, suggesting that the standard deviations may be overestimated in smaller samples, resulting in conservative confidence intervals.
This overestimation becomes less obvious as the sample size $N$ increases.

Tables~\ref{tab:lambda_pointwise_coverage} and~\ref{tab:b_pointwise_coverage} present the corresponding results for $\Lambda$ and $b$, respectively. The finite-sample performance of their corresponding estimators is slightly less stable than that of $\beta$. Nevertheless, most entries still attain empirical coverage probabilities close to the nominal $95\%$ level, although a few entries exhibit either over-coverage or under-coverage in finite samples.
This is possibly due to the increased computational complexity induced by the more complex model settings. 

\begin{table}[htbp]
\caption{{Pointwise coverage probabilities for entries of $b$ under $K = 2$, $N \in \{ 100, 150,200\}$ and $J \in \{8,10 \}$. The entry of $\mathrm{Row}j$ corresponds to the empirical coverage probability of $b_j$.}}
\label{tab:b_pointwise_coverage}
\begin{tabular}{lrrrrrr}
\toprule
 & \multicolumn{3}{c}{$J = 8$} & \multicolumn{3}{c}{$J = 10$} \\
 & $N = 100$ & $N = 150$ & $N = 200$ 
 & $N = 100$ & $N = 150$ & $N = 200$ \\
\midrule
Row1  & 0.950 & 0.895 & 0.790 & 0.939 & 0.959 & 0.930 \\
Row2  & 0.970 & 0.960 & 0.975 & 0.944 & 0.872 & 0.749 \\
Row3  & 0.985 & 0.955 & 0.985 & 0.995 & 0.964 & 0.950 \\
Row4  & 0.930 & 0.795 & 0.745 & 0.898 & 0.730 & 0.533 \\
Row5  & 0.990 & 0.960 & 0.940 & 0.995 & 1.000 & 0.935 \\
Row6  & 0.990 & 0.975 & 0.960 & 1.000 & 0.985 & 0.980 \\
Row7  & 0.975 & 0.920 & 0.870 & 0.964 & 0.923 & 0.829 \\
Row8  & 0.975 & 0.945 & 0.890 & 0.954 & 0.867 & 0.739 \\
Row9  & --    & --    & --    & 0.995 & 0.949 & 0.965 \\
Row10 & --    & --    & --    & 0.990 & 0.898 & 0.784 \\
\bottomrule
\end{tabular}
\end{table}

 Overall, the aforementioned results indicate that, when the number of LLM families $N$ is on the order of $10^2$, the proposed asymptotic standard errors and confidence intervals provide reliable inference for the model parameters, particularly for the primary parameter of interest, $\beta$.

}

\end{document}